\documentclass[aps,prl,longbibliography,twocolumn,showpacs,noeprint]{revtex4-2}

\usepackage{CJK}

\usepackage{graphicx}
\usepackage{amsmath}
\usepackage{dcolumn}
\usepackage{balance}

\usepackage{bm}
\usepackage[normalem]{ulem}
\usepackage{color}
\usepackage{hyperref}

\usepackage{epstopdf}
\newcommand{\bew}{\begin{widetext}}
\newcommand{\ew}{\end{widetext}}

\newcommand{\ii}{{\rm i}}
\newcommand{\bx}{\mathbf{x}}

\newcommand{\bp}{\mathbf{p}}
\newcommand{\bq}{\mathbf{q}}

\newcommand{\br}{\mathbf{r}}

\newcommand{\bff}{\mathbf{f}}

\newcommand{\bk}{\mathbf{k}}

\newcommand{\bh}{\mathbf{h}}
\newcommand{\bg}{\mathbf{g}}

\newcommand{\sep}{ \ \ \ , \ \ \ }

\newcommand{\beq}{\begin{equation}}
\newcommand{\eeq}{\end{equation}}
\newcommand{\beqn}{\begin{eqnarray}}
\newcommand{\eeqn}{\end{eqnarray}}
\newcommand{\pp}{\partial}
\newcommand{\dd}{{\rm d}}

\newcommand{\cO}{{\cal O}}

\newcommand{\la}{\langle}

\newcommand{\ra}{\rangle}

\newcommand{\vnab}{{\bf \nabla}}

\newcommand{\bl}{{\bf l}}

\allowdisplaybreaks[1]
\usepackage{amsfonts}

\usepackage{placeins}
\usepackage{float}
\usepackage[percent]{overpic}

\begin{document}
\title{New universality classes govern the critical and multicritical behavior of an active Ising model}
\author{Matthew Wong}
\email{matthew.wong20@imperial.ac.uk}
\address{Department of Bioengineering, Imperial College London, South Kensington Campus, London SW7 2AZ, U.K.}
\author{Chiu Fan Lee}
\email{c.lee@imperial.ac.uk}
\address{Department of Bioengineering, Imperial College London, South Kensington Campus, London SW7 2AZ, U.K.}
\date{\today}

	\begin{abstract}
The Ising model is one of the most well known models in statistical physics, with its critical behavior governed by 
the Wilson-Fisher universality class (UC). When active motility is incorporated into the Ising model by, e.g., dictating that the spins' directional movements follow their orientations, the spin number density necessarily constitutes a soft mode in the hydrodynamic description, and can therefore modify the scaling behavior of the system. Here, we show that this is indeed the case in a critical active Ising model  in which density can impede the system's collective motion. Specifically, we use  a perturbative  dynamic renormalization group method to the one-loop level to uncover three new UCs, one of which supersedes the Wilson-Fisher UC to become the generic UC that governs the critical behavior of the active Ising model.
	
	\end{abstract}

\maketitle

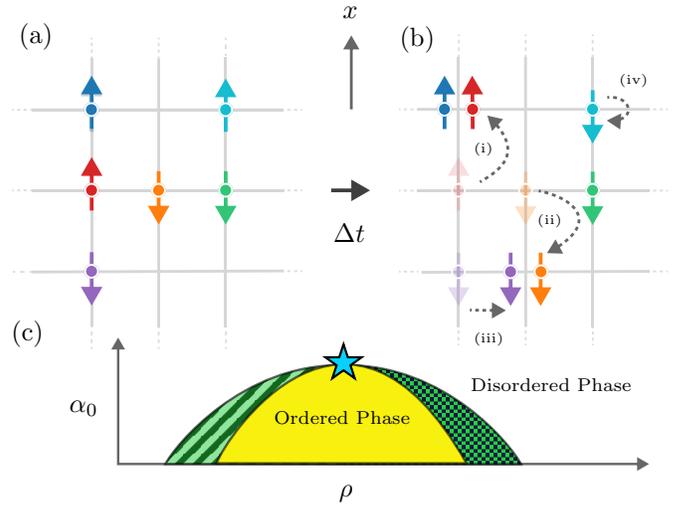
\begin{figure}
\hspace{-0cm}
\begin{picture}(0,0)
\put(-122,-4.5){\normalsize $(\rm a)$}
\put(22,-5.55){\normalsize $\rm(b)$}
\put(0.5,5){\normalsize $x$}
\put(-3,-80){\normalsize $\Delta t$}
\put(50,-46){\tiny$\rm (i)$}
\put(50,-118){\tiny$\rm (iii)$}
\put(74,-73){\tiny $\rm (ii)$}
\put(105,-20){\tiny$\rm (iv)$}
\end{picture}
\includegraphics[width=0.49\textwidth]{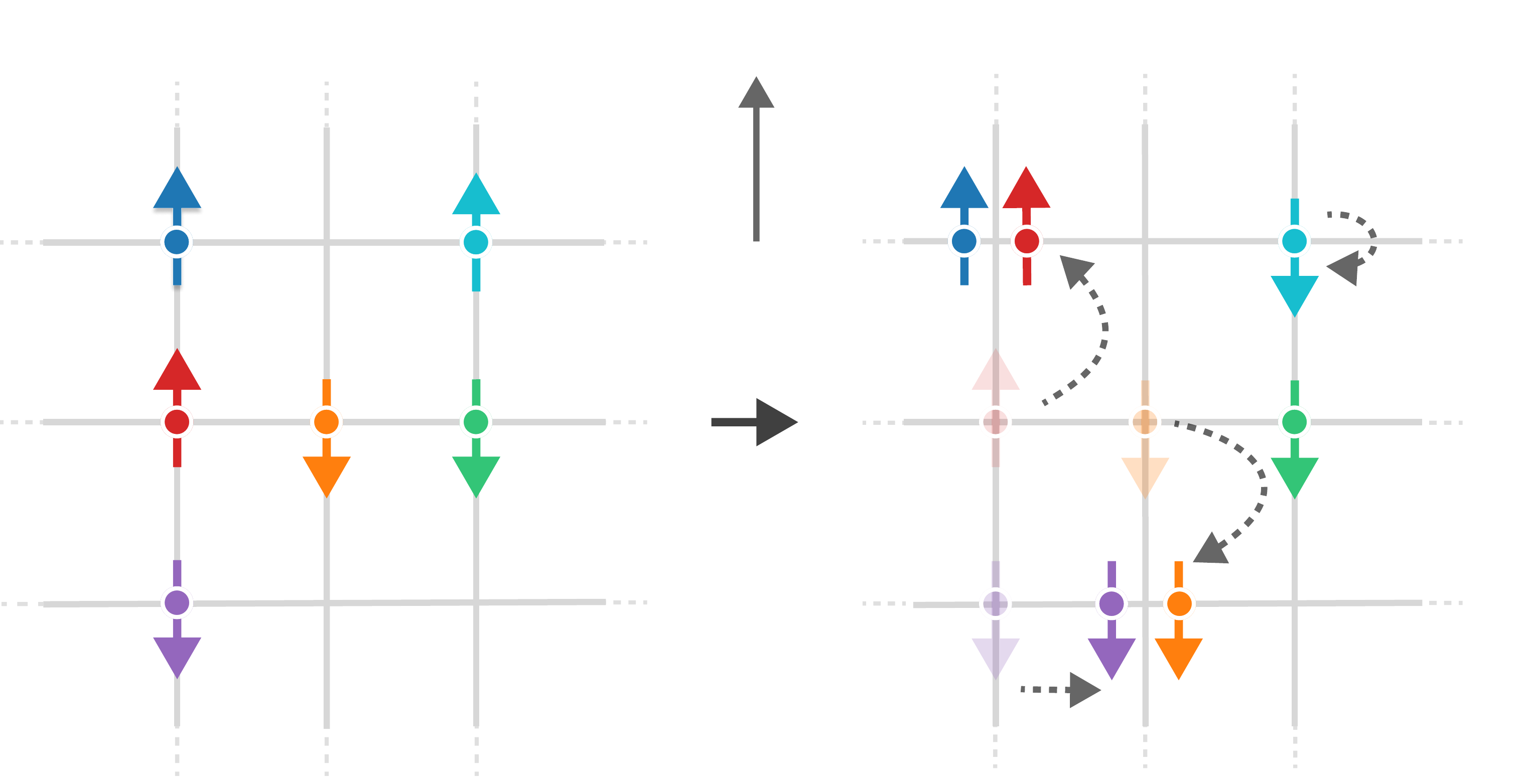}\\[-1.5ex]
\hspace{1.0cm}
\vspace{0.7cm}
\begin{picture}(0,0)
\put(-34,50){\normalsize $\rm (c)$}
\put(-12,23){\normalsize $\alpha_0$}
\put(90,-9){\normalsize $\rho$}
\end{picture}
\begin{overpic}[width=0.4\textwidth,percent]{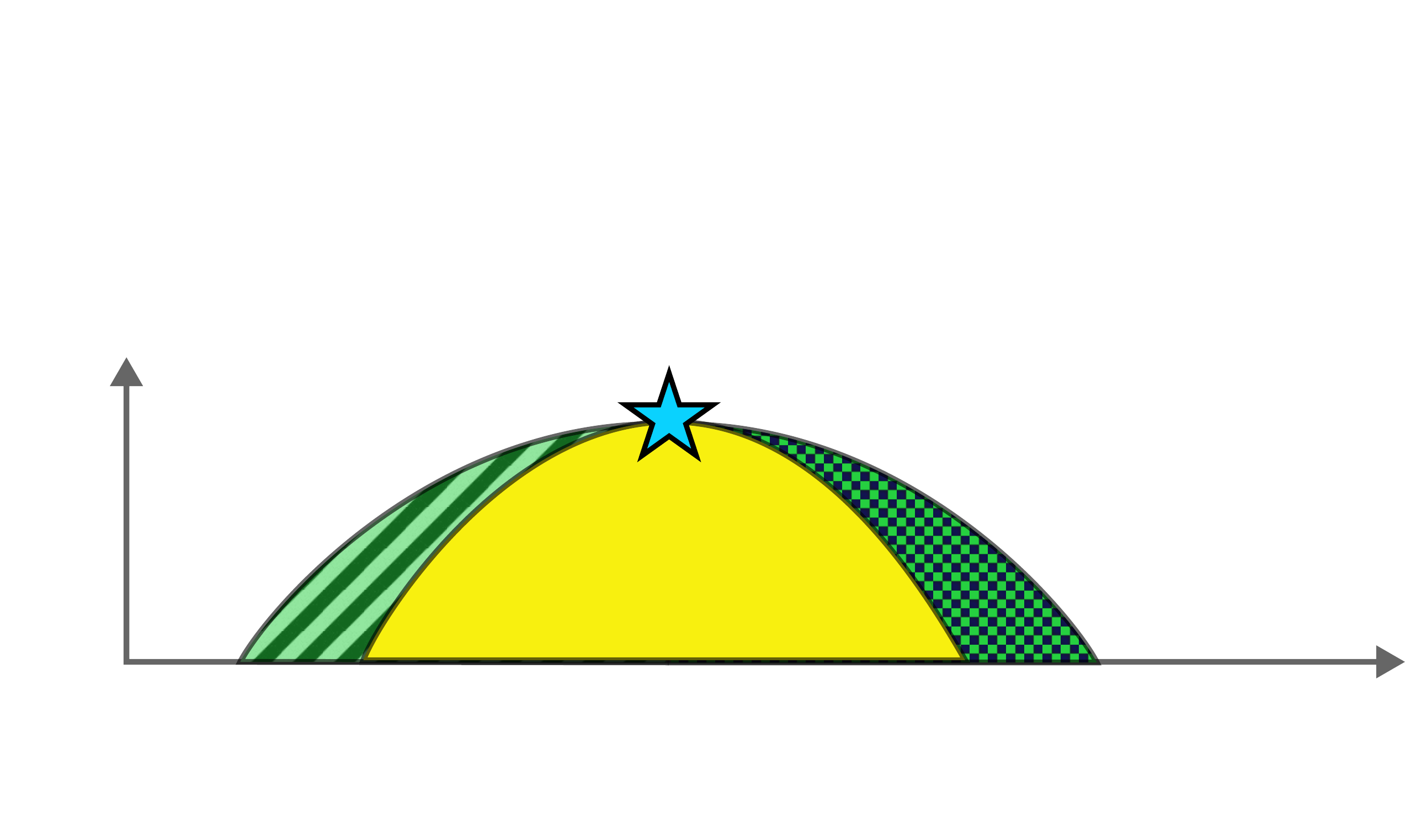}
\put(30.5,9){\scriptsize Ordered Phase}
\put(67,15){\scriptsize Disordered Phase}
\end{overpic}
\vspace{-0.3cm}
\caption{{\it A critical active Ising model.} (a) \& (b) In our active Ising model (AIM), spins' directions preferentially align with the vertical $x$-axis and they dictate the spins' directional motion (depicted in (i) \& (ii)). 
However, fluctuations can modify spins' directions, leading to the spins moving sideways (iii), and spin-flips (iv). Further, there is an disorientation of the spins' alignment  when the spin number becomes high, leading to the cease of collective movement (not shown).
(c) The phase diagram, in terms of $\rho$ and $\alpha_0$ (see  Eqs.~(\ref{eq:cont})--(\ref{eq:alpha})),  derived by using a Toner-Tu (TT) model with an easy-axis to describe the hydrodynamic behavior of the AIM  \cite{nesbitt_njp21,bertrand_prr22}. A critical point (blue star) emerges where the banding co-existence phase (green striped region) meets the re-verse banding co-existence phase (green meshed region) \cite{nesbitt_njp21}. Specifically, the critical point is attained by fine tuning the model parameters $\alpha_0$ and $\alpha_1$ to zero. 
}
\label{fig:cartoon}
\end{figure}

In statistical mechanics,  it could be argued that no critical phenomenon is more familiar than the critical behaviour of the equilibrium Ising model, which is governed by the Wilson-Fisher universality class (UC) \cite{wilson_prl72}. However, the living world extends itself out of equilibrium, with active motility being one of the most prominent manifestations of a plethora of non-equilibrium phenomena \cite{marchetti_rmp13,julicher_rpp18}. Active motility  can be easily incorporated into the Ising model by dictating that each individual spin's directional motion follows their spin's orientation (see Fig.~\ref{fig:cartoon}(a,b)). Indeed, the coupling between the spin's orientational space and the real space underlies much of the novel emergent physics uncovered in this type of active Ising models (AIMs) \cite{solon_prl13,solon_pre15,scandolo_epje23,bandyopadhyay_pre24}. However, in AIMs with a fixed number of moving spins, the spin number density field constitutes a soft (or gapless/massless) mode and is thus a crucial ingredient in the hydrodynamic description of the systems. Indeed, in the Vicsek simulation model \cite{vicsek_prl95}--an archetypal microscopic model of active matter systems, the presence of the density field eliminates the expected critical order-disorder transition by rendering the transition discontinuous \cite{gregoire_prl04,chate_pre08,bertin_pre06}. Fortunately, going beyond the Vicsek model, it was later realized that a critical order-disorder transition naturally emerges in active systems in which collective motion ceases when the density becomes high \cite{nesbitt_njp21,bertrand_prr22,jentsch_prr23}, as manifested
in, e.g., systems of motile agents with contact inhibition
of locomotion \cite{schnyder_scirep17}. The presence of such a ``rescued" critical transition is most naturally demonstrated using the Toner-Tu (TT) model \cite{toner_prl95,toner_pre98, toner_pre12,chen_a25}, which generically describes any polar active fluids.

Returning to our focus of active Ising models (AIMs), the corresponding hydrodynamic description is  most straightforwardly formulated using the TT model with an easy axis \cite{hohenberg_rmp77}, i.e., a spin's orientation preferentially aligns with a pre-determined direction. As such, the analysis on the existence of a critical order-disorder transition for the TT model \cite{nesbitt_njp21,bertrand_prr22} applies directly to the AIMs as well (see, e.g., \cite{agranov_njp24}). Fig.~\ref{fig:cartoon}(c) shows a typical phase diagram of the system around the critical region.

If the density field is ignored, the order-disorder critical transition in the AIM was found to be governed by the usual Wilson-Fisher UC \cite{bassler_prl94}. However, what impact the presence of the density field has remains unknown. Here, we solve this problem by using a dynamic renormalisation group (DRG) analysis together with the $\epsilon$-expansion method (to order $\epsilon$), and find that the Wilson-Fisher UC no longer describes the AIM at criticality. In fact, we find three new RG fixed points (FPs) (hence three new UCs), with one of these novel fixed points generically governing the critical behavior.

{\it Active Ising Model---}Just like interpreting the equilibrium Ising model as an $O(n)$ model with an easy axis (i.e., spins' directions tend to align along the easy-axis) \cite{hohenberg_rmp77}, we formulate our AIM by adding an easy-axis to the standard TT model in $d$ dimensions. 
By choosing the easy axis to be, without loss of generality, along the $x$-axis, and denoting the mass density field by  $\rho$ and the momentum density field by $\bg$ ($=g_x \hat{\bx} +\bg_\perp$), which  aligns with the local coarse-grained spin orientation, the equation of motions (EOM) of our AIM are:
\begin{align}
\label{eq:cont}
\partial_t \rho&+ \nabla \cdot \bg = 0 \ ,
\\
\label{eq:mom}
\partial_t g_x &+\frac{\lambda}{2} \pp_x g_x^2
+ \lambda_1
 (\bg_\perp \cdot \nabla_\perp ) g_x + \lambda_2 g_x (\nabla_\perp \cdot \bg_\perp)
 \\
\nonumber  
 & +\frac{\lambda_3}{2} \pp_x (|\bg_\perp|^2) =\left[-\alpha(\rho) -\beta g_x^2\right]g_x 
 \\
\nonumber 
& +\left(\mu_1 \pp_x^2 + \mu_1'\nabla_\perp^2 \right) g_x
 + \left[\pp_x \mu_2 (\pp_x g_x)+ \pp_x \mu_2' (\nabla_\perp \cdot \bg_\perp)\right]  
 \\
\nonumber 
& - \pp_x P^x(\rho) +\mathrm{h.o.t.}+f_x  \ ,
\\
\label{eq:mom2}
\partial_t \bg_\perp &=-\Gamma \bg_\perp  - { \nabla_\perp P^\perp(\rho)} +\mathrm{h.o.t.}+\bff^\perp  \ ,
\end{align}
where $\alpha$ and   $P^{\theta}$ ($\theta=x$ or $\perp$) are scalar functions of $\rho$:
\beq
\label{eq:alpha}
\alpha =\sum_{n\geq 0} \alpha_n (\rho-\rho_0)^n \sep
P^{\theta} =\sum_{n\geq 1} \kappa_n^{\theta} (\rho-\rho_0)^n  \ ,
\eeq
with $\rho_0$ being the mean density, and the noise term $\bff$ is Gaussian with vanishing mean and statistics:
\beqn
\langle f_x(\br,t) f_x(\br^\prime,t^\prime)  \rangle &=& 2D\delta(t-t^\prime)\delta^d(\br-\br^\prime)\ ,
\\
\langle f^\perp_i(\br,t) f^\perp_j(\br^\prime,t^\prime)  \rangle &=& 2D_\perp\delta^\perp_{ij}\delta(t-t^\prime)\delta^d(\br-\br^\prime)
 \ .
\eeqn
All other coefficients in the EOM are independent of the hydrodynamic fields, and ``h.o.t." denote higher-order terms that do not affect the hydrodynamic behavior of the system. Further, the super/sub-script ``$\perp$" generically denotes directions orthogonal to $\hat{\bx}$, e.g., $\nabla_\perp = \pp_y \hat{\bf y}+ \pp_z \hat{\bf z}$ for $d=3$ where $d$ denotes the spatial dimensions. 

Note that Eq.~(\ref{eq:cont}) models the first key ingredient of our AIM: the directional motion of a local mass element is dictated by its spin orientation.  The second key ingredient of our AIM is the existence of an easy axis, which is modeled by the damping term $(-\Gamma \bg_\perp)$ in Eq.~(\ref{eq:mom2}). 

We can now use Eq.~(\ref{eq:mom2}) to immediately solve for $\bg_\perp$:
\beq
\bg_\perp \simeq \Gamma^{-1} \left[- { \nabla_\perp P^\perp} +\mathrm{h.o.t.}+\bff^\perp\right] \ ,
\eeq
which we can then substitute into  Eqs.~(\ref{eq:cont}) \& (\ref{eq:mom}) to eliminate $\bg_\perp$, leading to our refined EOM:
\begin{align}
\label{eq:rho}
\partial_t \rho&= -\gamma \pp_x \phi +K \nabla^2_\perp \rho \ ,
\\\nonumber
\partial_t\phi&+ \frac{\lambda}{2}\partial_x \phi^2 =-\kappa_1 \pp_x \rho -\kappa_2 \pp_x \rho^2 + \mu_x\partial_x^2\phi + \mu_\bot\vnab_\bot^2\phi    
\\
&+\left(-\alpha_0-\alpha_1 \rho -\alpha_2 \rho^2 -\beta \phi^2\right)\phi +f_x
 \ ,
\label{eq:x}
\end{align}
where we have only shown the relevant terms (to be justified  {\it a posteriori}). To ease notation, we have also  made the following redefinitions: $\rho-\rho_0 \mapsto\rho $, $\phi=g_x$, $\mu_x=\mu_1+\mu_2$, $\mu_\perp=\mu_1'+\mu_2'$, $\kappa_{1,2} = \kappa^x_{1,2}$, and $K= \kappa^\perp_1 /\Gamma$.  Further, we have introduced the dimensionless parameter $\gamma$ in Eq.~(\ref{eq:rho}) for a more transparent  rescaling of our EOM later.

    \begin{table*}[t]
\renewcommand{\arraystretch}{1.2}
\begin{tabular}
{p{0.12\textwidth} p{0.1\textwidth} p{0.2\textwidth} p{0.1\textwidth}
p{0.13\textwidth} p{0.13\textwidth} p{0.08\textwidth}}
\hline
FP & $N_{\rm unst}$ & $[g_\beta^*, g_\lambda^*, g_{\kappa_2}^*, g_{\alpha_2}^*]$  & $z$ & $\zeta$ &  $ \chi_\phi/\chi_{\rho}$ & $ y_{\alpha_0}$ \\
\hline
\hline
  $\rm I$ \raisebox{-.2ex}{%
  \includegraphics[height=1.9ex]{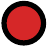}%
}  & $0$ & $\epsilon\left[\frac{2}{9}, 0, 0, \frac{2}{3}\right]$
&2
& $1$ & $-1+\frac{1}{2}\epsilon$  & $2-\frac{1}{3}\epsilon$ \\
  $\rm II$ \raisebox{-.2ex}{%
  \includegraphics[height=2ex]{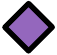}%
} & $1
$ & $\epsilon\left[\frac{4  }{9}, 0, 0,0\right]$ &2 & $1$ & $-1+\frac{1}{2}\epsilon$ & $2-\frac{1}{3}\epsilon$\\
  $\rm III$ \raisebox{-.2ex}{%
  \includegraphics[height=1.8ex]{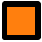}%
} & $2
$ & $\epsilon\left[0, \frac{16}{9}, 0,\frac{2}{3}\right]$ &2& $1+\frac{1}{6}\epsilon$ & $-1+\frac{5}{12}\epsilon$ &  $2-\frac{1}{6}\epsilon$\\
$\rm IV$ \raisebox{-.2ex}{%
  \includegraphics[height=2ex]{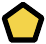}%
} & $3
$ & $\epsilon\left[0, \frac{32}{9}, 0, 0\right]$ &2 & $1+\frac{1}{3}\epsilon$ & $-1+\frac{1}{3}\epsilon$ & $2$\\
 $\rm V$ \raisebox{-.2ex}{%
  \includegraphics[height=2ex]{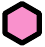}%
} & $3
$ & $\epsilon\left[0, 0, 0,\frac{4}{5}\right]$ &2& $1$ & $-1+\frac{1}{2}\epsilon$ & $2-\frac{1}{5}\epsilon$\\
 VI \raisebox{-.2ex}{%
  \includegraphics[height=1.8ex]{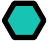}%
} & 3 & [0,0,0,0] &2& 1 & $-1+\frac{1}{2}\epsilon$ & 2 \\
\hline
\end{tabular}
\caption{{\it RG FPs, their stabilities, locations, \& critical exponents}.  Six FPs are obtained in our DRG calculation to order $\epsilon$. The symbols next to the Roman numerals are used in Fig.~\ref{fig:RGflow} to depict their locations schematically. The integral quantity $N_{\rm unst}$ denotes the unstable direction of each FP within the critical manifold ($\alpha_0=\alpha_1=0$). The FP locations are shown by the $g^*$'s.  The subsequent  4 columns show the numerical values of the critical exponents  to order $\epsilon$: $z$ (dynamic), $\zeta$ (anisotropic), $\chi_\phi/ \chi_\rho$ (roughness exponents for $\phi$ and $\rho$, respectively), and $y_{\alpha_0}$ (the exponent associated to the divergence of $\alpha_0$ under RG transformation).
As expected, we recover the Wilson-Fisher FP (II), the Hwa-Kardar FP (IV) first elucidated in the context of self-organized criticality \cite{hwa_prl89,hwa_pra92}, and the Gaussian FP (VI). As far as we are aware, the remaining 3 FPs are new. 
}
\label{tab}
\end{table*}

{\it Criticality---}At the mean-field level, the critical transition occurs when $\alpha_0 =0$. However, when the density mode is present, such a transition is generically pre-empted by a discontinuous transition that is manifested by the so-called  banding regime \cite{bertin_pre06}. But all is not lost: recent work has demonstrated that the critical transition can be rescued by setting {\it both} $\alpha_0$ and $\alpha_1$ in Eq.~(\ref{eq:x}) to zero \cite{nesbitt_njp21}. Namely, criticality of our AIM can be achieved by fine tuning two model parameters. A schematic phase diagram of such a scenario is depicted in Fig.~\ref{fig:cartoon}(c). We will now use a perturbative DRG method to study the universal behavior of this AIM at criticality.

{\it Upper critical dimension \& relevant terms---}For our DRG analysis, we first need to identify the 1) upper critical dimension $d_c$, and 2) the relevant nonlinear terms in the EOM close to $d_c$. To do so, we use the linearized EOM (\ref{eq:rho},\ref{eq:x}) to calculate the $\rho$-$\rho$ and $\phi$-$\phi$ correlation functions, which can be done by spatio-temporal Fourier transforming the linear EOM. We provide details of this standard procedure in \cite{kardar_b07,SM}, and just quote the results here:
\beqn \la \phi({\bf 0},0) \phi(\br,t)\ra &=&  \br_\perp^{2\chi^{\rm lin}_\phi} S_{\rm lin} \left(\frac{t}{\br_\perp^z}, \frac{x}{\br_\perp^\zeta} \right) \ ,
\\  \la \rho({\bf 0},0) \rho(\br,t)\ra &=&  \br_\perp^{2\chi^{\rm lin}_\rho} \frac{\gamma}{\kappa_1} S_{\rm lin} \left(\frac{t}{\br_\perp^z}, \frac{x}{\br_\perp^\zeta}\right) \ ,
   \eeqn
where  $S_{\rm lin}$ is the universal scaling functions at the linear level, and the values of the scaling exponents are:
\beq
\label{eq:lin}
z^{\rm lin} =2 \sep \zeta^{\rm lin} =1 \sep \chi_\rho^{\rm lin} =\chi_{\phi}^{\rm lin} =\frac{2-d}{2} \ .
\eeq

With the above scaling exponents, we can now use the standard power-counting method \cite{SM} to determine that 1) the upper critical dimension $d_c$ is 4, and 2) the relevant nonlinear terms are precisely those appearing in Eq.~(\ref{eq:x}). In particular, there are now four nonlinearities in the EOM, two of which  ($\kappa_2 \pp_x \rho^2$ and $\alpha_2 \rho^2 \phi$) result from the incorporation of the density field which, as we shall see, are crucial to the correct description of the critical behavior of the AIM.

{\it DRG analysis---}We now perform a DRG analysis together with the $\epsilon$-perturbation method, where $\epsilon=d_c-d = 4-d$. A RG transformation amounts to averaging over the fluctuations at the short wavelength regime and then incorporating their effects on the coefficients in the rescaled EOM \cite{forster_pra77,toner_b24}. 
As detailed in \cite{SM}, the fact that that both $\gamma$ and $\kappa_1$ are clearly divergent under the RG flow around the upper critical dimension $d_c$ (as seen, e.g., by their linear scaling dimension upon re-scaling the EOMs) helps to simplify our calculation. Indeed, a similar divergence has been successfully exploited by a recent study of the Vicsek flocking phase \cite{jentsch_prl24}.

We provide full details of our 1-loop calculation in \cite{SM}, and just quote the resulting RG flow equations at the one-loop ($\cO(\epsilon)$) level and at criticality (i.e., $\alpha_0=\alpha_1=0$):
\beqn
\frac{ \dd \ln \lambda}{\dd \ell} &=& z -\zeta + \chi_\phi- \frac{15}{8}g_{\beta}-\frac{3}{2}\frac{g_{\kappa_2}g_\beta}{g_\lambda}-\frac{3}{8}g_{\alpha_2}\ ,
\\
\frac{ \dd \ln \gamma}{\dd \ell} &=& z - \zeta + \chi_{\phi} -\chi_\rho\ ,
\\
\frac{ \dd \ln \kappa_{1}}{\dd \ell} &=& z - \zeta - \chi_\phi+ \chi_{\rho}\ ,
\\
\frac{ \dd \ln \kappa_{2}}{\dd \ell} &=& z- \zeta - \chi_\phi + 2\chi_{\rho}- \frac{1}{8}\frac{g_{\alpha_2}g_\beta}{g_{\kappa_2}} - \frac{3}{8}g_{\alpha_2}\ ,
\\
\frac{ \dd \ln \mu_x}{\dd \ell} &=& z- 2\zeta + \frac{3}{16}g_{\lambda} + \frac{3}{8}g_{\kappa_2}\ ,
\\
\frac{ \dd \ln \mu_\perp}{\dd \ell} &=& z-2\ , 
\\
\frac{ \dd \ln \alpha_2}{\dd \ell} &=& z + 2\chi_{\rho} - \frac{3}{4}g_\beta - \frac{5}{4}g_{\alpha_2}\ ,
\\
\frac{ \dd \ln \beta}{\dd \ell} &=& z+2\chi_\phi- \frac{9}{4}g_\beta - \frac{3}{4}g_{\alpha_2}\ ,
\\
\frac{ \dd \ln D}{\dd \ell} &=& z - \zeta- 2\chi_\phi - d+1\ ,
\\
\frac{ \dd \ln K}{\dd \ell} &=& z-2\ ,
\eeqn
where $\ell$ now corresponds to the level of ``coarse-graining", i.e., the larger the $\ell$, the more averaging has been done over the short wavelength regime. Further, we have defined the following  dimensionless coefficients:
\begin{align}
&g_\beta = \frac{\beta DS_{d-1}}{\mu_x^{1/2}\mu_\bot^{3/2}}\Lambda^{d-4}
\ \ , \ \ 
 g_\lambda=\frac{\lambda^2 DS_{d-1}}{\mu_x^{3/2}\mu_\bot^{3/2}}\Lambda^{d-4}\ ,
\\
&g_{\kappa_2}=\frac{\gamma\kappa_2\lambda DS_{d-1}}{\kappa_1\mu_x^{3/2}\mu_\bot^{3/2}}\Lambda^{d-4}
\ \ , \ \ 
 g_{\alpha_2}=\frac{\gamma\alpha_2 DS_{d-1}}{\kappa_1\mu_x^{1/2}\mu_\bot^{3/2}}\Lambda^{d-4}\ .
\end{align}

The universal behavior of the system as dictated by the RG fixed points can now be elucidated by analyzing the corresponding, much simpler, RG flow equations of the dimensionless coefficients $g$'s, which are
\beqn
\frac{\dd \ln g_\beta}{\dd\ell} &=& \epsilon-\frac{9}{4}g_\beta-\frac{3}{4}g_{\alpha_2}-\frac{3}{32}g_\lambda - \frac{3}{16}g_{\kappa_2}\ ,
\\
\nonumber
\frac{\dd \ln g_\lambda}{\dd\ell} &=& \epsilon-\frac{15}{4}g_{\beta}-3\frac{g_{\kappa_2} g_\beta}{g_\lambda}-\frac{3}{4}g_{\alpha_2} - \frac{9}{32}g_\lambda
\\
&&- \frac{9}{16}g_{\kappa_2}\ ,
\\
\nonumber
\frac{\dd \ln g_{\kappa_2}}{\dd \ell} &=& \epsilon-\frac{1}{8}\frac{g_\lambda g_{\alpha_2}}{g_{\kappa_2}}-\frac{3}{4}g_{\alpha_2}-\frac{15}{8}g_\beta- \frac{3}{2}\frac{g_{\kappa_2}g_\beta}{g_\lambda}
\\
&&- \frac{9}{32}g_\lambda-\frac{9}{16}g_{\kappa_2} \ ,
\\
\frac{\dd \ln g_{\alpha_2}}{\dd \ell} &=& \epsilon-\frac{3}{4}g_\beta -\frac{5}{4}g_{\alpha_2}-\frac{3}{32}g_\lambda-\frac{3}{16}g_{\kappa_2}\ .
\eeqn

{\it RG fixed points---}Equipped with the RG flow equations above, one can elucidate the corresponding flow behavior, locations of FPs and the their stabilities. Specifically, we obtain a total of 6 FPs, as summarized in Table I. In the table, FPs are ordered by their stability, with $N_{\rm unst}$ being the number of unstable directions within the critical manifold (defined by setting both  $\alpha_0$ \& $\alpha_1$ to zero). Crucially, we find that the generic critical behavior is governed by a new UC corresponding to FP 1. By further fine tuning  $\alpha_2$ to zero,  we recover the Wilson-Fisher UC (FP II). If we now fine tune both $\kappa_2$ and $\beta$ to zero, we uncover another new UC (FP III). Finally, by fine tuning 3 parameters to zero, we recover i) the UC (FP IV) first elucidated by Hwa and Kardar in the context of self-organized criticality \cite{hwa_prl89,hwa_pra92}, ii) a new UC (FP V) when $\beta$, $\lambda$, and $\kappa_2$ are zero and (iii) the Gaussian UC (FP VI) when $\beta$, $\lambda$, and $\alpha_2$ are zero.
The associated critical exponents are shown in Table I. We note that although some of the FPs share the same numerical values at the 1-loop level, it is natural to expect that these values will deviate when the perturbative  calculation goes to a higher order in $\epsilon$, or when a nonperturbative RG method is used  \cite{dupuis_physrep21}. For instance, this is shown to be the case in a recent study of multi-critical behavior of the TT model \cite{jentsch_prr23}.

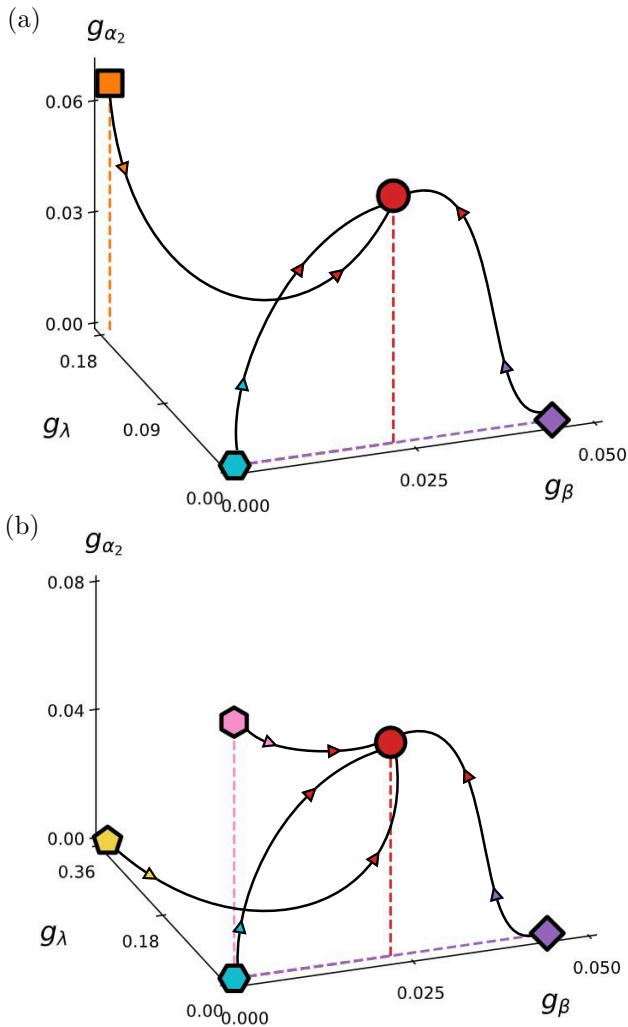
\begin{figure}[t]
   \centering
    \begin{picture}(0,0)
        \put(-10,185){\normalsize $(\rm a)$}
        \put(-11,-7){\normalsize $(\rm b)$}
    \end{picture}
    \includegraphics[width=0.9\linewidth]{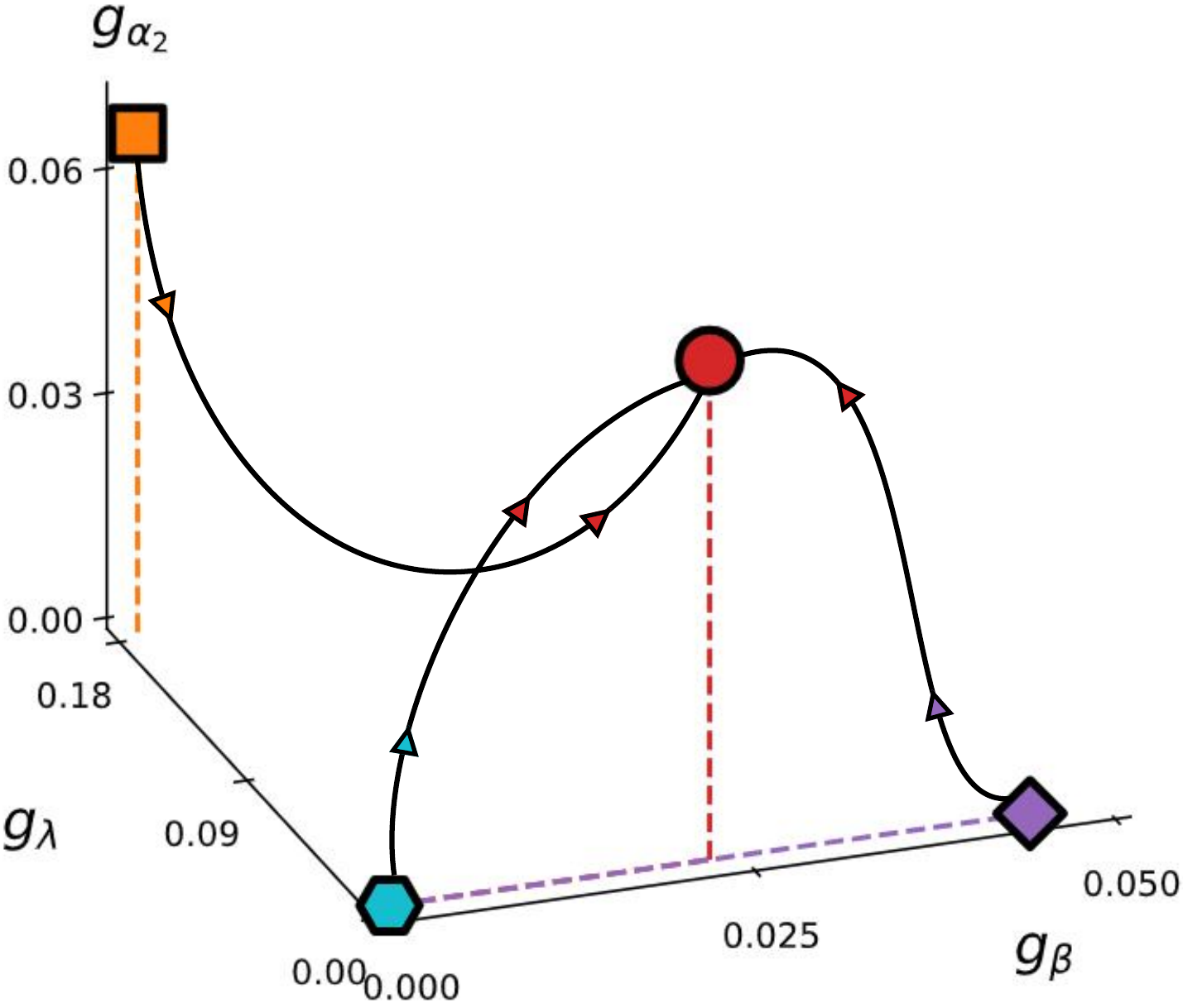}
    \\[0.5em]
    \includegraphics[width=0.9\linewidth]{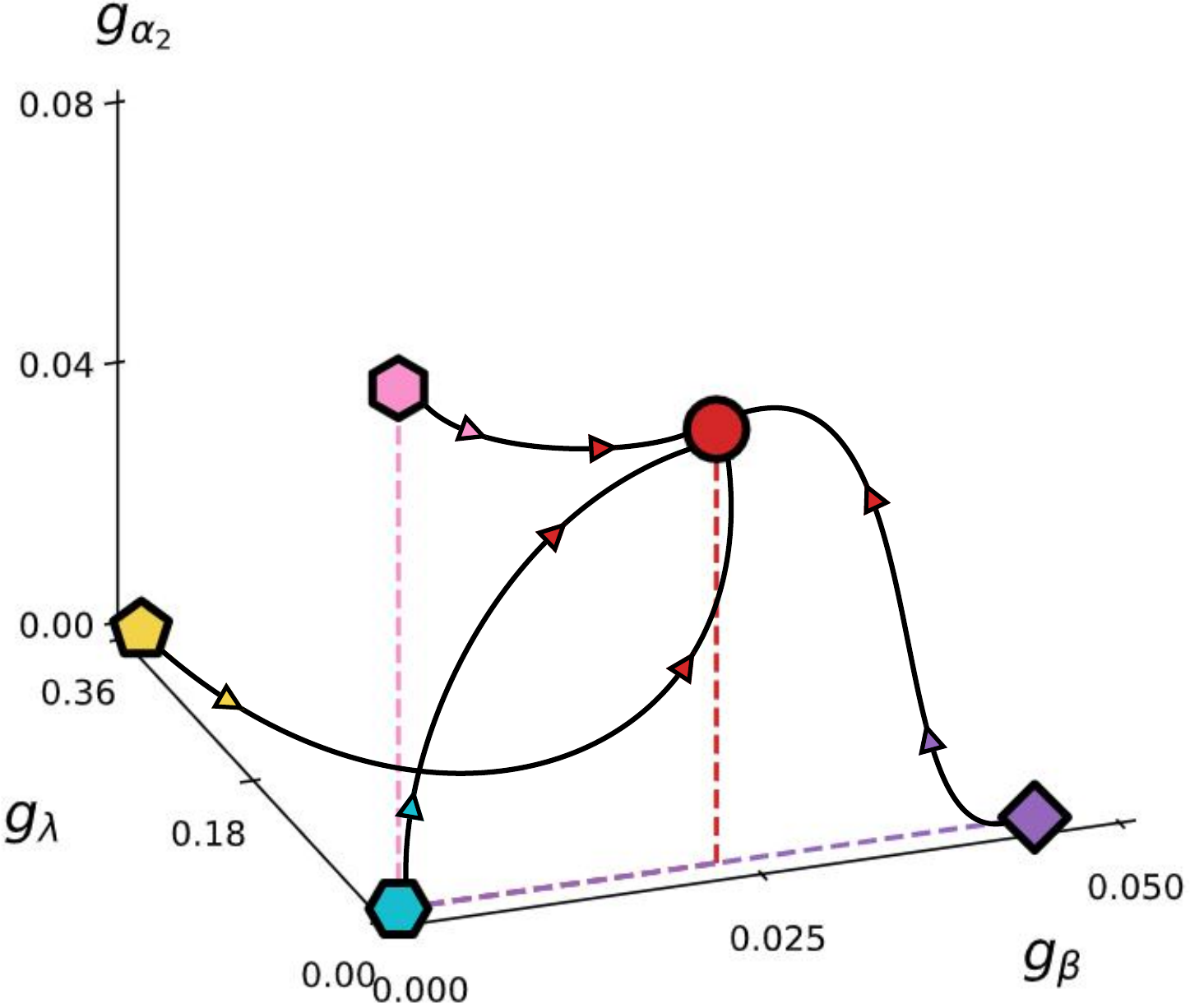}
    \\[1em]
    \caption{{\it Schematic RG flow diagrams depicting the flow relationships between the FPs ($\epsilon=0.1$).} Since there are  6 FPs (see Table I), we use  two separate figures for clarity: (a) contains the first  3 FPs together with the Gaussian FP (VI, blue hexagon), and (b) contains the last 3 FPs together with FP I (red circle) and the Wilson-Fisher FP (II, purple diamond). 
    }   
    \label{fig:RGflow}
\end{figure}

{\it Summary \& Outlook---}We have studied the critical order-disorder transition of an active Ising model (AIM), formulated as the Toner-Tu active fluid model with an easy axis. By using a perturbative DRG analysis to the 1-loop level, we have comprehensively elucidated  the diverse critical behavior around $d=4$: the upper critical dimension.
Crucially, we found that the model system contains 6 distinct universality classes, 3 of which are novel, including the one that generically describes  the critical behavior of this AIM. 
 We believe that an interesting future direction would be to verify the diverse UCs uncovered here using simulation methods. Moreover, quantitative refinement of the scaling exponents  can potentially be  performed using functional RG methods \cite{dupuis_physrep21}. The combination of these two directions will thus provide a fertile testing ground for further method development.

%

\widetext
\clearpage

\begin{center}
{\huge  {\bf Supplemental Material}}
\end{center}

\section{Active Ising model}
As discussed in the main text, our starting point is the TT model \cite{Stoner_prl95,Stoner_pre98, Stoner_pre12,Schen_a25} with an easy axis, arbitrarily chosen to be the $x$ axis, resulting in the reduced equations of motion (EOM) below:
\begin{align}
\label{eq:rho}
\partial_t \rho&= -\gamma\pp_x \phi + K \nabla^2_\perp \rho \ ,
\\
\label{eq:x}
\partial_t\phi&+ \frac{\lambda}{2}\partial_x \phi^2 =-\kappa_1 \pp_x \rho -\kappa_2 \pp_x \rho^2 + \mu_x\partial_x^2\phi + \mu_\bot\vnab_\bot^2\phi   +\left(-\alpha_0-\alpha_1 \rho -\alpha_2 \rho^2 -\beta \phi^2\right)\phi +f_x
 \ ,
\end{align}
where $\rho$ is the mass density field, $\phi$ is the $x$ component of the momentum density field, and  the noise term $f_x$ is Gaussian with vanishing mean and statistics:
\beq
\langle f_x(\br,t) f_x(\br^\prime,t^\prime)  \rangle = 2D\delta(t-t^\prime)\delta^d(\br-\br^\prime)
 \ .
\eeq

\section{Achieving criticality at the linear level}

Here, we review a liner stability analysis \cite{Snesbitt_njp21,Sbertrand_prr22} that leads us to identify the fine tuning needed to be done to achieve criticality.

By substituting the linear perturbations below 
\beq
\phi = \phi_{0} + \varepsilon\delta\phi e^{st - \ii\bp\cdot\br}
\sep\rho  = \varepsilon\rho e^{st - \ii\bp\cdot\br}\ ,
\eeq
to the linearized EOM (\ref{eq:rho},\ref{eq:x}), we have, to \(\mathcal{O}(\varepsilon)\),
\begin{align}
\label{cont}
&(s+K p_\bot^2)\rho - \ii p_x\gamma\delta\phi=0\ ,
\\
\label{mom}
&(\ii\kappa_1 p_x-\alpha_1 \phi_0)\rho-(s-\ii p_x \lambda\phi_0+\alpha_0+\mu_xp_x^2+\mu_\bot p_\bot^2)\delta \phi = 0\ .
\end{align}
Solving for the temporal eigenvalues from the above equations, we find the real parts are
\beq
{\rm Re}[s]  =
\begin{cases}
-\alpha_0 + \mathcal{O}(\bp^2)\ ,\\
\frac{\gamma p_x^2}{\alpha_0^3}\left[-\kappa_1\alpha_0^2+\lambda\alpha_0\alpha_1\phi_0^2+\gamma\alpha_1^2\phi_0^2\right]-K p_\bot^2 +\mathcal{O}(\bp^3)\ ,
\end{cases}
\eeq
where we can use $\phi_0= \sqrt{\frac{\alpha_0}{\beta}}$. 
When Re[$s]>0$, perturbations grow exponentially. The first eigenvalue branch \(-\alpha_0\) corresponds to the relaxational mode where quick relaxations occur when the momenta deviates from its mean field value \(\phi_0\) in the absence of spatial variations. The second eigenvalue highlights the hydrodynamic mode and the nature of instability in our system. To arrive at criticality, where the sign of  Re[$s$] flips, it is  necessary to tune both $\alpha_0$ and $\alpha_1$ to 0.

\section{Correlation functions at the linear level}
Our linearized EOM at criticality are
\begin{align}
\label{eq:linear}
\pp_t\rho &= -\gamma\pp_x\phi +  K\vnab_\bot^2\rho \ ,
\\
\nonumber
\partial_t\phi &= -\kappa_1\partial_{x}\rho + \mu_x\partial_x^2\phi + \mu_\bot\vnab^2_\bot\phi + f_x\ .
\end{align}

To calculate the correlation functions, we Fourier transform the above to get
\begin{align}
\label{eq:fourierlinear}
&(-\ii\omega+Kp_\bot^2)\rho +\ii
\gamma p_x\phi = 0\ ,
\\
\nonumber
&\ii\kappa_1 p_x\rho +(-\ii\omega+ \mu_x p_x^2 + \mu_\bot p_\bot^2)\phi = f_x\ ,
\end{align}

In matrix form, Eq.~(\ref{eq:fourierlinear}) reads
\beq
\begin{pmatrix}
    -\ii\omega+Kp_\bot^2 & \ii \gamma p_x\\
    \ii\kappa_1 p_x & -\ii\omega + \mu_xp_x^2 + \mu_\bot p_\bot^2 \\
\end{pmatrix}
\begin{pmatrix}
    \rho\\
    \phi\\
\end{pmatrix}=
\begin{pmatrix}
    0\\
    f_x
\end{pmatrix} 
\ .
\eeq
We now solve for $\rho$ and $\phi$ by inverting the dynamical matrix to arrive at
\beq
\begin{pmatrix}
    \rho\\
    \phi\\
\end{pmatrix}= \frac{-1}{[(\omega^2-\gamma\kappa_1 p_x^2-K p_\bot^2(\mu_xp_x^2+\mu_\bot p_\bot^2)) +\ii\omega(\mu_x p_x^2 + (\mu_\bot+K)p_\bot^2)]}
\begin{pmatrix}
    -\ii\omega+\mu_x p_x^2 + \mu_\bot p_\bot^2 & -\ii \gamma p_x \\
    -\ii\kappa_1p_x & -\ii\omega+K p_\bot^2 \\
\end{pmatrix}
\\
\nonumber
\begin{pmatrix}
   0\\
    f_x   
\end{pmatrix}
 \ .
\eeq

More specifically, we have
\beqn
\label{eq:prop}
\rho(\tilde{\bp}) &=& \frac{\ii \gamma p_x}{[(\omega^2-\gamma\kappa_1 p_x^2-Kp_\bot^2(\mu_xp_x^2+\mu_\bot p_\bot^2)) +\ii\omega(\mu_x p_x^2 + (\mu_\bot+K)p_\bot^2)]}f_x(\tilde{\bp}) \ ,
\\
\phi(\tilde{\bp}) &=& \frac{\ii\omega - Kp_\bot^2}{[(\omega^2-\gamma\kappa_1 p_x^2-Kp_\bot^2(\mu_xp_x^2+\mu_\bot p_\bot^2)) +\ii\omega(\mu_x p_x^2 + (\mu_\bot+K)p_\bot^2)]}f_x(\tilde{\bp}) \ ,
\eeqn
where \((\tilde{\bp}) \equiv (\bp,\omega)\).

In the hydrodynamic limit, we can eliminate  $K\mu_\bot p_\bot^4$ in the denominator above. Our ``bare" propagators, \(G_{\rho}, G_{\phi}\), are therefore
\begin{align}
\label{eq:rhoprop}
\rho(\tilde{\bp}) &= \frac{\ii\gamma p_x}{(\omega^2- (\gamma\kappa_1 + K\mu_x p_\perp^2)p_x^2) +\ii\omega(\mu_x p_x^2 +  (\mu_\bot +K)p_\bot^2)}f_x(\tilde{\bp})=G_\rho(\tilde{\bp}) f_x(\tilde{\bp})\ ,
\\
\label{eqn:xprop}
\phi(\tilde{\bp}) &= \frac{\ii\omega - Kp_\bot^2}{(\omega^2- (\gamma\kappa_1 + K\mu_x p_\perp^2) 
p_x^2) +\ii\omega(\mu_x p_x^2 +  (\mu_\bot +K) p_\bot^2)}f_x(\tilde{\bp})=G_\phi(\tilde{\bp}) f_x(\tilde{\bp}) \ .
\end{align}

To ease notation, we now define the quantities below
\beqn
\Pi(\bp) = \left[(\gamma\kappa_1+K\mu_xp_\bot^2)p_x^2+K\mu_\bot p_\bot^4\right]\sep \Gamma(\bp) = \mu_xp_x^2 + (\mu_\bot+K) p_\bot^2\sep\Delta(\bp) = 4\Pi(\bp) - \Gamma(\bp)^2
\ .
\eeqn

The correlation functions, in Fourier transformed space, are 
\beqn
\langle \rho(\tilde{\bp})\rho(\tilde{\bq})\rangle & =& 2DG_{\rho}(\tilde{\bp})G_{\rho}(\tilde{\bq})\delta^d(\bp+\bq)\delta(\omega+\nu)\ ,
\\
\langle \phi(\tilde{\bp})\phi(\tilde{\bq})\rangle & =& 2DG_{\phi}(\tilde{\bp})G_{\phi}(\tilde{\bq})\delta^d(\bp+\bq)\delta(\omega+\nu)\ .
\eeqn
The corresponding correlation functions in real space and time are thus 
\beqn
\label{eq:CFden}
\langle\rho(\br,t)\rho(\br',t')\rangle &=& 2D\int_{\tilde{\bp}}
 \frac{\gamma^2 p_x^2 e^{\ii\bp\cdot(\br-\br')-\ii \omega(t-t')}}{(\omega^2-\Pi(\bp))^2 +\omega^2\Gamma(\bp)^2}\ ,
 \\
 \label{eq:CFmom}
 \langle\phi(\br,t)\phi(\br',t')\rangle &=& 2D\int_{\tilde{\bp}}
 \frac{(\omega^2 + K^2p_\bot^4)e^{\ii\bp\cdot(\br-\br')-\ii \omega(t-t')}}{(\omega^2-\Pi(\bp))^2 +\omega^2\Gamma(\bp)^2}\ .
 \eeqn
 where $\int_{\tilde{\bp}} = \int\frac{d^d\bp d\omega}{(2\pi)^{d+1}}$.

\subsection{Equal-time correlation functions}
Focusing first on the $\phi$-$\phi$ correlation, the equal time correlation function is
\beqn
 \langle\phi(\br,t)\phi(\br',t)\rangle &=& 2D\int_{\tilde{\bp}}
 \frac{(\omega^2 + K^2p_\bot^4)e^{\ii\bp\cdot(\br-\br')}}{(\omega^2-\Pi(\bp))^2 +\omega^2\Gamma(\bp)^2}\ ,
 \\
  &=&  D\left[\int_{\bp} \frac{e^{\ii\bp\cdot(\br-\br')}}{\Gamma(\bp)} + \int_{\bp}\frac{ K^2p_\bot^4e^{\ii\bp\cdot(\br-\br')}}{\Pi(\bp)\Gamma(\bp)} \right]\ .
\eeqn
We now ignore the second term above as it is less singular than the first piece in the limit $\bp \rightarrow 0$.  
With the second term omitted, the form of the correlation function then reveals, by, e.g., comparing it to the standard critical $\cO(n)$ model at the linear level \cite{Skardar_b07},  that the roughness and anisotropic exponents for $\phi$ are
\beq
\chi_\phi^{\rm lin} = \frac{2-d}{2} \sep \zeta^{\rm lin} =1 \ .
\eeq

For the  $\rho$-$\rho$ correlation, the equal time correlation function is, after integrating out $\omega$,
\beq
 \langle\rho(\br,t)\rho(\br',t)\rangle =  D \int_{\bp}\frac{\gamma^2p_x^2 e^{\ii\bp\cdot(\br-\br')}}{\Pi(\bp)\Gamma(\bp)} \ .
\eeq
We now approximate $\Pi$ as $\gamma \kappa_1 p_x^2$ in the hydrodynamic limit. Hence, we have
\beq
 \langle\rho(\br,t)\rho(\br',t)\rangle = \frac{\gamma}{\kappa_1} \langle\phi(\br,t)\phi(\br',t)\rangle \ ,
 \eeq
 and  hence the roughness exponent for $\rho$ is
 \beq
 \chi_\rho^{\rm lin} = \frac{2-d}{2}  \ .
\eeq

\subsection{Equal-space correlation functions}
For the $\phi$-$\phi$ correlation, the equal space correlation function is
\beqn
\label{eq:phi_lin}
 \langle\phi(\br,t)\phi(\br,t')\rangle &=& 2D\int_{\tilde{\bp}}
 \frac{(\omega^2 + K^2p_\bot^4)e^{-\ii\omega(t-t')}}{(\omega^2-\Pi(\bp))^2 +\omega^2\Gamma(\bp)^2}\ .
\eeqn

Ignoring the $K^2p_\bot^4$  in the numerator for the time being, the real part of the correlation is, after integrating out $\omega$,
\beq
 D\int_{\bp}\cos\left(\frac{\Delta(\bp)^{1/2}(t-t')}{2}\right)\frac{e^{\frac{\Gamma(\bp)(t-t')}{2}}}{\Gamma(\bp)}\ .
\eeq
The form of the correlation function again reveals that the dynamic exponent is 
\beq
z^{\rm lin} =2 
\ .
\eeq

Moving onto the equal-space correlation for $\rho$, we have
\beqn
\label{eq:rho_lin}
 \langle\rho(\br,t)\rho(\br,t')\rangle &=& 2D\int_{\tilde{\bp}}
 \frac{\gamma^2 p_x^2 e^{-\ii\omega(t-t')}}{(\omega^2-\Pi(\bp))^2 +\omega^2\Gamma(\bp)^2}
 \\
&=& \frac{\gamma D}{\kappa_1}\int_{\bp}\cos\left(\frac{\Delta(\bp)^{1/2}(t-t')}{2}\right)\frac{e^{\frac{\Gamma(\bp)(t-t')}{2}}}{\Gamma(\bp)}\ \ ,
\eeqn
confirming again that $z^{\rm lin} =2 $. 

Comparing the form of (\ref{eq:rho_lin}) and (\ref{eq:phi_lin}), we can now justify the omission of $K^2p_\bot^4$   in the numerator when considering the $\phi$-$\phi$ correlation.

\section{Upper critical dimension $d_c$ \& relevant terms below $d_c$}

  If we rescale all terms (excluding $\alpha_0,\alpha_1$) in (\ref{eq:rho}) and (\ref{eq:x}) according to the respective relations: 
\beq
t\mapsto e^{z \ell}t \sep x \mapsto e^{\zeta \ell}x\sep   \br_\bot\mapsto e^{\ell}\br_\bot\sep\phi\mapsto e^{\chi_\phi \ell}\phi\sep\rho\mapsto e ^{\chi_{\rho} \ell}\rho\ ,
\eeq
\begin{align}
&e^{(\chi_{\rho}-z)\ell}\pp_t\rho = -e^{(\chi_\phi-\zeta)\ell}\gamma\pp_x\phi - e^{(\chi_\rho-2)\ell}K\vnab^2_\bot\rho\ ,
\\
&\pp_t\rho = -e^{(z-\zeta+\chi_\phi-\chi_\rho)\ell}\gamma\pp_x\phi - e^{(z-2)\ell}K\vnab^2_\bot\rho.
\end{align}
The noise term rescales as,
\beq
\langle f_x(\br,t)f_x(\br',t')\rangle\mapsto e^{{-(z+\zeta+d-1)\ell}}2D\delta^{d+1}(x-x',\br_\bot-\br_\bot',t-t')\ ,
\eeq
resulting in the rescaling of (\ref{eq:x}) as,
\beqn
e^{(\chi_\phi -z)\ell}\partial_t\phi &+& e^{(2\chi_\phi -\zeta)\ell}\frac{\lambda}{2} \partial_x\phi^2 = -e^{(\chi_{\rho}-\zeta)\ell}\kappa_1\partial_x\rho - e^{(2\chi_{\rho}-\zeta)\ell}\kappa_2\partial_x\rho^2 + e^{(\chi_\phi -2\zeta)\ell}\mu_x\partial^2_x\phi\
\\
&+& e^{(\chi_\phi -2)\ell}\mu_{\perp}\vnab^2_{\perp}\phi - e^{(2\chi_{\rho}+\chi_\phi) \ell}\alpha_2\rho^2\phi -  e^{3\chi_\phi \ell}\beta\phi^3 + e^{\frac{{-(z+\zeta+d-1)\ell}}{2}}f_x\ ,
\eeqn
which rearranged is,
\beqn
\partial_t\phi&+&e^{(\chi_\phi + z-\zeta )\ell}\frac{\lambda}{2} \partial_x\phi^2  =  -e^{(\chi_{\rho}-\chi_\phi +z-\zeta)\ell}\kappa_1\partial_x\rho - e^{(2\chi_{\rho}-\chi_\phi+z-\zeta)\ell}\kappa_2\partial_x\rho^2  +e^{(z -2\zeta)\ell}\mu_x\partial^2_x\phi \,
\\
&+&e^{(z -2)\ell}\mu_{\perp}\vnab^2_\bot\phi - e^{(2\chi_{\rho}+z)\ell}\alpha_2\rho^2\phi -  e^{(2\chi_\phi+z) \ell}\beta \phi^3  + e^{\frac{{(z-d-2\chi_\phi-\zeta+1)\ell}}{2}}f_x\ .
\eeqn
 If we also enforce the noise term to be invariant,
\beq
e^{\frac{{(z-d-2\chi_\phi-\zeta+1)\ell}}{2}}f_x\rightarrow e^{\frac{{(2-d-2\chi_\phi)\ell}}{2}}f_x\implies \chi_\phi^{ \rm lin}= \frac{2-d}{2}\ .
\\
\eeq
With our density correlation function derived result \(\chi^{\rm lin}_{\rho}= \frac{2-d}{2}\), Eq.~(\ref{eq:x}) at a linear level becomes
\beq
\partial_t\phi+ e^{(\frac{4-d}{2} )\ell}\frac{\lambda}{2} \partial_x\phi^2  =  -e^{\ell}\kappa_1\partial_x\rho - e^{(\frac{4-d}{2} )\ell}\kappa_2\partial_x\rho^2  +\mu_x\partial^2_x\phi \\
+\mu_{\perp}\vnab^2_\bot\phi - e^{(4-d) \ell}\alpha_2\rho^2\phi -  e^{(4-d)\ell}\beta \phi^3  +f_x.
\\
\eeq

  Non-linear terms are then only relevant for dimensions below an upper critical dimension \(d_c=4\).   Terms like \(\mu_3(\bg\cdot\vnab)^2\bg\), \(\bg(\bg\cdot\vnab)P_2\), \(\bg^4\), \(\rho^4\) and other higher order terms (h.o.t.) will vanish exponentially quicker than all terms in \ref{eq:x} under RG flow, thus validating why they are excluded from our EOM.

\section{Dynamic Renormalisation Group (DRG) Treatment}

  A DRG analysis together with a $\epsilon$-expansion method is now applied to Eq.~(4) to a one loop approximation. We follow the standard procedure of Fourier transforming Eq.~(4) and splitting the fields into small and large scale modes, at an arbitrary inverse length scale $\Lambda' = \Lambda e^{-l}$, which is a fraction of the physical cutoff $\Lambda$ scale that defines the smallest length scale of the system e.g. the smallest distance between our spin carrying particles:
\begin{align}
    \rho({\rm \bq},\omega) &= \rho_{>}({\rm \bq},\omega)+\rho_{<}({\rm \bq},\omega)\ ,\\
    \nonumber
    \phi({\rm \bq},\omega)&=\phi_{>}({\rm \bq},\omega)+\phi_{<}({\rm \bq},\omega)\ ,
\end{align}
such that $\rho_{>}({\rm \bq},\omega)=\rho({\rm \bq},\omega)$ if ${\rm |\bq|}>\Lambda'$ and  $\rho_{>}({\rm \bq},\omega)=0$ otherwise, with the same rules for $\phi({\rm \bq},\omega)$. 

  Small scale modes $\rho_>$ and $\phi_>$ are then eliminated by recursively inserting the formal solution for the small scale modes in terms of the large scales modes provided by the EOM below. 
\begin{align}
\phi(\tilde{\bp}) = G_\phi(\tilde{\bp})[ f_x(\tilde{\bp}) - \ii\frac{\lambda}{2}\int_{\tilde{\bq}}(p_x-q_x)\phi(\tilde{\bq})&\phi(\tilde{\bp}-\tilde{\bq}) - \ii\kappa_2\int_{\tilde{\bq}}(p_x-q_x)\rho(\tilde{\bq})\rho(\tilde{\bp}-\tilde{\bq}) - \alpha_2\int_{\tilde{\bq},\tilde{\bl}}\rho(\tilde{\bp}-\tilde{\bl})\rho(\tilde{\bl}-\tilde{\bq})\phi(\tilde{\bq})
\\
\nonumber
&-\beta\int_{\tilde{\bq},\tilde{\bl}}\phi(\tilde{\bp}-\tilde{\bl})\phi(\tilde{\bl}-\tilde{\bq})\phi(\tilde{\bq})]\ .
\end{align}

  The resulting hierarchy of terms can be truncated by taking the perturbative limit; considering the coefficient terms (vertex parameters) as small. Averaging this resultant expression, over the small scale noise terms $f_{x_>}$ left in the EOM, produces effective terms which contribute to the EOM's coefficients which are dependent on the coarse graining scale as $\Lambda^{-\epsilon}$. These terms are represented via Feynman diagrams which are constructed according to the number of loops desired.

\section{graphical notation and corrections}

  As well as the diagrammatic conventions specified in the MT, we also follow the same ``rules" where wavevectors are conserved at each vertex and closed loops imply an integration over an internal wavevector and frequency such as $\tilde{\bq}=(\bq,\Omega)$ and $\tilde{\bh}=(\bh,\nu)$. The notation for our diagrams are in Fig.~(\ref{fig:notation}) below,

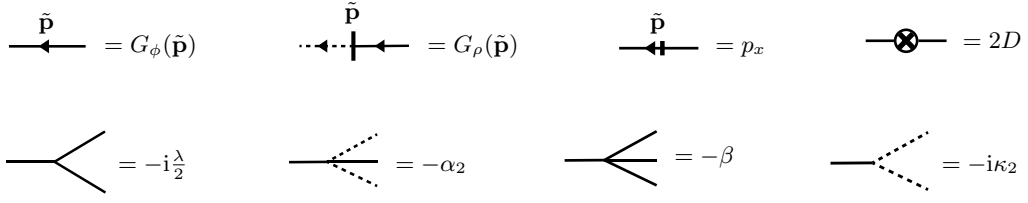
\begin{figure}[H]
    \centering
    \includegraphics[width=0.06\linewidth]{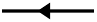} \raisebox{0.05cm}{ $=G_\phi(\tilde{\bp})$}
    \begin{picture}(0,0)
    \put(-63, 10){\small $\tilde{\bp}$}
    \end{picture}
    \hspace{1cm}
    \raisebox{-0.07cm}{\includegraphics[width=0.085\linewidth]{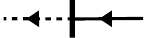}} \raisebox{0.05cm}{ $=G_\rho(\tilde{\bp})$}
    \begin{picture}(0,0)
    \put(-69, 14){\small $\tilde{\bp}$}
    \end{picture}
    \hspace{1cm}
    \includegraphics[width=0.06\linewidth]{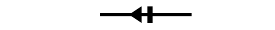} \raisebox{0.05cm}{ $=p_x$}
    \begin{picture}(0,0)
    \put(-47, 10){\small $\tilde{\bp}$}
    \end{picture}
    \hspace{1cm}
    \includegraphics[width=0.06\linewidth]{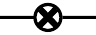} \raisebox{0.1cm}{ $=2D$}
    \\[0.8cm] 
    \includegraphics[width=0.08\linewidth]{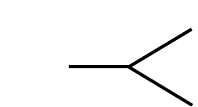} \raisebox{0.3cm}{$= -\ii\frac{\lambda}{2}$}
    \hspace{1cm}
    \includegraphics[width=0.08\linewidth]{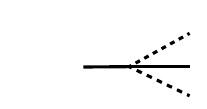} \raisebox{0.3cm}{$= -\alpha_2$} 
    \hspace{1cm}
    \includegraphics[width=0.08\linewidth]{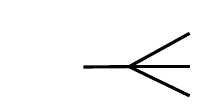} \raisebox{0.4cm}{$=-\beta$} 
    \hspace{1cm}
    \raisebox{-0.1cm}{\includegraphics[width=0.08\linewidth]{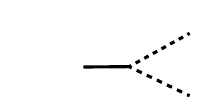}} \raisebox{0.3cm}{$=-\ii\kappa_2$}
    \vspace{0.2cm}
    \caption{Graphical notation for the Feynman diagrams.}
    \label{fig:notation}
\end{figure}

\subsection{2-vertex diagrams}
All 2-vertex diagrams that contribute to the graphical corrections of the coefficients in the EOM are shown in the figure below.

\begingroup
  \setlength{\abovedisplayskip}{4pt}
  \setlength{\abovedisplayshortskip}{4pt}
  \setlength{\belowdisplayskip}{4pt}
  \setlength{\belowdisplayshortskip}{4pt}
\begin{align}
  {\rm   Gr}^{(2)}_{\lambda}
  &=
-12\hspace{0.3em} \vcenter{\hbox{\raisebox{0.3\height}{\includegraphics[width=0.155\textwidth]{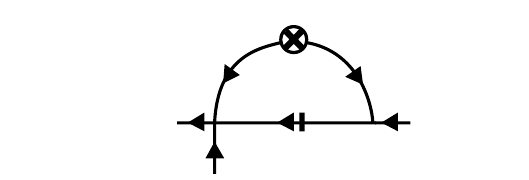}}}}
\begin{picture}(0,0)
\put(-47, 45){\small $I_{1.2}$}
\end{picture}
-6\;\vcenter{\hbox{\raisebox{0.3\height}{\includegraphics[width=0.16\textwidth]{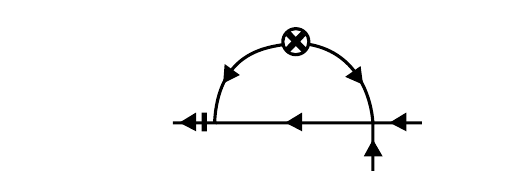}}}}
\begin{picture}(0,0)
\put(-47, 45){\small $I_{6}$}
\end{picture}
-6 \;\vcenter{\hbox{\raisebox{0.25\height}{\includegraphics[width=0.16\textwidth]{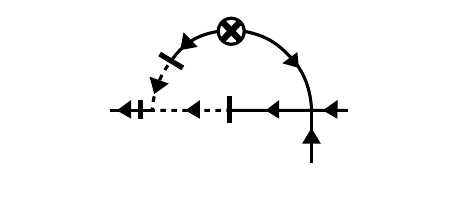}}}}
\begin{picture}(0,0)
\put(-46, 45){\small $I_{7}$}
\end{picture}
-4 \;\vcenter{\hbox{\raisebox{0.25\height}{\includegraphics[width=0.16\textwidth]{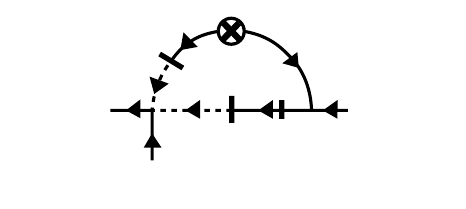}}}}
\begin{picture}(0,0)
\put(-48, 43){\small $I_{2.2}$}
\end{picture}
\\
{\rm   Gr}^{(2)}_{\alpha_2}
&=
-6\hspace{0.3em}\vcenter{\hbox{\raisebox{0.2\height}{\includegraphics[width=0.16\textwidth]{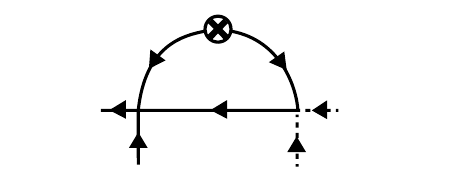}}}}
\begin{picture}(0,0)
\put(-47, 45){\small $I_{6}$}
\end{picture}
-4\hspace{0.2em} \;\vcenter{\hbox{\raisebox{0.2\height}{\includegraphics[width=0.16\textwidth]{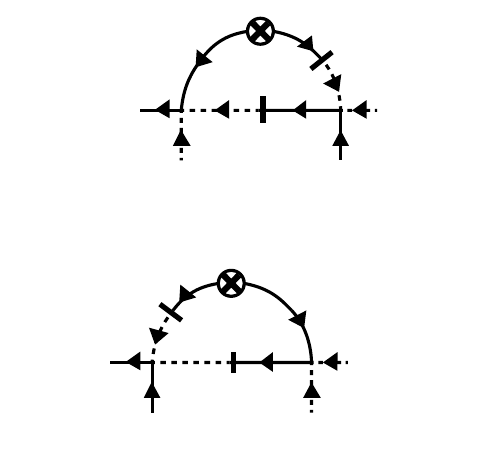}}}}
\begin{picture}(0,0)
\put(-46, 45){\small $I_{7}$}
\end{picture}
-4\hspace{0.2em}\;\vcenter{\hbox{\raisebox{0.25\height}{\includegraphics[width=0.155\textwidth]{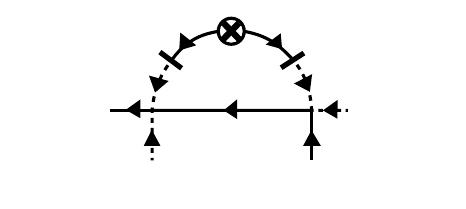}}}}
\begin{picture}(0,0)
\put(-45, 45){\small $I_{7}$}
\end{picture}
-2\hspace{0.15em}\;\vcenter{\hbox{\raisebox{0.25\height}{\includegraphics[width=0.16\textwidth]{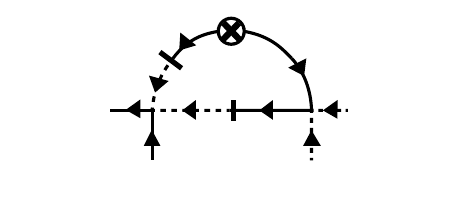}}}}
\begin{picture}(0,0)
\put(-46, 45){\small $I_{7}$}
\end{picture}
\\
{\rm   Gr}^{(2)}_{\kappa_2}
&=
-2\hspace{0.3em}\vcenter{\hbox{\raisebox{0.3\height}{\includegraphics[width=0.17\textwidth]{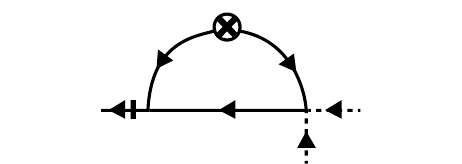}}}}
\begin{picture}(0,0)
\put(-48, 45){\small $I_{6}$}
\end{picture}
-4\hspace{0.2em}\;\vcenter{\hbox{\raisebox{0.3\height}{\includegraphics[width=0.16\textwidth]{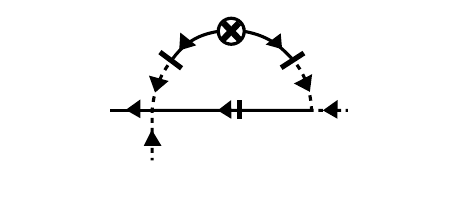}}}}
\begin{picture}(0,0)
\put(-46, 45){\small $I_{3}$}
\end{picture}
-4\hspace{0.2em}\;\vcenter{\hbox{\raisebox{0.25\height}{\includegraphics[width=0.16\textwidth]{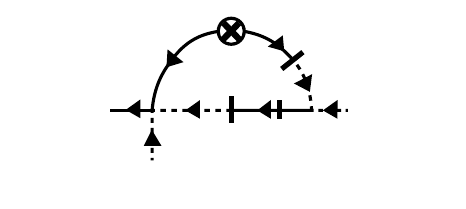}}}}
\begin{picture}(0,0)
\put(-46, 45){\small $I_{4}$}
\end{picture}
-2\hspace{0.2em}\;\vcenter{\hbox{\raisebox{0.25\height}{\includegraphics[width=0.16\textwidth]{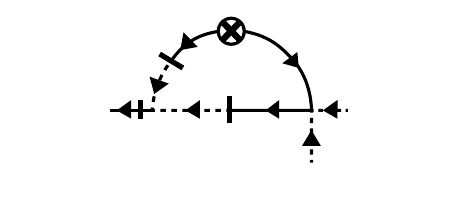}}}}
\begin{picture}(0,0)
\put(-47, 45){\small $I_{7}$}
\end{picture}
\\
{\rm   Gr}^{(2)}_{\beta}
&=
-18\hspace{0.3em}\vcenter{\hbox{\raisebox{0.3\height}{\includegraphics[width=0.16\textwidth]{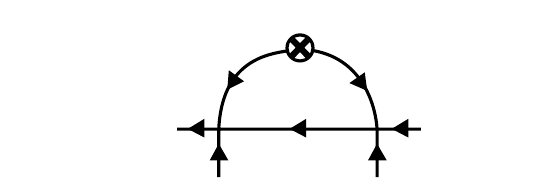}}}}
\begin{picture}(0,0)
\put(-45, 45){\small $I_{6}$}
\end{picture}
-6\hspace{0.2em}\;\vcenter{\hbox{\raisebox{0.25\height}{\includegraphics[width=0.16\textwidth]{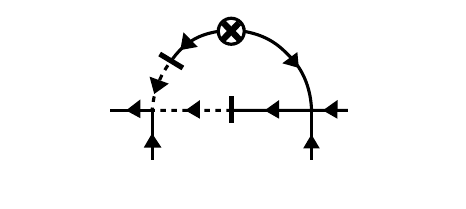}}}}
\begin{picture}(0,0)
\put(-45, 45){\small $I_{7}$}
\end{picture}
\\
\vspace{1cm}
{\rm   Gr}^{(2)}_{\kappa_1}
&=
-4\hspace{0.3em}\vcenter{\hbox{\raisebox{0.75\height}{\includegraphics[width=0.16\textwidth]{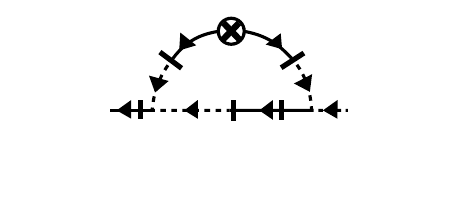}}}}
\begin{picture}(0,0)
\put(-45, 45){\small $I_{5}$}
\end{picture}
-4 \hspace{0.3em}\vcenter{\hbox{\raisebox{0.75\height}{\includegraphics[width=0.16\textwidth]{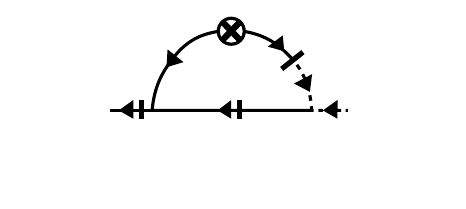}}}}
\begin{picture}(0,0)
\put(-50, 45){\small $=0$}
\end{picture}
\\
{\rm   Gr}^{(2)}_{\mu_x}
&=
+4\hspace{0.3em}\vcenter{\hbox{\raisebox{0.75\height}{\includegraphics[width=0.16\textwidth]{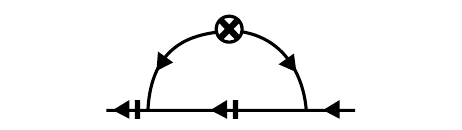}}}}
\begin{picture}(0,0)
\put(-50, 45){\small $I_{1.1}$}
\end{picture}
+4\hspace{0.3em}\vcenter{\hbox{\raisebox{0.75\height}{\includegraphics[width=0.16\textwidth]{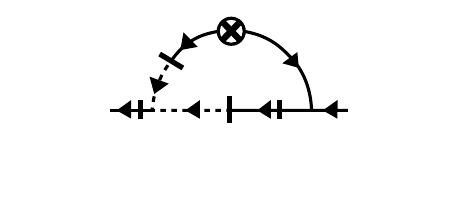}}}}
\begin{picture}(0,0)
\put(-50, 45){\small $I_{2.1}$}
\end{picture}
\end{align}
\endgroup

The dimensionless $ $Gr which is directly added onto the flow equations can be obtained from the integral results by factoring out the appropriate parameter which it contributes to as well as any dimensionful quantities such as $d\ell$ and $\int_{\tilde{\bh}}\phi(\tilde{\bk}-\tilde{\bh})\phi(\tilde{\bh})$. As such any further reference to $I_i$ below will automatically assume this.

Overall, the 2-vertex diagram contributions can be structured as,
\beqn
 {\rm Gr}^{(2)}_{\lambda} &=& -12\beta I_{1.2} - 6\beta I_6 - 12\frac{\kappa_2\beta}{\lambda}I_7 - 4\alpha_2 I_{2.2}\ ,
\\
 {\rm Gr}^{(2)}_{\alpha_2} &=& -6\beta I_6 - 10\alpha_2 I_7\ ,
\\
 {\rm Gr}^{(2)}_{\kappa_2} &=& -\frac{\beta\lambda}{\kappa_2}I_6 - 4\alpha_2 I_3 - 4\alpha_2 I_4 - 2\alpha_2 I_7\ ,
\\
 {\rm Gr}^{(2)}_{\beta} &=& -18\beta I_6 - 6\alpha_2 I_7\ ,
\\
 {\rm Gr}^{(2)}_{\kappa_1} &=& -4\frac{\kappa_2^2}{(-\ii\kappa_1)}I_5\ ,
\\
 {\rm Gr}^{(2)}_{\mu_x} &=& \frac{\lambda^2}{\mu_x}I_{1.1} + \frac{2\kappa_2\lambda}{\mu_x}I_{2.1}.
\eeqn
where
\beqn
I_{1.1} &=&  \left[\frac{3}{32}\frac{D}{\mu_x^{1/2}\mu_\bot^{3/2}}\frac{S_{d-1}}{(2\pi)^{d-1}}\Lambda^{d-4}d\ell+\frac{3}{32}\frac{KD}{\mu_x^{1/2}\mu_\bot^{5/2}}\frac{S_{d-1}}{(2\pi)^{d-1}}\Lambda^{d-4}d\ell\right]\ ,
\\
I_{1.2} &=& \left[\frac{3}{64}\frac{D}{\mu_x^{1/2}\mu_\bot^{3/2}}\frac{S_{d-1}}{(2\pi)^{d-1}}\Lambda^{d-4}d\ell+\frac{3}{64}\frac{KD}{\mu_x^{1/2}\mu_\bot^{5/2}}\frac{S_{d-1}}{(2\pi)^{d-1}}\Lambda^{d-4}d\ell\right]\ ,
\\
I_{2.1} &=&  \left[\frac{3}{32}\frac{\gamma D}{\kappa_1\mu_x^{1/2}\mu_\bot^{3/2}}\frac{S_{d-1}}{(2\pi)^{d-1}}\Lambda^{d-4}d\ell+\frac{3}{32}\frac{\gamma KD}{\kappa_1\mu_x^{1/2}\mu_\bot^{5/2}}\frac{S_{d-1}}{(2\pi)^{d-1}}\Lambda^{d-4}d\ell\right]\ ,
\\
I_{2.2} &=& \left[\frac{3}{64}\frac{\gamma D}{\kappa_1\mu_x^{1/2}\mu_\bot^{3/2}}\frac{S_{d-1}}{(2\pi)^{d-1}}\Lambda^{d-4}d\ell+\frac{3}{64}\frac{\gamma KD}{\kappa_1\mu_x^{1/2}\mu_\bot^{5/2}}\frac{S_{d-1}}{(2\pi)^{d-1}}\Lambda^{d-4}d\ell\right]\ ,
\\
I_{3} &=& \left[-\frac{1}{64}\frac{\gamma D}{\kappa_1\mu_x^{1/2}\mu_\bot^{3/2}}\frac{S_{d-1}}{(2\pi)^{d-1}}\Lambda^{d-4}d\ell+\frac{3}{64}\frac{\gamma KD}{\kappa_1\mu_x^{1/2}\mu_\bot^{5/2}}\frac{S_{d-1}}{(2\pi)^{d-1}}\Lambda^{d-4}d\ell\right]\ ,
\\
I_{4} &=& \left[\frac{3}{64}\frac{\gamma D}{\kappa_1\mu_x^{1/2}\mu_\bot^{3/2}}\frac{S_{d-1}}{(2\pi)^{d-1}}\Lambda^{d-4}d\ell-\frac{3}{64}\frac{\gamma KD}{\kappa_1\mu_x^{1/2}\mu_\bot^{5/2}}\frac{S_{d-1}}{(2\pi)^{d-1}}\Lambda^{d-4}d\ell\right]\ ,
\\
I_5 &=& \frac{1}{4}\frac{\ii\gamma D}{\kappa_1^2\mu_x^{1/2}\mu_\bot^{1/2}}\frac{S_{d-1}}{(2\pi)^{d-1}}\Lambda^{d-2}d\ell\ ,
\\
I_{6} &=& \frac{1}{8}\frac{D}{\mu_x^{1/2}\mu_\bot^{3/2}}\frac{S_{d-1}}{(2\pi)^{d-1}}\Lambda^{d-4}d\ell\ ,
\\
I_{7} &=& \frac{1}{8}\frac{\gamma D}{\kappa_1\mu_x^{1/2}\mu_\bot^{3/2}}\frac{S_{d-1}}{(2\pi)^{d-1}}\Lambda^{d-4}d\ell\ .
\eeqn
We will show how these $I$'s are computed in the Appendix \ref{app2-vertex}.

\subsection{3-vertex diagrams}
All 3-vertex diagrams that contribute to the graphical corrections of the coefficients in the EOM are shown in the figure below.

\vspace*{-\parskip}%
\vspace*{-\abovedisplayskip}%
\begin{align}
{\rm   Gr}^{(3)}_{\lambda}
&=
-4\hspace{0.3em}\vcenter{\hbox{\raisebox{0.1\height}{\includegraphics[width=0.16\textwidth]{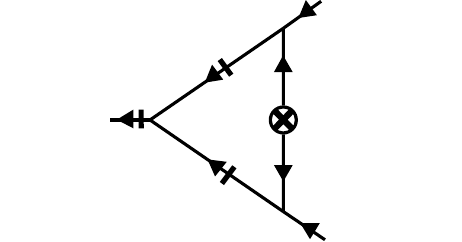}}}}
+4\hspace{0.3em}\;\vcenter{\hbox{\raisebox{0.1\height}{\includegraphics[width=0.16\textwidth]{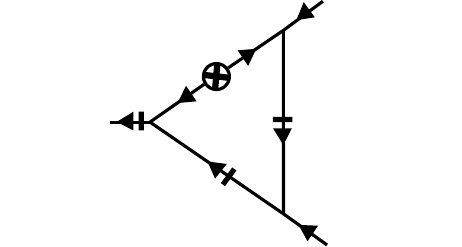}}}}
+4\hspace{0.3em}\;\vcenter{\hbox{\raisebox{0.1\height}{\includegraphics[width=0.16\textwidth]{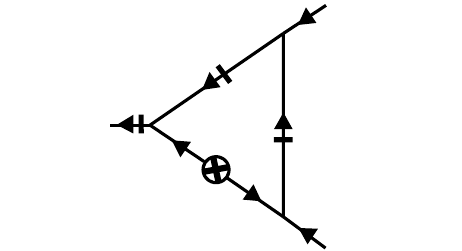}}}}
\\
\nonumber
&-4\hspace{0.3em}\;\vcenter{\hbox{\raisebox{0.1\height}{\includegraphics[width=0.17\textwidth]{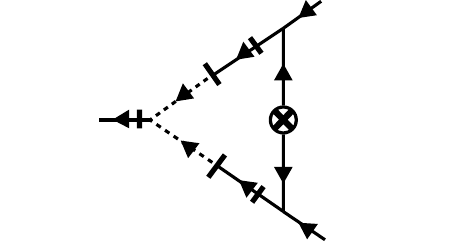}}}}
+4\hspace{0.3em}\;\vcenter{\hbox{\raisebox{0.1\height}{\includegraphics[width=0.17\textwidth]{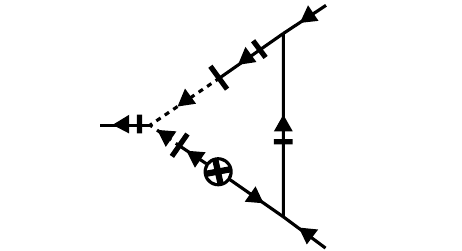}}}}
+4\hspace{0.3em}\;\vcenter{\hbox{\raisebox{0.1\height}{\includegraphics[width=0.17\textwidth]{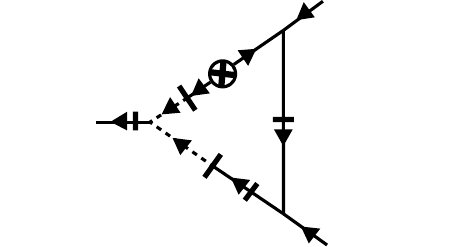}}}}
\\
\\
\nonumber
{\rm   Gr}^{(3)}_{\alpha_2}
&=
-4\hspace{0.3em}\vcenter{\hbox{\raisebox{0.1\height}{\includegraphics[width=0.17\textwidth]{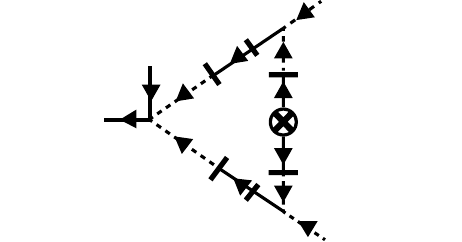}}}}
+4\hspace{0.3em}\;\vcenter{\hbox{\raisebox{0.1\height}{\includegraphics[width=0.17\textwidth]{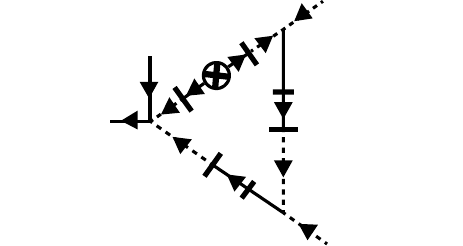}}}}
+4\hspace{0.3em}\;\vcenter{\hbox{\raisebox{0.1\height}{\includegraphics[width=0.17\textwidth]{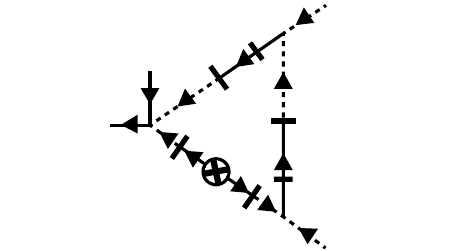}}}}
\\
\nonumber
&-12\hspace{0.3em}\;\vcenter{\hbox{\raisebox{0.1\height}{\includegraphics[width=0.17\textwidth]{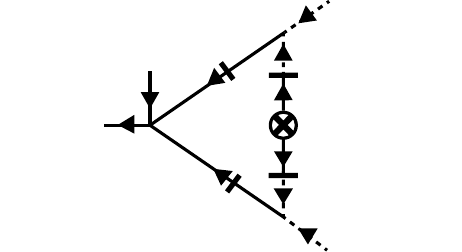}}}}
+12\hspace{0.3em}\;\vcenter{\hbox{\raisebox{0.1\height}{\includegraphics[width=0.165\textwidth]{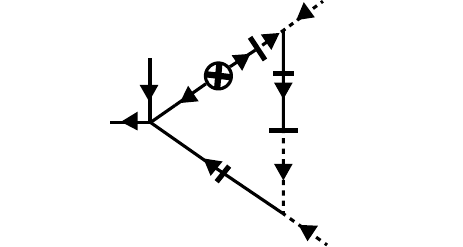}}}}
+12\hspace{0.2em}\;\vcenter{\hbox{\raisebox{0.1\height}{\includegraphics[width=0.17\textwidth]{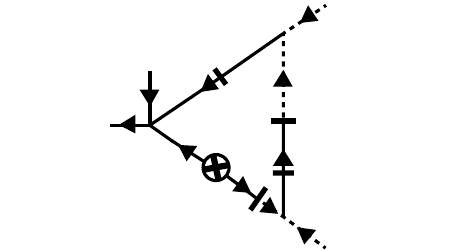}}}}
\\
\nonumber
&-8\hspace{0.3em}\;\vcenter{\hbox{\raisebox{0.1\height}{\includegraphics[width=0.17\textwidth]{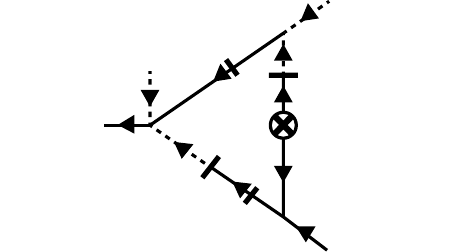}}}}
+8\hspace{0.3em}\;\vcenter{\hbox{\raisebox{0.1\height}{\includegraphics[width=0.17\textwidth]{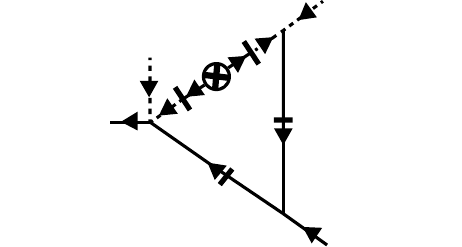}}}}
+8\hspace{0.3em}\;\vcenter{\hbox{\raisebox{0.1\height}{\includegraphics[width=0.17\textwidth]{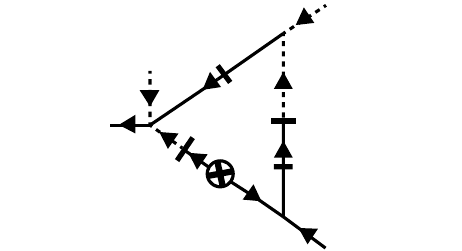}}}}
\\
\nonumber
&-8\hspace{0.3em}\;\vcenter{\hbox{\raisebox{0.1\height}{\includegraphics[width=0.17\textwidth]{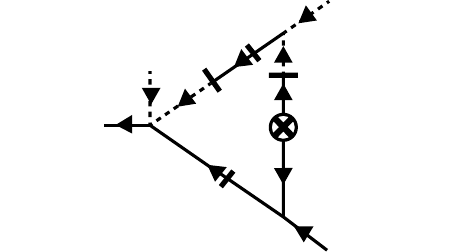}}}}
+8\hspace{0.3em}\;\vcenter{\hbox{\raisebox{0.1\height}{\includegraphics[width=0.17\textwidth]{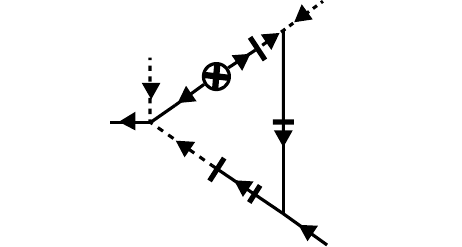}}}}
+8\hspace{0.3em}\;\vcenter{\hbox{\raisebox{0.1\height}{\includegraphics[width=0.17\textwidth]{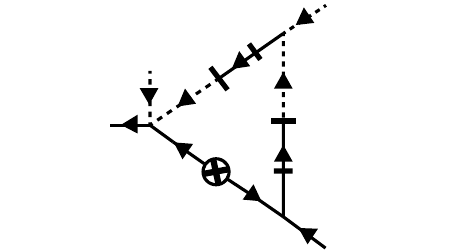}}}}
&&
\end{align}

\begin{align}
{\rm   Gr}^{(3)}_{\beta}
&=
-12\hspace{0.3em}\vcenter{\hbox{\raisebox{0.1\height}{\includegraphics[width=0.16\textwidth]{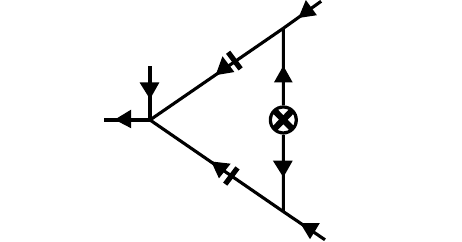}}}}
+12\hspace{0.3em}\;\vcenter{\hbox{\raisebox{0.1\height}{\includegraphics[width=0.16\textwidth]{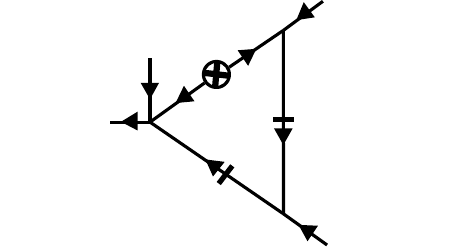}}}}
+12\hspace{0.3em}\;\vcenter{\hbox{\raisebox{0.1\height}{\includegraphics[width=0.155\textwidth]{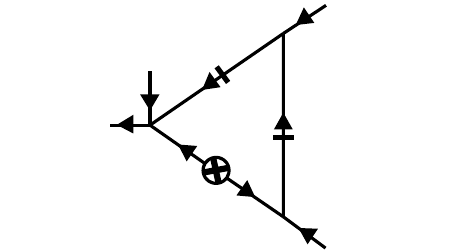}}}}
\\
\nonumber
&-8\hspace{0.3em}\;\vcenter{\hbox{\raisebox{0.1\height}{\includegraphics[width=0.16\textwidth]{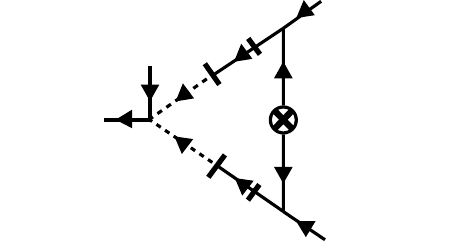}}}}
+8\hspace{0.3em}\;\vcenter{\hbox{\raisebox{0.1\height}{\includegraphics[width=0.155\textwidth]{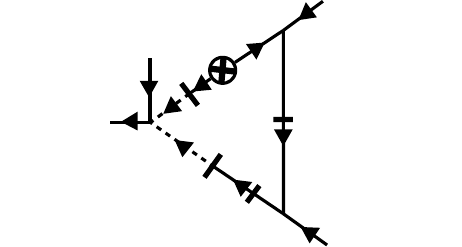}}}}
+8\hspace{0.3em}\;\vcenter{\hbox{\raisebox{0.1\height}{\includegraphics[width=0.16\textwidth]{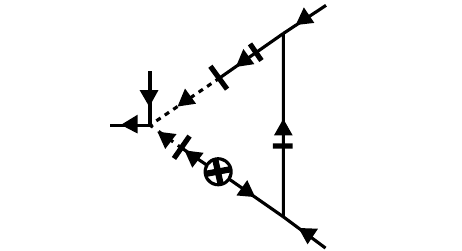}}}}
\\
{\rm   Gr}^{(3)}_{\kappa_2}
&=
-8\hspace{0.3em}\vcenter{\hbox{\raisebox{0.1\height}{\includegraphics[width=0.16\textwidth]{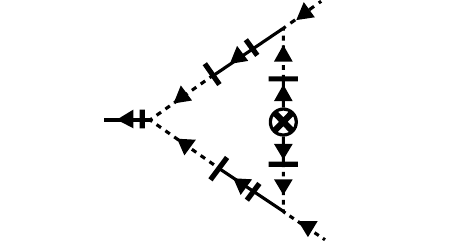}}}}
+8\hspace{0.3em}\;\vcenter{\hbox{\raisebox{0.1\height}{\includegraphics[width=0.16\textwidth]{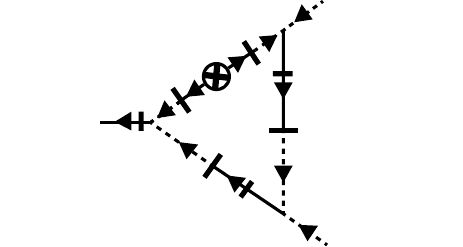}}}}
+8\hspace{0.3em}\;\vcenter{\hbox{\raisebox{0.1\height}{\includegraphics[width=0.155\textwidth]{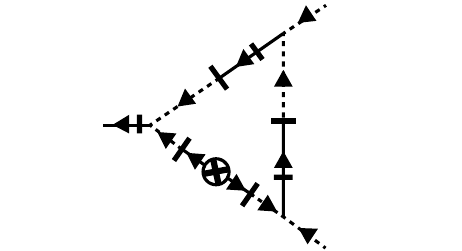}}}}
\\
\nonumber
&-8\hspace{0.3em}\;\vcenter{\hbox{\raisebox{0.1\height}{\includegraphics[width=0.16\textwidth]{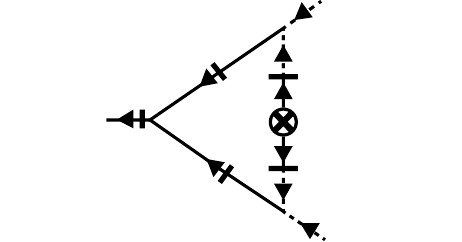}}}}
+8\hspace{0.3em}\;\vcenter{\hbox{\raisebox{0.1\height}{\includegraphics[width=0.16\textwidth]{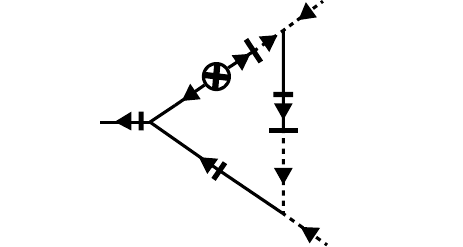}}}}
+8\hspace{0.3em}\;\vcenter{\hbox{\raisebox{0.1\height}{\includegraphics[width=0.16\textwidth]{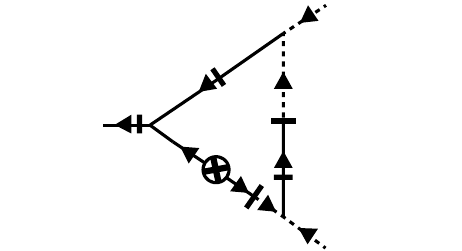}}}}
&&
\end{align}

As we show in Appendix \ref{app3-vertex}, the overall contributions from these 3-vertex diagrams are null. Specifically, we have
\beqn
 {\rm Gr}^{(3)}_{\lambda} &=& -4\lambda^2(I_8 + 2I_9) -4\kappa_2\lambda(I_{10}+ 2I_{11}) = 0\ ,
\\
 {\rm Gr}^{(3)}_{\alpha_2} &=& -4\kappa_2^2(I_{12} + 2I_{13}) -12\frac{\beta\kappa_2^2}{\alpha_2}(I_8 + 2I_9) - 8\kappa_2\lambda({\rm I}_8+ 2{\rm I}_9) =0 \ ,
\\
 {\rm Gr}^{(3)}_{\beta} &=&  -12\lambda^2(I_8 + 2I_9) - 8\frac{\alpha_2\lambda^2}{\beta}(I_{10}+2I_{11}) =0 \ ,
\\
 {\rm Gr}^{(3)}_{\kappa_2} &=& -8\kappa_2^2(I_{12} + 2I_{13}) - 8\kappa_2\lambda(I_{10}+2I_{11}) =0\ .
\eeqn

\section{RG flow equations}

  By taking the natural logarithm, followed by a derivative w.r.t $\ell$, of (\ref{eq:x}) renormalised parameters, we derive the flow equations with their diagrammatic contributions:
\beqn
\label{eq:l}
\frac{ \dd \ln \lambda}{\dd \ell} &=& z -\zeta + \chi_\phi- \frac{9}{16}(g_{\beta}+g_{\beta K})-\frac{3}{4}g_\beta - \frac{3}{2}\frac{g_{\kappa_2}g_\beta}{g_\lambda} - \frac{3}{16}(g_{\alpha_2}+g_{\alpha_2 K}) \ ,
\\
\label{eq:g}
\frac{ \dd \ln \gamma}{\dd \ell} &=& z - \zeta + \chi_{\phi} -\chi_\rho\ ,
\\
\label{eq:k1}
\frac{ \dd \ln \kappa_{1}}{\dd \ell} &=& z - \zeta - \chi_\phi+ \chi_{\rho}\ ,
\\
\label{eq:k2}
\frac{ \dd \ln \kappa_{2}}{\dd \ell} &=& z- \zeta - \chi_\phi + 2\chi_{\rho}- \frac{1}{8}\frac{g_\lambda g_{\alpha_2}}{g_{\kappa_2}}  - \frac{1}{16}(-g_{\alpha_2}+3g_{\alpha_2 K})- \frac{1}{16}(g_{\alpha_2}-g_{\alpha_2 K}) -\frac{1}{4}g_{\alpha_2}\ ,
\\
\label{eq:mx}
\frac{ \dd \ln \mu_x}{\dd \ell} &=& z- 2\zeta +  \frac{3}{32}(g_{\lambda}+g_{\lambda K}) + \frac{3}{16}(g_{\kappa_2}+g_{\kappa_2K})\ ,
\\
\label{eq:mp}
\frac{ \dd \ln \mu_\perp}{\dd \ell} &=& z-2\ , 
\\
\label{eq:a2}
\frac{ \dd \ln \alpha_2}{\dd \ell} &=& z + 2\chi_{\rho} - \frac{3}{4}g_\beta - \frac{5}{4}g_{\alpha_2}\ ,
\\
\label{eq:b}
\frac{ \dd \ln \beta}{\dd \ell} &=& z+2\chi_\phi- \frac{9}{4}g_\beta - \frac{3}{4}g_{\alpha_2}\ ,
\\
\label{eq:D}
\frac{ \dd \ln D}{\dd \ell} &=& z - \zeta- 2\chi_\phi - d+1\ ,
\\
\label{eq:K}
\frac{ \dd \ln K}{\dd \ell} &=& z-2\ .
\eeqn

There are ultimately only 4 unique correction couplings
\beqn
g_\beta &=& \frac{\beta D}{\mu_x^{1/2}\mu_\bot^{3/2}}\frac{S_{d-1}}{(2\pi)^{d-1}}\Lambda^{d-4}\sep g_\lambda=\frac{\lambda^2 D}{\mu_x^{3/2}\mu_\bot^{3/2}}\frac{S_{d-1}}{(2\pi)^{d-1}}\Lambda^{d-4}\ ,
\\
g_{\kappa_2}&=&\frac{\gamma\kappa_2\lambda D}{\kappa_1\mu_x^{3/2}\mu_\bot^{3/2}}\frac{S_{d-1}}{(2\pi)^{d-1}}\Lambda^{d-4}\sep g_{\alpha_2}=\frac{\gamma\alpha_2 D}{\kappa_1\mu_x^{1/2}\mu_\bot^{3/2}}\frac{S_{d-1}}{(2\pi)^{d-1}}\Lambda^{d-4}\ ,
\eeqn
with the realization that
\beqn
g_{\beta K} &=& \frac{\beta KD}{\mu_x^{1/2}\mu_\bot^{5/2}}\frac{S_{d-1}}{(2\pi)^{d-1}}\Lambda^{d-4}\sep g_{\lambda K}=\frac{\lambda^2 KD}{\mu_x^{3/2}\mu_\bot^{5/2}}\frac{S_{d-1}}{(2\pi)^{d-1}}\Lambda^{d-4}\ ,
\\
g_{\kappa_2 K}&=&\frac{\gamma\kappa_2\lambda KD}{\kappa_1\mu_x^{3/2}\mu_\bot^{5/2}}\frac{S_{d-1}}{(2\pi)^{d-1}}\Lambda^{d-4}\sep g_{\alpha_2 K}=\frac{\gamma\alpha_2 KD}{\kappa_1\mu_x^{1/2}\mu_\bot^{5/2}}\frac{S_{d-1}}{(2\pi)^{d-1}}\Lambda^{d-4}\ ,
\eeqn
are effectively redundant as their flow equation structures are identical to the former couplings.

We can now absorb them into the former couplings such that $g_\beta+g_{\beta K}\rightarrow 2g_\beta$, $g_{\alpha_2}-g_{\alpha_2 K}=0$ and etc, and the revised RG flow equations are shown in the MT.

The RG flow equations in terms of the dimensionless coefficients are:
\beqn
\frac{\dd \ln g_\beta}{\dd\ell} &=& \epsilon-\frac{9}{4}g_\beta-\frac{3}{4}g_{\alpha_2}-\frac{3}{32}g_\lambda - \frac{3}{16}g_{\kappa_2}\ ,
\\
\frac{\dd \ln g_\lambda}{\dd \ell} &=& \epsilon-\frac{15}{4}g_{\beta}-3\frac{g_{\kappa_2} g_\beta}{g_\lambda}-\frac{3}{4}g_{\alpha_2} - \frac{9}{32}g_\lambda-\frac{9}{16}g_{\kappa_2}\ ,
\\
\frac{\dd \ln g_{\kappa_2}}{\dd \ell} &=& \epsilon-\frac{1}{8}\frac{g_\lambda g_{\alpha_2}}{g_{\kappa_2}}- \frac{3}{4}g_{\alpha_2}- \frac{15}{8}g_\beta -\frac{3}{2}\frac{g_{\kappa_2}g_\beta}{g_\lambda} -  \frac{9}{32}g_\lambda-\frac{9}{16}g_{\kappa_2}\ ,
\\
\frac{\dd \ln g_{\alpha_2}}{\dd \ell} &=& \epsilon-\frac{3}{4}g_\beta -\frac{5}{4}g_{\alpha_2}- \frac{3}{32}g_\lambda-\frac{3}{16}g_{\kappa_2}\ .
\eeqn

\section{RG fixed points}

i.e

\beqn
\frac{\dd g_\beta}{\dd \ell} &=& g_\beta^*\left(\epsilon-\frac{9}{4}g_\beta^*-\frac{3}{4}g_{\alpha_2}^*-\frac{3}{32}g_\lambda - \frac{3}{16}g_{\kappa_2}^*\right)=0\ ,
\\
\frac{\dd g_\lambda}{\dd \ell} &=& g_\lambda^*\left(\epsilon-\frac{15}{4}g_{\beta}^*-3\frac{g_{\kappa_2}^* g_\beta^*}{g_\lambda^*}-\frac{3}{4}g_{\alpha_2}^* -\frac{9}{32}g_\lambda^*-\frac{9}{16}g_{\kappa_2}^*\right)=0\ ,
\\
\frac{\dd g_{\kappa_2}}{\dd\ell} &=& g_{\kappa_2}^*\left(\epsilon-\frac{1}{8}\frac{g_\lambda^* g_{\alpha_2}^*}{g_{\kappa_2}^*}-\frac{3}{4}g_{\alpha_2}^*-\frac{15}{8}g_\beta^* -\frac{3}{2}\frac{g_{\kappa_2^*}g_\beta^*}{g_\lambda^*} - \frac{9}{32}g_\lambda^*-\frac{9}{16}g_{\kappa_2^*}\right)=0\ ,
\\
\frac{\dd g_{\alpha_2}}{\dd\ell} &=& g_{\alpha_2}^*\left(\epsilon-\frac{3}{4}g_\beta^* -\frac{5}{4}g_{\alpha_2}^*-\frac{3}{32}g_\lambda^*-\frac{3}{16}g_{\kappa_2}^*\right) =0\ .
\eeqn

Solving all four ODE simultaneously indicates the RG flow drives $\lambda$ and $\kappa_2$ to zero almost immediately. Once this happens, $\lambda$ and $\kappa_2$ switch off meaning we return to the diagrammatics and remove their contributions, such that the generic behaviour is governed by the remaining flow equations,

\beqn
\frac{\dd g_\beta}{\dd \ell} &=& g_\beta^*\left(\epsilon-\frac{9}{4}g_\beta^*-\frac{3}{4}g_{\alpha_2}^*\right)=0\ ,
\\
\frac{\dd g_{\alpha_2}}{\dd\ell} &=& g_{\alpha_2}^*\left(\epsilon-\frac{3}{4}g_\beta^* -\frac{5}{4}g_{\alpha_2}^*\right) =0\ .
\eeqn

The available solutions are then

\beq
(g_\beta^*,g_\lambda^*, g_{\kappa_2}^*, g_{\alpha_2}^*) = \left(0, 0, 0, 0\right)\sep  \left(\frac{2}{9}\epsilon, 0, 0, \frac{2}{3}\epsilon\right)\sep \left(\frac{4}{9}\epsilon, 0, 0, 0\right)\ .
\eeq

These are in the order of the Gaussian FP, the novel generic FP and the Wilson Fischer FP.

  By tuning our model parameters, we can determine its multicritical nature and locate new UCs. N.B. after each tuning, we must return to the diagrammatics and eliminate the respective contributions carefully.

  By tuning only $\beta=0$, we ultimately have the flow equations,

\beqn
\frac{\dd \ln g_\lambda}{\dd \ell} &=& \epsilon-\frac{3}{4}g_{\alpha_2} - \frac{9}{32}g_\lambda-\frac{9}{16}g_{\kappa_2}\ ,
\\
\frac{\dd \ln g_{\kappa_2}}{\dd \ell} &=& \epsilon-\frac{1}{8}\frac{g_\lambda g_{\alpha_2}}{g_{\kappa_2}}-\frac{3}{4}g_{\alpha_2}  - \frac{9}{32}g_\lambda-\frac{9}{16}g_{\kappa_2}\ ,
\\
\frac{\dd \ln g_{\alpha_2}}{\dd \ell} &=& \epsilon -\frac{5}{4}g_{\alpha_2}-\frac{3}{32}g_\lambda-\frac{3}{16}g_{\kappa_2}\ ,
\eeqn

The solutions of just these two ODEs reveal an inability for the couplings to stabilise nor take physical values by fine tuning only $\beta$.

As such, we require an additional fine tuning of $\kappa_2 $, if we try this,

\beqn
\frac{\dd \ln g_\lambda}{\dd \ell} &=& \epsilon-\frac{3}{4}g_{\alpha_2} - \frac{9}{32}g_\lambda\ ,
\\
\frac{\dd \ln g_{\alpha_2}}{\dd \ell} &=& \epsilon -\frac{5}{4}g_{\alpha_2}-\frac{3}{32}g_\lambda\ ,
\eeqn

They yield the solutions,

\beqn
\label{eq:bk0}
(g_\lambda^*, g_{\alpha_2}^*) =
\begin{cases}
(0,0)\ ,\\
(0,\frac{4}{5}\epsilon)\ ,\\
(\frac{16}{9}\epsilon,\frac{2}{3}\epsilon)\ ,\\
(\frac{32}{9}\epsilon, 0)\ .
\end{cases}
\eeqn

It is realized that even by fine tuning other parameters, we have determined all possible solutions to our system.

The stabilities of all FPs can now be determined by linear stability analyses around the FPs' locations, and the results are shown in Table 1 in the MT.

\section{Evaluation of scaling exponents}
\subsection{FP I}
We observe the generic behaviour of the model to force $\lambda$ and $\kappa_2$ to 0 upon coarse graining, after which there are only 2 non nonlinearities remaining.

The remaining constrained flow equations obeyed are then
\beqn
\frac{ \dd \ln \mu_x}{\dd \ell} &=& z -2\zeta\ ,
\\
\frac{ \dd \ln \alpha_2}{\dd \ell} &=& z + 2\chi_{\rho} - \frac{3}{4}g_\beta - \frac{5}{4}g_{\alpha_2}\ ,
\\
\frac{ \dd \ln \beta}{\dd \ell} &=& z+2\chi_\phi- \frac{9}{4}g_\beta - \frac{3}{4}g_{\alpha_2}\ ,
\\
\frac{ \dd \ln D}{\dd \ell} &=& z - \zeta - 2\chi_\phi + \epsilon -3\ ,
\\
\frac{\dd\ln \alpha_0}{\dd\ell} &=& z + \frac{1}{4\alpha_0}(2\mu_\bot\Lambda^2 - \alpha_0)(3g_\beta + g_{\alpha_2})\ .
\eeqn

Using the coupling coordinates $\epsilon\left[\frac{2}{9},0,0,\frac{2}{3}\right]$, we can fix the values of the raw parameters i.e $\mu_x, D, \beta, \alpha_2 \neq 0$

\beqn
z+ 2\chi_\rho/\chi_\phi - \epsilon = 0\sep z-2\zeta = 0\sep z - \zeta - 2\chi_\phi + \epsilon-3 =0\ .
\eeqn

The critical exponents are
\beqn
z=2\sep\zeta=1\sep\chi_\phi=\chi_\rho = -1 + \frac{1}{2}\epsilon\ ,
\eeqn
Similarly with $\alpha_0 \neq 0 $, the instability exponent can be derived as,
\beqn
2 + \frac{\epsilon}{3\alpha_0^*}(2\mu_\bot\Lambda^2 - \alpha_0^*)=0\sep \implies \alpha_0^* \approx [-\frac{1}{3}\epsilon + \mathcal{O}(\epsilon^2)]\mu_\bot\Lambda^2 \ .
\eeqn
The exponential runaway from a fixed point embedded in the critical manifold is characterised by $e^{y_{\alpha_0}d\ell}$, where $y_{\alpha_0}$ is termed as the instability exponent.

Taking $\alpha_0$'s flow equation

\beqn
\frac{\dd \alpha_0}{\dd\ell} = z\alpha_0 + \frac{1}{4}(2\mu_\bot\Lambda^2 - \alpha_0)(3g_\beta + g_{\alpha_2})\ ,
\eeqn

If we perturb the value of $\alpha_0 = \alpha_0^* + \delta\alpha_0$, and analyse the growth of the perturbation, we are able to quantitatively describe the runwaway from the critical surface.

The perturbation's growth follows

\beqn
\frac{\dd\delta\alpha_0}{\dd\ell} &=& \left(2 - \frac{1}{4}(3g_\beta + g_{\alpha_2})\right)\delta\alpha_0\ ,
\\
\frac{\dd\delta\alpha_0}{\dd\ell} &=& \left(2 - \frac{1}{3}\epsilon\right)\delta\alpha_0\ ,
\eeqn

where we define $y_{\alpha_0} = 2 - \frac{1}{4}(3g_\beta + g_{\alpha_2})$ such that for FP I 

\beqn
y_{\alpha_0} =2-\frac{1}{3}\epsilon\ .
\eeqn

\subsection{FP II}

FP II occurs when we tune $\alpha_2$ to 0. The surviving flow equations are

\beqn
\frac{ \dd \ln \mu_x}{\dd \ell} &=& z -2\zeta\ ,
\\
\frac{ \dd \ln \beta}{\dd \ell} &=& z+2\chi_\phi- \frac{9}{4}g_\beta\ ,
\\
\frac{ \dd \ln D}{\dd \ell} &=& z - \zeta - 2\chi_\phi + \epsilon -3\ ,
\\
\frac{\dd\ln \alpha_0}{\dd\ell} &=& z + \frac{1}{4\alpha_0}(2\mu_\bot\Lambda^2 - \alpha_0)(3g_\beta + g_{\alpha_2})\ .
\eeqn

With the coupling coordinates $\epsilon\left[\frac{4}{9},0,0,0\right]$, the constraints set are

\beqn
z -2=0\sep z-2\zeta=0\sep z+ 2\chi_\phi - \epsilon=0\ .
\eeqn

As a result, its critical exponents are the same
\beqn
z = 2\sep\zeta=1\sep\chi_\phi=\chi_\rho = -1 + \frac{1}{2}\epsilon\ .
\eeqn

Its stability exponent is similarly

\beq
\alpha_0^* \approx \left[-\frac{\epsilon}{3}+ \mathcal{O}(\epsilon^2)\right]\mu_\bot\Lambda^2\ .
\eeq

With the growth equation

\beqn
\frac{\dd\delta\alpha_0}{\dd\ell} &=& \left(2 - \frac{1}{4}(3g_\beta + g_{\alpha_2})\right)\delta\alpha_0\ ,
\\
\frac{\dd\delta\alpha_0}{\dd\ell} &=& \left(2 - \frac{1}{3}\epsilon\right)\delta\alpha_0\ .
\eeqn

The instability exponent is
\beqn
y_{\alpha_0} = 2 - \frac{1}{3}\epsilon\ .
\eeqn

\subsection{FP III}

FP III occurs when we tune both $\beta$ and $\kappa_2$ to 0. The surviving flow equations are

\beqn
\frac{\dd\lambda}{\dd\ell} &=& z - \zeta + \chi_\phi - \frac{3}{8}g_{\alpha_2}\ ,
\\
\frac{\dd\mu_x}{\dd\ell} &=& z - 2\zeta + \frac{3}{16}g_\lambda\ ,
\\
\frac{\dd\alpha_2}{\dd\ell} &=& z + 2\chi_\rho - \frac{5}{4}g_{\alpha_2}\ ,
\\
\frac{ \dd \ln D}{\dd \ell} &=& z - \zeta - 2\chi_\phi + \epsilon -3\ ,
\\
\frac{\dd\ln \alpha_0}{\dd\ell} &=& z + \frac{1}{4\alpha_0}(2\mu_\bot\Lambda^2 - \alpha_0)(3g_\beta + g_{\alpha_2})\ .
\eeqn

With the coupling coordinates $\epsilon\left[0,\frac{16}{9}, 0, \frac{2}{3}\right]$, the constraints set are

\beq
z-\zeta+\chi_\phi - \frac{1}{4}\epsilon = 0\sep z- 2\zeta + \frac{1}{3}\epsilon=0\sep z+ 2\chi_\rho -\frac{5}{6}\epsilon=0\sep z - \zeta - 2\chi_\phi + \epsilon -3 =0\ .
\eeq

The critical exponents are thus
\beqn
z = 2\sep \zeta = 1+ \frac{1}{6}\epsilon\sep \chi_\phi=\chi_\rho = -1 + \frac{5}{12}\epsilon\ .
\eeqn

The instability exponent can be derived as

\beqn
2\alpha_0^* + \frac{\epsilon}{6}(2\mu_\bot\Lambda^2-\alpha_0^*) = 0\sep \alpha_0^* = \left[-\frac{1}{6}\epsilon + \mathcal{O}(\epsilon^2)\right]\mu_\bot\Lambda^2\ .
\eeqn

With the growth equation

\beqn
\frac{\dd\delta\alpha_0}{\dd\ell} &=& \left(2 - \frac{1}{4}(3g_\beta + g_{\alpha_2})\right)\delta\alpha_0\ ,
\\
\frac{\dd\delta\alpha_0}{\dd\ell} &=& \left(2 - \frac{1}{6}\epsilon\right)\delta\alpha_0\ .
\eeqn

The instability exponent is
\beqn
y_{\alpha_0} = 2 - \frac{1}{6}\epsilon\ .
\eeqn

\subsection{FP IV}
FP IV occurs when we tune $\beta, \kappa_2, \alpha_2$ to zero. The surviving flow equations are,

\beqn
\frac{\dd\mu_x}{\dd\ell} &=& z - 2\zeta + \frac{3}{16}\epsilon\ ,
\\
\frac{\dd\lambda}{\dd\ell} &=& z -\zeta + \chi_\phi\ ,
\\
\frac{ \dd \ln D}{\dd \ell} &=& z - \zeta - 2\chi_\phi + \epsilon -3\ ,
\\
\frac{\dd\ln \alpha_0}{\dd\ell} &=& z + \frac{1}{4\alpha_0}(2\mu_\bot\Lambda^2 - \alpha_0)(3g_\beta + g_{\alpha_2})\ .
\eeqn

With the coupling coordinates $\epsilon\left[0,\frac{32}{9},0,0\right]$, the constraints set are,

\beqn
z - 2\zeta + \frac{2}{3}\epsilon =0\sep  z -\zeta + \chi_\phi=0\sep z - \zeta - 2\chi_\phi + \epsilon -3=0\ .
\eeqn

The critical exponents are thus,

\beqn
z =2\sep \zeta = 1+ \frac{1}{3}\epsilon\sep\chi_\phi=\chi_\rho=-1+\frac{1}{3}\epsilon\ .
\eeqn

Given both $g_\beta$ and $g_{\alpha_2}$ are 0, the instability exponent is immediately realised as \beqn
y_{\alpha_0} = 2\ .
\eeqn

\subsection{FP V}

FP V occurs when we tune $\beta, \kappa_2, \lambda$ to zero. The surviving flow equations are,

\beqn
\frac{\dd\mu_x}{\dd\ell} &=& z - 2\zeta\ ,
\\
\frac{\dd\alpha_2}{\dd\ell} &=& z + 2\chi_\rho - \frac{5}{4}g_{\alpha_2}\ ,
\\
\frac{ \dd \ln D}{\dd \ell} &=& z - \zeta - 2\chi_\phi + \epsilon -3\ .
\eeqn

With the coupling coordinates $\epsilon\left[0,0,0,\frac{4}{5}\right]$, the constraints set are,

\beqn
z-2 = 0\sep z-2\zeta=0\sep z+2\chi_\rho-\epsilon=0\sep z - \zeta - 2\chi_\phi + \epsilon -3 =0\ .
\eeqn

The critical exponents are thus,

\beqn
z=2\sep \zeta =1 \sep \chi_\phi=\chi_\rho=-1+\frac{1}{2}\epsilon\ .
\eeqn

The instability exponent can be derived as,
\beqn
2\alpha_0^* + \frac{\epsilon}{5}(2\mu_\bot\Lambda^2-\alpha_0^*) = 0\sep \alpha_0 = \left[-\frac{1}{5}\epsilon+\mathcal{O}(\epsilon^2)\right]\mu_\bot\Lambda^2
\eeqn

With the growth equation,

\beqn
\frac{\dd\delta\alpha_0}{\dd\ell} &=& \left(2 - \frac{1}{4}(3g_\beta + g_{\alpha_2})\right)\delta\alpha_0\ ,
\\
\frac{\dd\delta\alpha_0}{\dd\ell} &=& \left(2 - \frac{1}{5}\epsilon\right)\delta\alpha_0\ .
\eeqn

The instability exponent is,
\beqn
y_{\alpha_0} = 2 - \frac{1}{5}\epsilon\ .
\eeqn

\section{Calculation of the critical exponent $y_{\alpha_0}$}
This section involves solving the $\beta$ and $\alpha_2$ bubble diagrams, in order to determine the instability of the FPs on the critical surface. To do this we need to alter the propagator such that it includes $\alpha_0$,

In Fourier space, the full linear equations are,
\beq
\begin{pmatrix}
    -\ii\omega+Kp_\bot^2 & \ii \gamma p_x\\
    \ii\kappa_1 p_x & -\ii\omega + \mu_xp_x^2 + \mu_\bot p_\bot^2 + \alpha_0 \\
\end{pmatrix}
\begin{pmatrix}
    \rho\\
    \phi\\
\end{pmatrix}=
\begin{pmatrix}
    0\\
    f_x
\end{pmatrix}\ ,
\eeq
\beqn
\begin{pmatrix}
    \rho\\
    \phi\\
\end{pmatrix}
&=& \frac{-1}{\omega^2-\Pi(\bp)+\ii\omega\Gamma(\bp)}
\begin{pmatrix}
    -\ii\omega+\mu_x p_x^2 + \mu_\bot p_\bot^2+ \alpha_0 & -\ii \gamma p_x \\
    -\ii\kappa_1p_x & -\ii\omega+K p_\bot^2 \\
\end{pmatrix}
\begin{pmatrix}
   0\\
    f_x   
\end{pmatrix}\ .
\eeqn

The altered propagator becomes,

\beqn
{G}_\phi(\bp) &=& \frac{\ii\omega-Kp_\bot^2}{\omega^2-\Pi(\bp)+\ii\omega\Gamma(\bp)}\ .
\eeqn

We define the altered quantities

\beqn
\Pi(\bp) = (\gamma\kappa_1+K\mu_xp_\bot^2) p_x^2 - (\alpha_0-\mu_\bot p_\bot^2)Kp_\bot^2\sep \Gamma(\bp) = \mu_x p_x^2 + (\mu_\bot+K) p_\bot^2+\alpha_0 \ .
\eeqn

\begin{figure}[H]
    \centering
    \includegraphics[width=0.17\linewidth]{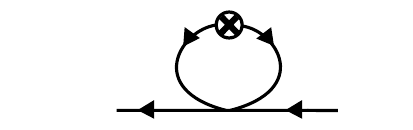}
    \caption{$\beta$ bubble diagram}
\end{figure}

This diagram represents a contribution to $\delta(\pp_t\phi)$ to $\pp_t\phi$ given by,
\beqn
&=& -3\beta\int_{\bq,\Omega}\frac{2D(\Omega^2+K^2q_\bot^4)}{(\Omega^2-\Pi(\bq))^2 +\Omega^2\Gamma(\bq)^2}\ .
\eeqn
 Admitting the treatment justified a posteriori: $\mu_\bot+K\rightarrow\mu_\bot$, the first term in the numerator contributes
\beqn
&=& -3\beta\int_{\bq,\Omega}\frac{2D\Omega^2}{(\Omega^2-\Pi(\bq))^2 +\Omega^2\Gamma(\bq)^2}\ ,
\\
&=& (-3\beta)2D\int_{\bq}\frac{1}{2}\frac{1}{\Gamma(\bq)}\ ,
\\
&=& (-3\beta)2D\int_{\bq_\bot}\frac{1}{4}\frac{1}{\mu_x^{1/2}(\mu_\bot q_\bot^2+\alpha_0)^{1/2}}\ ,
\\
&=& -3\beta D\frac{S_{d-1}}{(2\pi)^{d-1}}\frac{1}{2\mu_x^{1/2}}\int \dd q_\bot q_\bot^{d-2}\frac{1}{(\mu_\bot q_\bot^2+\alpha_0)^{1/2}}\ .
\eeqn

We admit a neat integral trick using a Taylor expansion given an infinitesimal $\dd\ell$ which simplifies the problem to
\beqn
 &=& -\frac{3}{2}\frac{\beta D}{\mu_x^{1/2}}\frac{S_{d-1}}{(2\pi)^{d-1}}\frac{\Lambda^{d-1} d\ell}{\sqrt{\mu_\bot\Lambda^2+\alpha_0}}\ .
\eeqn

The second term in the numerator contributes
\beqn
&=& (-3\beta)2D\int_{\bq,\Omega}\frac{K^2q_\bot^4}{(\Omega^2-\Pi(\bq))^2 +\Omega^2\Gamma(\bq)^2}\ ,
\\
&=& (-3\beta)2D\int_{\bq}\frac{1}{2}\frac{K^2q_\bot^4}{\Pi(\bq)\Gamma(\bq)}\ .
\eeqn

This term is irrelevant due to the divergence of the denominator under RG flow. Therefore
\beqn
\delta\alpha_0 = -\frac{S_{d-1}}{(2\pi)^{d-1}}\frac{3\beta D}{2\mu_x^{1/2}}\frac{\Lambda^{d-1} d\ell}{\sqrt{\mu_\bot\Lambda^2+\alpha_0}}\ .
\eeqn

To realise the contribution to the $\alpha_0$ flow equation, we perform an expansion assuming small $\alpha_0$,
\beqn
\left(\mu_\bot\Lambda^2 + \alpha_0\right)^{-1/2}\approx \frac{1}{\mu_\bot^{1/2}\Lambda}\left(1-\frac{\alpha_0}{2\mu_\bot\Lambda^2}\right)\ .
\eeqn

We then factor out $\alpha_0$
\beqn
\alpha_0\left[\frac{1}{\mu_\bot^{1/2}\Lambda}\left(\frac{1}{\alpha_0}-\frac{1}{2\mu_\bot\Lambda^2}\right)\right] = \alpha_0\left[\frac{2\mu_\bot\Lambda^2 - \alpha_0}{2\alpha_0\mu_\bot^{3/2}\Lambda^3}\right]\ .
\eeqn

Therefore, by removing $-\alpha_0$, this diagram's contribution is
\beqn
&=& \frac{3\beta D}{2\mu_x^{1/2}}\frac{S_{d-1}}{(2\pi)^{d-1}}\Lambda^{d-4}\frac{2\mu_\bot\Lambda^2 - \alpha_0}{2\alpha_0\mu_\bot^{3/2}} = \frac{3g_\beta}{4\alpha_0}(2\mu_\bot\Lambda^2 - \alpha_0)\ .
\eeqn

\begin{figure}[H]
    \centering
    \includegraphics[width=0.15\linewidth]{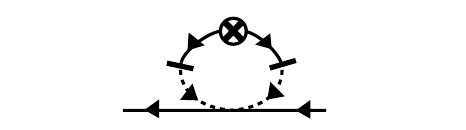}
    \caption{$\alpha_2$ bubble diagram}
\end{figure}
This diagram represents a contribution to $\delta(\pp_t\phi)$ to $\pp_t\phi$ given by
\beqn
&=& -\alpha_22D\int_{\bq,\Omega}\frac{\gamma^2q_x^2}{(\Omega^2-\Pi(\bq))^2 +\Omega^2\Gamma(\bq)^2}\ ,
\\
&\approx&-\alpha_22D\int_{\bq}\frac{1}{2}\frac{\gamma^2}{\gamma\kappa_1+K\mu_xq_\bot^2}\frac{1}{\Gamma(\bq)}\ ,
\\
&=& -\frac{1}{2}\frac{\gamma\alpha_2 D}{\kappa_1\mu_x^{1/2}}\frac{S_{d-1}}{(2\pi)^{d-1}}\int dq_\bot q_\bot^{d-2} \frac{1}{\sqrt{\mu_\bot q_\bot^2 + \alpha_0}}\left[1+\frac{K\mu_xq_\bot^2}{\gamma\kappa_1}\right]^{-1}\ ,
\\
&=& -\frac{\gamma\alpha_2 D}{2\kappa_1\mu_x^{1/2}}\frac{S_{d-1}}{(2\pi)^{d-1}}\frac{\Lambda^{d-1} d\ell}{\sqrt{\mu_\bot\Lambda^2 + \alpha_0}}\ ,
\eeqn

where higher order terms will be irrelevant due to the divergence of the denominator under RG flow.

Therefore removing $-\alpha_0$  results in its overall contribution as
\beqn
&=& \frac{\gamma\alpha_2 D}{4\kappa_1\mu_x^{1/2}\mu_\bot^{3/2}}\frac{S_{d-1}}{(2\pi)^{d-1}}\Lambda^{d-4}\frac{2\mu_\bot\Lambda^2 - \alpha_0}{\alpha_0}\ ,
\\
&=& \frac{g_{\alpha_2}}{4\alpha_0}(2\mu_\bot\Lambda^2 - \alpha_0)\ .
\eeqn

Therefore after adding contributions from both diagrams, the flow equation can be written as
\beqn
\frac{\dd\ln \alpha_0}{\dd\ell} = z + \frac{1}{4\alpha_0}(2\mu_\bot\Lambda^2 - \alpha_0)(3g_\beta + g_{\alpha_2})\ .
\eeqn

\appendix

\section{Calculation of  2-vertex diagrams}

\label{app2-vertex}
As a reminder, our integral is of the form:
\beq
\int_{\tilde{\bq}}= S_{d-1}\int^{\Lambda}_{\Lambda e^{-d\ell}}\frac{dq_\bot q^{d-2}_\bot}{(2\pi)^{d-1}} \int^{+\infty}_{-\infty}\frac{dq_x}{2\pi}\int^{+\infty}_{-\infty}\frac{d\Omega}{2\pi}\ ,
\eeq
where
\beq
S_{d-1}=\frac{2\pi^{\frac{d-1}{2}}}{\Gamma(\frac{d-1}{2})}\ ,
\eeq
is the surface area of the $d-1$ dimensional unit sphere.

We evaluate the frequency and angular integrals analytically via Cauchy's theorem. This is applied by selecting the upper region of the complex plane enclosing only the positive poles of the propagators. Prior to any integration, we assess which diagrammatics the integral contributes to, and only after can we prescribe the appropriate hydrodynamic limit.

\begin{figure}[H]
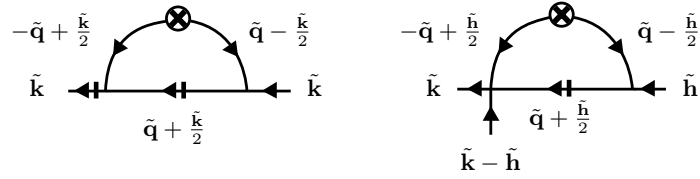

\begin{picture}(0,0)
\put(95, 3){\small $\tilde{\bk}$}
\put(-10, 3){\small $\tilde{\bk}$}
\put(73, 25){\small $\tilde{\bq} - \frac{\tilde{\bk}}{2}$}
\put(-17, 25){\small $-\tilde{\bq} + \frac{\tilde{\bk}}{2}$}
\put(33, -12){\small $\tilde{\bq} + \frac{\tilde{\bk}}{2}$}
\put(140, 3){\small $\tilde{\bk}$}
\put(237, 3){\small $\tilde{\bh}$}
\put(153, -25){\small $\tilde{\bk}-\tilde{\bh}$}
\put(179, -8){\small $\tilde{\bq}+\frac{\tilde{\bh}}{2}$}
\put(220, 25){\small $\tilde{\bq}-\frac{\tilde{\bh}}{2}$}
\put(130, 25){\small $-\tilde{\bq}+\frac{\tilde{\bh}}{2}$}
\end{picture}
\vspace{0.3cm}
    \centering
    \includegraphics[width=0.17\linewidth]{2LL-M1G.pdf}
     \hspace{1.8cm}
    \raisebox{-0.43cm}{
    \includegraphics[width=0.16\linewidth]{2BL-L1G.pdf}}
    \vspace{0.2cm}
    \caption{Symmeterised internal momenta loops.}
    \label{fig:symm}
\end{figure}

Further, the internal momenta are symmetrized as shown in Fig.~\ref{fig:symm}.

There are multiple variations of internal loops, each corresponding to a unique integral calculation. We can always trivially take the hydrodynamic limit of the external frequency immediately by setting $\omega=0$, but depending on the diagram, we cannot do so with the external momenta as we need to extract a $k_x$ from each branch. Having taken care to symmeterise the integrand, as seen in (\ref{fig:symm}), and chosen $\bk$ to represent the external momenta, they take the following forms below,

We make the following  simplification:
\beqn
\Gamma(\bq) = \mu_xq_x^2+(\mu_\bot+K)q_\bot^2 = \mu_xq_x^2 +\mu_\bot q_\bot^2\ ,
\eeqn
which is justified a posteriori as $\mu_\bot,K$ behave identically from an RG perspective.

\subsection{Variation I}

\beqn
I_1 &=& \int_{\bq,\Omega}2D\left(\frac{k_x}{2}+q_x\right)G_\phi\left(\frac{\bk}{2}+\bq, \frac{\omega}{2}+\Omega\right)G_\phi\left(\bq-\frac{\bk}{2},\Omega - \frac{\omega}{2}\right)G_\phi\left(\frac{\bk}{2}-\bq, \frac{\omega}{2}-\Omega\right)\ ,
\eeqn

We set $\omega=0$ and the integrand simplifies as,

\beqn
&=&\frac{(\ii\Omega -Kq_\bot^2)\left(\frac{k_x}{2}+q_x\right)}{[\Omega^2 -  \Pi(\frac{\bk}{2}+\bq)^2+ \ii\Omega\Gamma(\frac{\bk}{2}+\bq)]}\frac{\ii\Omega-Kq_\bot^2}{[\Omega^2 - \Pi(\bq-\frac{\bk}{2})^2 + \ii\Omega\Gamma(\bq-\frac{\bk}{2})]}\frac{-\ii\Omega-Kq_\bot^2}{[\Omega^2 - \Pi(\frac{\bk}{2}-\bq)^2 - \ii\Omega\Gamma(\frac{\bk}{2}-\bq)]}\ .
\eeqn

The numerator expands to
\beqn
&=& \left(\frac{k_x}{2}+q_x\right)\left(\ii\Omega - Kq_\bot^2\right)\left(\ii\Omega - Kq_\bot^2\right)\left(-\ii\Omega - Kq_\bot^2\right)\ ,
\\
&=& \left(\frac{k_x}{2}+q_x\right)\left(\ii\Omega^3-K\Omega^2q_\bot^2 + \ii K^2\Omega q_\bot^4 -K^3 q_\bot^6 \right)\ ,
\eeqn

Because we are also dealing with contributions to $\pp_x$ or $\pp_x^2$, these correspond to $\lambda$, $\kappa_2$ or $\mu_x$, we need to reproduce $\mathcal{O}(k_x)$ terms from the integrand. 

Having realised the $Kq_\bot^2$ terms do not contribute to the $\mathcal{O}(k_x)$ terms from the calculation in variation VII, the $\mathcal{O}(k_x)$ terms provide
\beqn
&=& 2Dk_x\int_{\bq,\Omega}\frac{1}{2}\frac{\Omega^4\Gamma(\bq)}{[(\Omega^2 - \Pi(\bq))^2 + \Omega^2\Gamma(\bq)^2][(\Omega^2 - \Pi(\bq)^2 + \Omega^2\Gamma(\bq)^2]}\ ,
\\
&=& 2Dk_x\int_{\bq}\frac{1}{8}\frac{1}{\Gamma(\bq)^2}\ ,
\\
&=& k_x\frac{1}{16}\frac{D}{\mu_x^{1/2}\mu_\bot^{3/2}}\frac{S_{d-1}}{(2\pi)^{d-1}}\Lambda^{d-4}d\ell\ .
\eeqn

For $\mathcal{O}(k_x^0)$ terms in the numerator, we want to extract $\mathcal{O}(k_x)$ terms from their denominator, as such we need to expand every denominator to $\mathcal{O}(k_x)$,

\beqn
\frac{1}{[\Omega^2 - \Pi(\frac{\bk}{2}+\bq)^2+ \ii\Omega\Gamma(\frac{\bk}{2}+\bq)]}&\approx& \frac{1}{[\Omega^2 - \Pi(\bq) + \ii\Omega\Gamma(\bq)]} + \frac{k_xq_x\left[(\gamma\kappa_1+K\mu_xq_\bot^2)-\ii\mu_x\Omega\right]}{[(\Omega^2 - \Pi(\bq) + \ii\Omega\Gamma(\bq))^2]}\ ,
\\
\frac{1}{[\Omega^2 -\Pi(\bq-\frac{\bk}{2})^2 + \ii\Omega\Gamma(\bq-\frac{\bk}{2})]}&\approx& \frac{1}{[\Omega^2 - \Pi(\bq) + \ii\Omega\Gamma(\bq)]}+\frac{k_xq_x\left[-(\gamma\kappa_1+K\mu_xq_\bot^2)+\ii\mu_x\Omega\right]}{[(\Omega^2 - \Pi(\bq) + \ii\Omega\Gamma(\bq))^2]}\ ,
\\
\frac{1}{[\Omega^2 -\Pi(\frac{\bk}{2}-\bq)^2 - \ii\Omega\Gamma(\frac{\bk}{2}-\bq)]}&\approx&  \frac{1}{[\Omega^2 - \Pi(\bq) - \ii\Omega\Gamma(\bq)]}-\frac{k_xq_x\left[(\gamma\kappa_1+K\mu_xq_\bot^2)+\ii\mu_x\Omega\right]}{[(\Omega^2 - \Pi(\bq) - \ii\Omega\Gamma(\bq))^2]}\ .
\eeqn

The only $\mathcal{O}(k_x)$ term is thus,

\beqn
\label{eq:denom}
&=& -\frac{k_xq_x\left[(\gamma\kappa_1+K\mu_xq_\bot^2)+\ii\mu_x\Omega\right]}{[(\Omega^2 - \Pi(\bq))^2 + \Omega^2\Gamma(\bq)^2][(\Omega^2 - \Pi(\bq))^2 + \Omega^2\Gamma(\bq)^2]}
\eeqn 

For the $q_x$ terms, we multiply the expanded denominator as seen in (\ref{eq:denom}) with the original numerator such that the integrand becomes,

\beqn
&=&\frac{k_xq_x^2\left[\mu_x\Omega^4+K(\gamma\kappa_1+K\mu_xq_\bot^2)\Omega^2 q_\bot^2 + \mu_xK^2\Omega^2 q_\bot^4 +K^3(\gamma\kappa_1+K\mu_xq_\bot^2)q_\bot^6\right]}{[(\Omega^2 - \Pi(\bq))^2 + \Omega^2\Gamma(\bq)^2][(\Omega^2 - \Pi(\bq))^2 + \Omega^2\Gamma(\bq)^2]}\ ,
\eeqn

where we have neglected any odd $\Omega$ terms as they will evaluate to 0.

The first term in the numerator evaluates as

\beqn
&=& 2Dk_x\int_{\bq,\Omega}\frac{\mu_x\Omega^4q_x^2}{[(\Omega^2 - \Pi(\bq))^2 + \Omega^2\Gamma(\bq)^2][(\Omega^2 - \Pi(\bq)^2 + \Omega^2\Gamma(\bq)^2]}\ ,
\\
&=& 2Dk_x\int_{\bq}\frac{1}{4}\frac{\mu_x q_x^2}{\Gamma(\bq)^3}\ ,
\\
&=& k_x\frac{1}{32}\frac{D}{\mu_x^{1/2}\mu_\bot^{3/2}}\frac{S_{d-1}}{(2\pi)^{d-1}}\Lambda^{d-4}d\ell\ .
\eeqn

The second term evaluates as
\beqn
&=& 2Dk_x\int_{\bq,\Omega}\frac{K(\gamma\kappa_1+K\mu_xq_\bot^2)\Omega^2 q_x^2q_\bot^2 }{[(\Omega^2 - \Pi(\bq))^2 + \Omega^2\Gamma(\bq)^2][(\Omega^2 - \Pi(\bq))^2 + \Omega^2\Gamma(\bq)^2]}\ ,
\\
&=& 2Dk_x\int_{\bq}\frac{1}{4}\frac{ K(\gamma\kappa_1+K\mu_xq_\bot^2)q_x^2 q_\bot^2}{\Pi(\bq)\Gamma(\bq)^3}\ ,
\\
&\approx& 2Dk_x\int_{\bq}\frac{1}{4}\frac{Kq_\bot^2}{\Gamma(\bq)^3}\ ,
\\
&=& k_x\frac{3}{32}\frac{KD}{\mu_x^{1/2}\mu_\bot^{5/2}}\frac{S_{d-1}}{(2\pi)^{d-1}}\Lambda^{d-4}d\ell.
\eeqn

The third term evaluates as 

\beqn
&=& 2Dk_x\int_{\bq,\Omega}\frac{\mu_xK^2\Omega^2q_x^2q_\bot^4 }{[(\Omega^2 -\Pi(\bq))^2 + \Omega^2\Gamma(\bq)^2][(\Omega^2 - \Pi(\bq))^2 + \Omega^2\Gamma(\bq)^2]}\ ,
\\
&=& 2Dk_x\int_{\bq}\frac{1}{4}\frac{\mu_xK^2q_x^2q_\bot^4}{\Pi(\bq)\Gamma(\bq)^3}\ ,
\\
&\approx& 2Dk_x\int_{\bq}\frac{1}{4}\frac{\mu_xK^2q_\bot^4}{\gamma\kappa_1+K\mu_xq_\bot^2}\frac{1}{\Gamma(\bq)^3}\ ,
\\
&=& k_x\frac{1}{32}\frac{K^2D}{\gamma\kappa_1\mu_\bot^{5/2}}\Lambda^dd\ell\ .
\eeqn

This term is irrelevant as it does not produce $\mathcal{O}(\Lambda^{d-4})$ terms and will have a divergence in its denominator under RG flow.

The fourth term evaluates as
\beqn
&=&2Dk_x\int_{\bq,\Omega}\frac{K^3(\gamma\kappa_1+K\mu_xq_\bot^2)q_x^2q_\bot^6}{[(\Omega^2 - \Pi(\bq))^2 + \Omega^2\Gamma(\bq)^2][(\Omega^2 - \Pi(\bq))^2 + \Omega^2\Gamma(\bq)^2]}\ ,
\\
&=& 2Dk_x\int_{\bq,\Omega}\frac{K^3(\gamma\kappa_1+K\mu_xq_\bot^2)q_x^2q_\bot^6}{4}\left[\frac{\Pi(\bq)+\Gamma(\bq)^2}{\Pi(\bq)^3\Gamma(\bq)^3}\right]\ ,
\\
&\approx&2Dk_x\int_{\bq,\Omega}\frac{K^3q_\bot^6}{4}\left[\frac{\Pi(\bq)+\Gamma(\bq)^2}{\Pi(\bq)^2\Gamma(\bq)^3}\right]\ .
\eeqn

This term is irrelevant as it does not produce $\mathcal{O}(\Lambda^{d-4})$ terms and will have a divergence in its denominator udner RG flow.

Together the $q_x$ terms evaluate to

\beqn
k_x \left[\frac{1}{32}\frac{D}{\mu_x^{1/2}\mu_\bot^{3/2}}\frac{S_{d-1}}{(2\pi)^{d-1}}\Lambda^{d-4}d\ell+\frac{3}{32}\frac{KD}{\mu_x^{1/2}\mu_\bot^{5/2}}\frac{S_{d-1}}{(2\pi)^{d-1}}\Lambda^{d-4}d\ell\right]\ ,
\eeqn

such that after combining the $k_x$ term, the overall integral evaluates to

\beqn
\int_{\bq,\Omega} = k_x \left[\frac{3}{32}\frac{D}{\mu_x^{1/2}\mu_\bot^{3/2}}\frac{S_{d-1}}{(2\pi)^{d-1}}\Lambda^{d-4}d\ell+\frac{3}{32}\frac{KD}{\mu_x^{1/2}\mu_\bot^{5/2}}\frac{S_{d-1}}{(2\pi)^{d-1}}\Lambda^{d-4}d\ell\right]\ .
\eeqn

There are several diagrammatics which have two external legs with this internal structure. Having made a change of variable $k_x\rightarrow h_x$, they have the form,

\beqn
&=& \left[\frac{3}{32}\frac{D}{\mu_x^{1/2}\mu_\bot^{3/2}}\frac{S_{d-1}}{(2\pi)^{d-1}}\Lambda^{d-4}d\ell+\frac{3}{32}\frac{KD}{\mu_x^{1/2}\mu_\bot^{5/2}}\frac{S_{d-1}}{(2\pi)^{d-1}}\Lambda^{d-4}d\ell\right]\int_{\bh}h_x\phi(\bh)\phi(\bk-\bh)\ ,
\\
&=& k_x\left[\frac{3}{32}\frac{D}{\mu_x^{1/2}\mu_\bot^{3/2}}\frac{S_{d-1}}{(2\pi)^{d-1}}\Lambda^{d-4}d\ell+\frac{3}{32}\frac{KD}{\mu_x^{1/2}\mu_\bot^{5/2}}\frac{S_{d-1}}{(2\pi)^{d-1}}\Lambda^{d-4}d\ell\right]\int_{\bh}\phi(\bh)\phi(\bk-\bh)\ ,
\eeqn

We shall name the two sub variations of this integral as,

\beqn
I_{1.1} &=&  \left[\frac{3}{32}\frac{D}{\mu_x^{1/2}\mu_\bot^{3/2}}\frac{S_{d-1}}{(2\pi)^{d-1}}\Lambda^{d-4}d\ell+\frac{3}{32}\frac{KD}{\mu_x^{1/2}\mu_\bot^{5/2}}\frac{S_{d-1}}{(2\pi)^{d-1}}\Lambda^{d-4}d\ell\right]\ ,
\\
I_{1.2} &=& \left[\frac{3}{64}\frac{D}{\mu_x^{1/2}\mu_\bot^{3/2}}\frac{S_{d-1}}{(2\pi)^{d-1}}\Lambda^{d-4}d\ell+\frac{3}{64}\frac{KD}{\mu_x^{1/2}\mu_\bot^{5/2}}\frac{S_{d-1}}{(2\pi)^{d-1}}\Lambda^{d-4}d\ell\right]\ .
\eeqn

\subsection{Variation II}

The next variation is,

\beqn
I_2 &=& \int_{\bq,\Omega}2D\left(\frac{k_x}{2}+q_x\right)G_\rho\left(\frac{\bk}{2}+\bq, \frac{\omega}{2}+\Omega\right)G_\phi\left(\bq-\frac{\bk}{2},\Omega - \frac{\omega}{2}\right)G_\rho\left(\frac{\bk}{2}-\bq, \frac{\omega}{2}-\Omega\right)\ ,
\eeqn

We set $\omega=0$ and the integrand simplifies as,

\beqn
&=& \frac{\ii\gamma(\frac{k_x}{2}+q_x)^2}{[\Omega^2 - \Pi(\frac{\bk}{2}+\bq)^2 + \ii\Omega\Gamma(\frac{\bk}{2}+\bq)]}\cdot\frac{\ii\Omega - Kq_\bot^2}{[\Omega^2 - \Pi(\bq-\frac{\bk}{2})^2 + \ii\Omega\Gamma(\bq-\frac{\bk}{2})]}\frac{\ii\gamma(\frac{k_x}{2}-q_x)}{[\Omega^2 - \Pi(\frac{\bk}{2}-\bq)^2- \ii\Omega\Gamma(\frac{\bk}{2}-\bq)]}\ .
\eeqn

Because we are dealing with contributions to $\pp_x$ or $\pp_x^2$, we also need to reproduce $\mathcal{O}(k_x)$ terms from the integrand.

Expanding to $\mathcal{O}(k_x)$ and $\mathcal{O}(k_x^0)$ terms, the numerator evaluates to,

\beqn
&=& -\gamma^2\left(\frac{k_x}{2}+q_x\right)^2\left(\frac{k_x}{2}-q_x\right)(\ii\Omega-Kq_\bot^2)\ ,
\\
&=&\gamma^2\left(\frac{k_x}{2}q_x^2+q_x^3\right)(\ii\Omega-Kq_\bot^2)
\eeqn

The $\mathcal{O}(k_x)$ integrand evaluates to
\beqn
&=& 2Dk_x\int_{\bq,\Omega}\frac{\gamma^2q_x^2}{2}\frac{\Omega^2\Gamma(\bq)- Kq_\bot^2(\Omega^2-\Pi(\bq))}{[(\Omega^2 - \Pi(\bq))^2 + \Omega^2\Gamma(\bq)^2][(\Omega^2 - \Pi(\bq))^2 + \Omega^2\Gamma(\bq)^2]}\ .
\eeqn

The first term in the numerator evaluates to

\beqn
&=& 2Dk_x\int_{\bq}\frac{1}{8}\frac{\gamma^2q_x^2 }{\Pi(\bq)\Gamma(\bq)^2}\ ,
\\
&\approx& 2Dk_x\int_{\bq_\bot}\frac{1}{16}\frac{1}{\gamma\kappa_1+K\mu_xq_\bot^2}\frac{\gamma^2}{\mu_x^{1/2}\mu_\bot^{3/2}q_\bot^3}\ ,
\\
&=&  2Dk_x\int_{\bq_\bot}\frac{1}{16}\frac{\gamma}{\kappa_1\mu_x^{1/2}\mu_\bot^{3/2}q_\bot^3}\left[1+\frac{K\mu_x q_\bot^2}{\gamma\kappa_1}\right]^{-1}\ ,
\\
&=& k_x\frac{1}{16}\frac{\gamma D}{\kappa_1\mu_x^{1/2}\mu_\bot^{3/2}}\frac{S_{d-1}}{(2\pi)^{d-1}}\Lambda^{d-4}d\ell + {\rm h.o.t}\ .
\eeqn

The second term in the numerator evaluates to

\beqn
&=& -2Dk_x\int_{\bq}\frac{\gamma^2Kq_x^2q_\bot^2}{8}\left[\frac{1}{\Pi(\bq)\Gamma^3(\bq)}-\frac{\Pi(\bq)^2+\Pi(\bq)\Gamma(\bq)^2}{\Pi(\bq)^3\Gamma(\bq)^3}\right]\ ,
\\
&=& 2Dk_x\int_{\bq}\frac{\gamma^2Kq_x^2q_\bot^2}{8}\frac{1}{\Pi(\bq)^2\Gamma(\bq)}\ .
\eeqn

This term is irrelevant as it does not produce $\mathcal{O}(\Lambda^{d-4})$ terms.

From the $\mathcal{O}(k_x^0)$ term in the numerator, we want to extract $\mathcal{O}(k_x)$ terms. As such, we need to expand every denominator to $\mathcal{O}(k_x)$, which is the result derived in (\ref{eq:denom}).

For the $q_x^3$ terms, we multiply the original numerator with this result such that the integrand becomes,

\beqn
&=& \frac{k_x\gamma^2q_x^4\left[\mu_x\Omega^2+K(\gamma\kappa_1+K\mu_xq_\bot^2)q_\bot^2)\right]}{[(\Omega^2 - \Pi(\bq))^2 + \Omega^2\Gamma(\bq)^2][(\Omega^2 - \Pi(\bq))^2 + \Omega^2\Gamma(\bq)^2]}\ 
\eeqn

The first term in the numerator evaluates to
\beqn
&=& 2Dk_x\int_{\bq}\frac{1}{4}\frac{\gamma^2\mu_x q_x^4}{\Pi(\bq)\Gamma(\bq)^3}\ ,
\\
&\approx& 2Dk_x\int_{\bq}\frac{1}{4}\frac{\gamma^2\mu_x q_x^2}{\gamma\kappa_1+Kq_\bot^2}\frac{1}{\Gamma(\bq)^3}\ ,
\\
&=& k_x\frac{1}{32}\frac{\gamma D}{\kappa_1\mu_x^{1/2}\mu_\bot^{3/2}}\frac{S_{d-1}}{(2\pi)^{d-1}}\Lambda^{d-4}d\ell+{\rm h.o.t}\ .
\eeqn

The second term in the numerator evaluates to

\beqn
&=& 2Dk_x\int_{\bq}\frac{\gamma^2K(\gamma\kappa_1+K\mu_xq_\bot^2)q_x^4q_\bot^2}{4}\left[\frac{\Pi(\bq)+\Gamma(\bq)^2}{\Pi(\bq)^3\Gamma(\bq)^3}\right]\ ,
\\
&=&  2Dk_x\int_{\bq}\frac{\gamma^2K(\gamma\kappa_1+K\mu_xq_\bot^2)q_x^4q_\bot^2}{4}\left[\frac{1}{\Pi(\bq)^2\Gamma(\bq)^3}+\frac{1}{\Pi(\bq)^3\Gamma(\bq)}\right]\ ,
\\
&\approx&  2Dk_x\int_{\bq}\frac{\gamma^2Kq_\bot^2}{4}\frac{1}{\gamma\kappa_1+K\mu_xq_\bot^2}\left[\frac{1}{\Gamma(\bq)^3}+\frac{1}{\Pi(\bq)\Gamma(\bq)}\right]\ ,
\\
&=& k_x\frac{3}{32}\frac{\gamma KD}{\kappa_1\mu_x^{1/2}\mu_\bot^{5/2}}\frac{S_{d-1}}{(2\pi)^{d-1}}\Lambda^{d-4}d\ell+{\rm h.o.t}\ .
\eeqn

Combining these results, the integral evaluates to

\beqn
\int_{\bq,\Omega} &=&  k_x\left[\frac{3}{32}\frac{\gamma D}{\kappa_1\mu_x^{1/2}\mu_\bot^{3/2}}\frac{S_{d-1}}{(2\pi)^{d-1}}\Lambda^{d-4}d\ell+\frac{3}{32}\frac{\gamma KD}{\kappa_1\mu_x^{1/2}\mu_\bot^{5/2}}\frac{S_{d-1}}{(2\pi)^{d-1}}\Lambda^{d-4}d\ell\right]\ .
\eeqn

There are several diagrammatics which have two external legs with this internal structure. Having made a change of variable $k_x\rightarrow h_x$, they have the form,

\beqn
&=& \left[\frac{3}{32}\frac{\gamma D}{\kappa_1\mu_x^{1/2}\mu_\bot^{3/2}}\frac{S_{d-1}}{(2\pi)^{d-1}}\Lambda^{d-4}d\ell+\frac{3}{32}\frac{\gamma KD}{\kappa_1\mu_x^{1/2}\mu_\bot^{5/2}}\frac{S_{d-1}}{(2\pi)^{d-1}}\Lambda^{d-4}d\ell\right]\int_{\bh}h_x\phi(\bh)\phi(\bk-\bh)\ ,
\\ 
&=& k_x\left[\frac{3}{64}\frac{\gamma D}{\kappa_1\mu_x^{1/2}\mu_\bot^{3/2}}\frac{S_{d-1}}{(2\pi)^{d-1}}\Lambda^{d-4}d\ell+\frac{3}{64}\frac{\gamma KD}{\kappa_1\mu_x^{1/2}\mu_\bot^{5/2}}\frac{S_{d-1}}{(2\pi)^{d-1}}\Lambda^{d-4}d\ell\right]\int_{\bh}\phi(\bh)\phi(\bk-\bh)\ .
\eeqn

We shall name the two sub variations of this integral as,

\beqn
I_{2.1} &=&  \left[\frac{3}{32}\frac{\gamma D}{\kappa_1\mu_x^{1/2}\mu_\bot^{3/2}}\frac{S_{d-1}}{(2\pi)^{d-1}}\Lambda^{d-4}d\ell+\frac{3}{32}\frac{\gamma KD}{\kappa_1\mu_x^{1/2}\mu_\bot^{5/2}}\frac{S_{d-1}}{(2\pi)^{d-1}}\Lambda^{d-4}d\ell\right]\ ,
\\
I_{2.2} &=&\left[\frac{3}{64}\frac{\gamma D}{\kappa_1\mu_x^{1/2}\mu_\bot^{3/2}}\frac{S_{d-1}}{(2\pi)^{d-1}}\Lambda^{d-4}d\ell+\frac{3}{64}\frac{\gamma KD}{\kappa_1\mu_x^{1/2}\mu_\bot^{5/2}}\frac{S_{d-1}}{(2\pi)^{d-1}}\Lambda^{d-4}d\ell\right]\ .
\eeqn

\subsection{Variation III}

The next variation is,

\beqn
I_3 = \int_{\bq,\Omega}2D\left(\frac{k_x}{2}+q_x\right)G_\phi\left(\frac{\bk}{2}+\bq, \frac{\omega}{2}+\Omega\right)G_\rho\left(\bq-\frac{\bk}{2},\Omega - \frac{\omega}{2}\right)G_\rho\left(\frac{\bk}{2}-\bq, \frac{\omega}{2}-\Omega\right)\ ,
\eeqn

We set $\omega=0$ and the integrand simplifies as,

\beqn
\frac{(\ii\Omega-Kq_\bot^2)(\frac{k_x}{2}+q_x)}{[\Omega^2 - \Pi(\frac{\bk}{2}+\bq)^2 + \ii\Omega\Gamma(\frac{\bk}{2}+\bq)]}\frac{\ii\gamma(q_x-\frac{k_x}{2})}{[\Omega^2 - (\bq-\frac{\bk}{2})^2 + \ii\Omega\Gamma(\bq-\frac{\bk}{2})]}\frac{\ii\gamma(\frac{k_x}{2}-q_x)}{[\Omega^2 - \Pi(\frac{\bk}{2}-\bq)^2 - \ii\Omega\Gamma(\frac{\bk}{2}-\bq)]}\ .
\eeqn

Expanding the numerator and extracting only $\mathcal{O}(k_x)$ and $\mathcal{O}(k_x^0)$ terms,

\beqn
&=& \gamma^2 \left(\frac{k_x}{2}+q_x\right)\left(\frac{k_x}{2}-q_x\right)^2(\ii\Omega-Kq_\bot^2) = \gamma^2\left(-\frac{k_xq_x^2}{2} +q_x^3\right)(\ii\Omega-Kq_\bot^2)\ .
\eeqn

The $\mathcal{O}(k_x)$ terms in the numerator evaluate to

\beqn
&=& 2Dk_x\int_{\bq,\Omega}\frac{\gamma^2q_x^2}{2}\frac{-\Omega^2\Gamma(\bq)+Kq_\bot^2(\Omega^2-\Pi(\bq))}{[(\Omega^2 - \Pi(\bq))^2 + \Omega^2\Gamma(\bq)^2][(\Omega^2 - \Pi(\bq))^2 + \Omega^2\Gamma(\bq)^2]}\ .
\eeqn

The first term in the numerator evaluates to

\beqn
&=&-2Dk_x\int_{\bq}\frac{\gamma^2 q_x^2}{8}\frac{1}{\Pi(\bq)\Gamma(\bq)^2}\ ,
\\
&\approx& -2Dk_x\int_{\bq}\frac{\gamma^2}{8}\frac{1}{\gamma\kappa_1+K\mu_xq_\bot^2}\frac{1}{\Gamma(\bq)^2}\ ,
\\
&=& -k_x\frac{1}{16}\frac{\gamma D}{\kappa_1\mu_x^{1/2}\mu_\bot^{3/2}}\frac{S_{d-1}}{(2\pi)^{d-1}}\Lambda^{d-4}d\ell+{\rm h.o.t}\ .
\eeqn

The second term in the numerator evaluates to

\beqn
&=&2Dk_x\int_{\bq}\frac{Kq_x^2q_\bot^2}{8}\left[\frac{1}{\Pi(\bq)\Gamma(\bq)^3} - \frac{\Pi(\bq)^2+\Pi(\bq)\Gamma(\bq)^2}{\Pi(\bq)^3\Gamma(\bq)^3}\right]\ ,
\\
&=&-2Dk_x\int_{\bq}\frac{Kq_x^2q_\bot^2}{8}\frac{1}{\Pi(\bq)^2\Gamma(\bq)}\ .
\eeqn

This term is irrelevant as it does not produce $\mathcal{O}(\Lambda^{d-4})$ terms.

From the $\mathcal{O}(k_x^0)$ term in the numerator, we want to extract $\mathcal{O}(k_x)$ terms. As such, we need to expand every denominator to $\mathcal{O}(k_x)$, which is the result derived in (\ref{eq:denom}).

For the $q_x^3$ terms, we multiply the original numerator with this result such that the integrand becomes,

\beqn
&=& \frac{k_x\gamma^2q_x^4\left[\mu_x\Omega^2+K(\gamma\kappa_1+K\mu_xq_\bot^2)q_\bot^2)\right]}{[(\Omega^2 - \Pi(\bq)^2 + \Omega^2\Gamma(\bq)^2][(\Omega^2 - \Pi(\bq))^2 + \Omega^2\Gamma(\bq)^2]}\ .
\eeqn

The first term in the numerator evaluates to
\beqn
&=& 2Dk_x\int_{\bq}\frac{1}{4}\frac{\gamma^2\mu_x q_x^4}{\Pi(\bq)\Gamma(\bq)^3}\ ,
\\
&\approx& 2Dk_x\int_{\bq}\frac{1}{4}\frac{\gamma^2\mu_x q_x^2}{\gamma\kappa_1+Kq_\bot^2}\frac{1}{\Gamma(\bq)^3}\ ,
\\
&=& k_x\frac{1}{32}\frac{\gamma D}{\kappa_1\mu_x^{1/2}\mu_\bot^{3/2}}\frac{S_{d-1}}{(2\pi)^{d-1}}\Lambda^{d-4}d\ell+{\rm h.o.t}\ .
\eeqn

The second term in the numerator evaluates to

\beqn
&=& 2Dk_x\int_{\bq}\frac{\gamma^2K(\gamma\kappa_1+K\mu_xq_\bot^2)q_x^4q_\bot^2}{4}\left[\frac{\Pi(\bq)+\Gamma(\bq)^2}{\Pi(\bq)^3\Gamma(\bq)^3}\right]\ ,
\\
&=&  2Dk_x\int_{\bq}\frac{\gamma^2K(\gamma\kappa_1+K\mu_xq_\bot^2)q_x^4q_\bot^2}{4}\left[\frac{1}{\Pi(\bq)^2\Gamma(\bq)^3}+\frac{1}{\Pi(\bq)^3\Gamma(\bq)}\right]\ ,
\\
&\approx&  2Dk_x\int_{\bq}\frac{\gamma^2Kq_\bot^2}{4}\frac{1}{\gamma\kappa_1+K\mu_xq_\bot^2}\left[\frac{1}{\Gamma(\bq)^3}+\frac{1}{\Pi(\bq)\Gamma(\bq)}\right]\ ,
\\
&=& k_x\frac{3}{32}\frac{\gamma KD}{\kappa_1\mu_x^{1/2}\mu_\bot^{5/2}}\frac{S_{d-1}}{(2\pi)^{d-1}}\Lambda^{d-4}d\ell+{\rm h.o.t}\ .
\eeqn

Combining all contributions, the integral evaluates to

\beqn
&=& k_x\left[-\frac{1}{32}\frac{\gamma D}{\kappa_1\mu_x^{1/2}\mu_\bot^{3/2}}\frac{S_{d-1}}{(2\pi)^{d-1}}\Lambda^{d-4}d\ell+\frac{3}{32}\frac{\gamma KD}{\kappa_1\mu_x^{1/2}\mu_\bot^{5/2}}\frac{S_{d-1}}{(2\pi)^{d-1}}\Lambda^{d-4}d\ell\right]\ .
\eeqn

Having made a change of variable $k_x\rightarrow h_x$, this integral is associated with two external density legs of the structure,

\beqn
&=& k_x\left[-\frac{1}{32}\frac{\gamma D}{\kappa_1\mu_x^{1/2}\mu_\bot^{3/2}}\frac{S_{d-1}}{(2\pi)^{d-1}}\Lambda^{d-4}d\ell+\frac{3}{32}\frac{\gamma KD}{\kappa_1\mu_x^{1/2}\mu_\bot^{5/2}}\frac{S_{d-1}}{(2\pi)^{d-1}}\Lambda^{d-4}d\ell\right]\int_{\tilde{\bh}}h_x\rho(\bh)\rho(\bk-\bh)\ ,
\\
&=&k_x\left[-\frac{1}{64}\frac{\gamma D}{\kappa_1\mu_x^{1/2}\mu_\bot^{3/2}}\frac{S_{d-1}}{(2\pi)^{d-1}}\Lambda^{d-4}d\ell+\frac{3}{64}\frac{\gamma KD}{\kappa_1\mu_x^{1/2}\mu_\bot^{5/2}}\frac{S_{d-1}}{(2\pi)^{d-1}}\Lambda^{d-4}d\ell\right]\int_{\tilde{\bh}}\rho(\bh)\rho(\bk-\bh)\ .
\eeqn

Therefore
\beqn
I_3 =\left[-\frac{1}{64}\frac{\gamma D}{\kappa_1\mu_x^{1/2}\mu_\bot^{3/2}}\frac{S_{d-1}}{(2\pi)^{d-1}}\Lambda^{d-4}d\ell+\frac{3}{64}\frac{\gamma KD}{\kappa_1\mu_x^{1/2}\mu_\bot^{5/2}}\frac{S_{d-1}}{(2\pi)^{d-1}}\Lambda^{d-4}d\ell\right]\ .
\eeqn

\subsection{Variation IV}

The next variation is, 

\beqn
I_4 = \int_{\bq,\Omega}2D\left(\frac{k_x}{2}+q_x\right)G_\rho\left(\frac{\bk}{2}+\bq, \frac{\omega}{2}+\Omega\right)G_\rho\left(\bq-\frac{\bk}{2},\Omega - \frac{\omega}{2}\right)G_\phi\left(\frac{\bk}{2}-\bq, \frac{\omega}{2}-\Omega\right)\ ,
\eeqn

We set $\omega=0$ and the integrand simplifies as,
\beqn
\frac{\ii\gamma(\frac{k_x}{2}+q_x)^2}{[\Omega^2 - \Pi(\frac{\bk}{2}+\bq)^2 + \ii\Omega\Gamma(\frac{\bk}{2}+\bq)]}\frac{\ii\gamma(q_x-\frac{k_x}{2})}{[\Omega^2 - \Pi(\bq-\frac{\bk}{2})^2 + \ii\Omega\Gamma(\bq-\frac{\bk}{2})]}\frac{-\ii\Omega-
Kq_\bot^2}{[\Omega^2 - \Pi(\frac{\bk}{2}-\bq)^2 - \ii\Omega\Gamma(\frac{\bk}{2}-\bq)]}\ .
\eeqn

Expanding the numerator and extracting only $\mathcal{O}(k_x)$ and $\mathcal{O}(k_x^0)$ terms,

\beqn
&=& \gamma^2\left(\frac{k_x}{2}+q_x\right)^2\left(\frac{k_x}{2}
-q_x\right)(-\ii\Omega - Kq_\bot^2)=\gamma^2\left(-\frac{k_xq_x^2}{2}-q_x^3\right)(-\ii\Omega-Kq_\bot^2)\ ,
\eeqn

The $\mathcal{O}(k_x)$ integrand evaluates to
\beqn
&=& 2Dk_x\int_{\bq,\Omega}\frac{\gamma^2q_x^2}{2}\frac{\Omega^2\Gamma(\bq)+ Kq_\bot^2(\Omega^2-\Pi(\bq))}{[(\Omega^2 - \Pi(\bq))^2 + \Omega^2\Gamma(\bq)^2][(\Omega^2 - \Pi(\bq))^2 + \Omega^2\Gamma(\bq)^2]}\ .
\eeqn

The first term in the numerator evaluates to

\beqn
&=& 2Dk_x\int_{\bq}\frac{1}{8}\frac{\gamma^2q_x^2 }{\Pi(\bq)\Gamma(\bq)^2}\ ,
\\
&=& 2Dk_x\int_{\bq_\bot}\frac{1}{16}\frac{1}{\gamma\kappa_1+K\mu_xq_\bot^2}\frac{\gamma^2}{\mu_x^{1/2}\mu_\bot^{3/2}q_\bot^3}\ ,
\\
&=&  2Dk_x\int_{\bq_\bot}\frac{1}{16}\frac{\gamma}{\kappa_1\mu_x^{1/2}\mu_\bot^{3/2}q_\bot^3}\left[1+\frac{K\mu_x q_\bot^2}{\gamma\kappa_1}\right]^{-1}\ ,
\\
&=& k_x\frac{1}{16}\frac{\gamma D}{\kappa_1\mu_x^{1/2}\mu_\bot^{3/2}}\frac{S_{d-1}}{(2\pi)^{d-1}}\Lambda^{d-4}d\ell+{\rm h.o.t}\ .
\eeqn

The second term in the numerator evaluates to

\beqn
&=& 2Dk_x\int_{\bq}\frac{\gamma^2Kq_x^2q_\bot^2}{8}\left[\frac{1}{\Pi(\bq)\Gamma^3(\bq)}-\frac{\Pi(\bq)^2+\Pi(\bq)\Gamma(\bq)^2}{\Pi(\bq)^3\Gamma(\bq)^3}\right]\ ,
\\
&=& -2Dk_x\int_{\bq}\frac{\gamma^2Kq_x^2q_\bot^2}{8}\frac{1}{\Pi(\bq)^2\Gamma(\bq)}\ .
\eeqn

This term is irrelevant as it does not produce $\mathcal{O}(\Lambda^{d-4})$ terms.

From the $\mathcal{O}(k_x^0)$ term in the numerator, we want to extract $\mathcal{O}(k_x)$ terms. As such, we need to expand every denominator to $\mathcal{O}(k_x)$, which is the result derived in (\ref{eq:denom}).

For the $q_x^3$ terms, we multiply the original numerator with this result such that the integrand becomes,

\beqn
&=& \frac{k_x\gamma^2q_x^4\left[\mu_x\Omega^2-K(\gamma\kappa_1+K\mu_xq_\bot^2)q_\bot^2\right]}{[(\Omega^2 - \Pi(\bq))^2 + \Omega^2\Gamma(\bq)^2][(\Omega^2 - \Pi(\bq))^2 + \Omega^2\Gamma(\bq)^2]}\ 
\eeqn

The first term in the numerator evaluates to
\beqn
&=& 2Dk_x\int_{\bq}\frac{1}{4}\frac{\gamma^2\mu_x q_x^4}{\Pi(\bq)\Gamma(\bq)^3}\ ,
\\
&\approx& 2Dk_x\int_{\bq}\frac{1}{4}\frac{1}{\gamma\kappa_1+Kq_\bot^2}\frac{\gamma^2\mu_x q_x^2}{\Gamma(\bq)^3}\ ,
\\
&=& k_x\frac{1}{32}\frac{\gamma D}{\kappa_1\mu_x^{1/2}\mu_\bot^{3/2}}\frac{S_{d-1}}{(2\pi)^{d-1}}\Lambda^{d-4}d\ell+{\rm h.o.t}\ .
\eeqn

The second term in the numerator evaluates to

\beqn
&=& -2Dk_x\int_{\bq}\frac{\gamma^2K(\gamma\kappa_1+K\mu_xq_\bot^2)q_x^4q_\bot^2}{4}\left[\frac{\Pi(\bq)+\Gamma(\bq)^2}{\Pi(\bq)^3\Gamma(\bq)^3}\right]\ ,
\\
&=&  -2Dk_x\int_{\bq}\frac{\gamma^2K(\gamma\kappa_1+K\mu_xq_\bot^2)q_x^4q_\bot^2}{4}\left[\frac{1}{\Pi(\bq)^2\Gamma(\bq)^3}+\frac{1}{\Pi(\bq)^3\Gamma(\bq)}\right]\ ,
\\
&\approx&  -2Dk_x\int_{\bq}\frac{\gamma^2Kq_\bot^2}{4}\frac{1}{\gamma\kappa_1+K\mu_xq_\bot^2}\left[\frac{1}{\Gamma(\bq)^3}+\frac{1}{\Pi(\bq)\Gamma(\bq)}\right]\ ,
\\
&=& -k_x\frac{3}{32}\frac{\gamma KD}{\kappa_1\mu_x^{1/2}\mu_\bot^{5/2}}\frac{S_{d-1}}{(2\pi)^{d-1}}\Lambda^{d-4}d\ell+{\rm h.o.t}\ .
\eeqn

Combining all contributions, the integral evaluates to

\beqn
&=& k_x\left[\frac{3}{32}\frac{\gamma D}{\kappa_1\mu_x^{1/2}\mu_\bot^{3/2}}\frac{S_{d-1}}{(2\pi)^{d-1}}\Lambda^{d-4}d\ell - \frac{3}{32}\frac{\gamma KD}{\kappa_1\mu_x^{1/2}\mu_\bot^{5/2}}\frac{S_{d-1}}{(2\pi)^{d-1}}\Lambda^{d-4}d\ell\right]\ .
\eeqn

Summing all the components and noting this integral is often associated with external legs of the structure where we make a change of variable $k_x\rightarrow h_x$,

\beqn
&=& \left[\frac{3}{32}\frac{\gamma D}{\kappa_1\mu_x^{1/2}\mu_\bot^{3/2}}\frac{S_{d-1}}{(2\pi)^{d-1}}\Lambda^{d-4}d\ell - \frac{3}{32}\frac{\gamma KD}{\kappa_1\mu_x^{1/2}\mu_\bot^{5/2}}\frac{S_{d-1}}{(2\pi)^{d-1}}\Lambda^{d-4}d\ell\right]\int_{\tilde{\bh}}h_x\rho(\bh)\rho(\bk-\bh)\ ,
\\
&=& k_x\left[\frac{3}{64}\frac{\gamma D}{\kappa_1\mu_x^{1/2}\mu_\bot^{3/2}}\frac{S_{d-1}}{(2\pi)^{d-1}}\Lambda^{d-4}d\ell - \frac{3}{64}\frac{\gamma KD}{\kappa_1\mu_x^{1/2}\mu_\bot^{5/2}}\frac{S_{d-1}}{(2\pi)^{d-1}}\Lambda^{d-4}d\ell\right]\int_{\tilde{\bh}}\rho(\bh)\rho(\bk-\bh)\ .
\eeqn

Therefore, 

\beqn
I_4 &=& \left[\frac{3}{64}\frac{\gamma D}{\kappa_1\mu_x^{1/2}\mu_\bot^{3/2}}\frac{S_{d-1}}{(2\pi)^{d-1}}\Lambda^{d-4}d\ell - \frac{3}{64}\frac{\gamma KD}{\kappa_1\mu_x^{1/2}\mu_\bot^{5/2}}\frac{S_{d-1}}{(2\pi)^{d-1}}\Lambda^{d-4}d\ell\right]\ .
\eeqn

\subsection{Variation V}

\beqn 
I_5 &=& \int_{\bq,\Omega}2D \left(\frac{k_x}{2}+q_x\right)G_\rho\left(\frac{\bk}{2}+\bq, \frac{\omega}{2}+\Omega\right)G_\rho\left(\bq-\frac{\bk}{2},\Omega - \frac{\omega}{2}\right)G_\rho\left(\frac{\bk}{2}-\bq, \frac{\omega}{2}-\Omega\right)\ .
\eeqn

Expanding the numerator and producing $\mathcal{O}(k_x)$ and $\mathcal{O}(k_x^0)$ terms,

\beqn
&=& -\ii\gamma^3\left(\frac{k_x}{2}+q_x\right)^2\left(q_x-\frac{k_x}{2}\right)\left(-q_x+\frac{k_x}{2}\right) = \ii\gamma^3\left(\frac{k_x}{2}+q_x\right)^2\left(\frac{k_x}{2}-q_x\right)^2 = \ii\gamma^3q_x^4\ ,
\eeqn

From the $\mathcal{O}(k_x^0)$ term in the numerator, we want to extract $\mathcal{O}(k_x)$ terms. As such, we need to expand every denominator to $\mathcal{O}(k_x)$, which is the result derived in (\ref{eq:denom}).

For the $q_x^3$ terms, we multiply the original numerator with this result such that the integrand becomes,

\beqn
&=& -k_x\frac{\ii\gamma^3q_x^5((\gamma\kappa_1+K\mu_xq_\bot^2)+\ii\Omega\mu_x)}{[(\Omega^2 - \Pi(\bq))^2 + \Omega^2\Gamma(\bq)^2][(\Omega^2 - \Pi(\bq))^2 + \Omega^2\Gamma(\bq)^2]}=0\ .
\eeqn

This integral vanishes due to its odd $q_x$ numerator.

The second term in the numerator is odd in $\Omega$ and vanishes, such that the integral reduces to,

\beqn
&=& 2Dk_x\int_{\bq,\Omega}-\frac{\ii\gamma^4\kappa_1q_x^5}{[(\Omega^2 - \gamma\kappa_1q_x^2)^2 + \Omega^2\Gamma(\bq)^2][(\Omega^2 - \gamma\kappa_1q_x^2)^2 + \Omega^2\Gamma(\bq)^2]}\ .
\eeqn

If instead we do not expand the numerator, the integral is,

\beqn
&=& 2D\int_{\bq,\Omega}\frac{\ii\gamma^3q_x^4(\Omega^2-\Pi(\bq))}{[(\Omega^2 - \gamma\kappa_1q_x^2)^2 + \Omega^2\Gamma(\bq)^2][(\Omega^2 - \gamma\kappa_1q_x^2)^2 + \Omega^2\Gamma(\bq)^2]}\ ,
\\
&=& \frac{1}{4}\frac{\ii\gamma D}{\kappa_1^2\mu_x^{1/2}\mu_\bot^{1/2}}\frac{S_{d-1}}{(2\pi)^{d-1}}\Lambda^{d-2}d\ell\ .
\eeqn

This integral will be irrelevant as it does not produce $\mathcal{O}(\Lambda^{d-4})$ terms.

\subsection{Variation VI}

\beqn 
I_6 &=& \int_{\bq,\Omega}2D G_\phi\left(\frac{\bk}{2}+\bq, \frac{\omega}{2}+\Omega\right)G_\phi\left(\bq-\frac{\bk}{2},\Omega - \frac{\omega}{2}\right)G_\phi\left(\frac{\bk}{2}-\bq, \frac{\omega}{2}-\Omega\right)\ ,
\eeqn

We can set both $\omega=0$ and $\bk=0$ inside the integrand, such that it simplifies to,

\beqn
\frac{\ii\Omega - Kq_\bot^2}{[\Omega^2 -\Pi(\bq) + \ii\Omega\Gamma(\bq)]}\frac{\ii\Omega-Kq_\bot^2}{[\Omega^2 - \Pi(\bq) + \ii\Omega\Gamma(\bq)]}\frac{-\ii\Omega-Kq_\bot^2}{[\Omega^2 -  \Pi(\bq) - \ii\Omega\Gamma(\bq)]}\ .
\eeqn

The numerator expands
\beqn
(\ii\Omega - Kq_\bot^2)(\ii\Omega - Kq_\bot^2)(-\ii\Omega - Kq_\bot^2) = \ii\Omega^3 - K\Omega^2q_\bot^2 + \ii K^2\Omega q_\bot^4 - K^3q_\bot^6\ .
\eeqn

If we multiply the numerator by $[\Omega^2 - \Pi(\bq) - \ii\Omega\Gamma(\bq)]$ whilst dividing it at the same time by the same quantity, the integrand becomes

\beqn
\frac{\Omega^4\Gamma(\bq) -(\Omega^2-\Pi(\bq) )K\Omega^2q_\bot^2 + K^2\Omega^2q_\bot^4\Gamma(\bq) - (\Omega^2-\Pi(\bq))K^3q_\bot^6}{[(\Omega^2 - \Pi(\bq))^2 + \Omega^2\Gamma(\bq)^2][(\Omega^2 - \Pi(\bq))^2 + \Omega^2\Gamma(\bq)^2]}\ ,
\eeqn

where we have ignored all odd $\Omega$ terms which vanish in the $\Omega$ integral.

The first term in the numerator evaluates to,

\beqn
&=&2D\int_{\bq,\Omega}\frac{ \Omega^4\Gamma(\bq)}{[(\Omega^2 -  \Pi(\bq))^2 + \Omega^2\Gamma(\bq)^2][(\Omega^2 -  \Pi(\bq))^2 + \Omega^2\Gamma(\bq)^2]}\ ,
\\
&=& 2D\int_{\bq}\frac{1}{4}\frac{1}{\Gamma(\bq)^2}\ ,
\\
\label{eq:OG}
&=& \frac{1}{8}\frac{D}{\mu_x^{1/2}\mu_\bot^{3/2}}\frac{S_{d-1}}{(2\pi)^{d-1}}\Lambda^{d-4}d\ell\ .
\eeqn

The second term in the numerator evaluates to 

\beqn
&=&-2D\int_{\bq,\Omega}\frac{ (\Omega^2-\Pi(\bq))K\Omega^2q_\bot^2}{[(\Omega^2 - \Pi(\bq))^2 + \Omega^2\Gamma(\bq)^2][(\Omega^2 - \Pi(\bq))^2 + \Omega^2\Gamma(\bq)^2]}\ ,
\\
&=&-2D\int_{\bq}\frac{1}{4}\left[\frac{ Kq_\bot^2}{\Gamma(\bq)^3}-\frac{ Kq_\bot^2}{\Gamma(\bq)^3}\right]\ ,
\\
&=& 0\ .
\eeqn

The third term in the numerator evaluates as

\beqn
&=&2D\int_{\bq,\Omega}\frac{K^2\Omega^2q_\bot^4\Gamma(\bq)}{[(\Omega^2 - \Pi(\bq))^2 + \Omega^2\Gamma(\bq)^2][(\Omega^2 - \Pi(\bq))^2 + \Omega^2\Gamma(\bq)^2]}\ ,
\\
&=& 2D\int_{\bq}\frac{1}{4}\frac{ K^2q_\bot^4}{\Pi(\bq)\Gamma(\bq)^2}\ .
\eeqn

This integral will be irrelevant as it does not produce $\mathcal{O}(\Lambda^{d-4})$ terms.

The fourth term in the numerator evaluates as

\beqn
&=&2D\int_{\bq,\Omega}\frac{(\Omega^2-\Pi(\bq))K^3q_\bot^6}{[(\Omega^2 - \Pi(\bq))^2 + \Omega^2\Gamma(\bq)^2][(\Omega^2 - \Pi(\bq))^2 + \Omega^2\Gamma(\bq)^2]}\ ,
\\
&=& 2D\int_{\bq}\frac{K^3q_\bot^6}{4}\left[\frac{ 1}{\Pi(\bq)\Gamma(\bq)^3}-\left[\frac{\Pi(\bq)+\Gamma(\bq)^2}{\Pi(\bq)^2\Gamma(\bq)^3}\right]\right]\ ,
\\
&=& -2D\int_{\bq}\frac{1}{4}\frac{K^3q_\bot^6}{\Pi(\bq)^2\Gamma(\bq)}\ .
\eeqn

This integral will be irrelevant as it does not produce $\mathcal{O}(\Lambda^{d-4})$ terms.

Therefore the overall integral evaluates as 
\beqn
I_6 = \frac{1}{8}\frac{D}{\mu_x^{1/2}\mu_\bot^{3/2}}\frac{S_{d-1}}{(2\pi)^{d-1}}\Lambda^{d-4}\ .
\eeqn

\subsection{Variation VII}

\beqn 
I_7 &=& \int_{\bq,\Omega}2D G_\phi\left(\frac{\bk}{2}+\bq, \frac{\omega}{2}+\Omega\right)G_\rho\left(\bq-\frac{\bk}{2},\Omega - \frac{\omega}{2}\right)G_\rho\left(\frac{\bk}{2}-\bq, \frac{\omega}{2}-\Omega\right)\ ,
\\
&=&  \int_{\bq,\Omega}2D G_\rho\left(\frac{\bk}{2}+\bq, \frac{\omega}{2}+\Omega\right)G_\rho\left(\bq-\frac{\bk}{2},\Omega - \frac{\omega}{2}\right)G_\phi\left(\frac{\bk}{2}-\bq, \frac{\omega}{2}-\Omega\right)\ ,
\\
&=&   \int_{\bq,\Omega}2D G_\rho\left(\frac{\bk}{2}+\bq, \frac{\omega}{2}+\Omega\right)G_\phi\left(\bq-\frac{\bk}{2},\Omega - \frac{\omega}{2}\right)G_\rho\left(\frac{\bk}{2}-\bq, \frac{\omega}{2}-\Omega\right)\ .
\eeqn

We can set both $\omega=0$ and $\bk=0$ inside the integrand, such that it simplifies significantly.

For the first and third row, the numerator evaluates to

\beqn
\frac{\gamma^2 q_x^2(\ii\Omega - Kq_\bot^2)}{[\Omega^2 - \Pi(\bq) + \ii\Omega\Gamma(\bq)]}\frac{1}{[\Omega^2 - \Pi(\bq) + \ii\Omega\Gamma(\bq)]}\frac{1}{[\Omega^2 - \Pi(\bq) - \ii\Omega\Gamma(\bq)]}\ .
\eeqn

If we multiply the numerator by $[\Omega^2 - \Pi(\bq) - \ii\Omega\Gamma(\bq)]$ whilst dividing it at the same time by the same quantity, the integrand becomes

\beqn
\frac{\gamma^2q_x^2\left[\Omega^2\Gamma(\bq)-Kq_\bot^2(\Omega^2-\Pi(\bq))\right]}{[(\Omega^2 - \Pi(\bq))^2 + \Omega^2\Gamma(\bq)^2][(\Omega^2 - \Pi(\bq))^2 + \Omega^2\Gamma(\bq)^2]}\ ,
\eeqn

where we have ignored all odd $\Omega$ terms as they will vanish in the integral.

The first term in the numerator evaluates to

\beqn
&=& 2D\int_{\bq}\frac{1}{4}\frac{\gamma^2q_x^2}{\Pi(\bq)\Gamma(\bq)^2}\ ,
\\
&=& 2D\int_{\bq}\frac{1}{4}\frac{1}{\gamma\kappa_1+K\mu_xq_\bot^2}\frac{\gamma^2}{\Gamma(\bq)^2}\ ,
\\
&=& \frac{1}{8}\frac{\gamma D}{\kappa_1\mu_x^{1/2}\mu_\bot^{3/2}}\frac{S_{d-1}}{(2\pi)^{d-1}}\Lambda^{d-4}d\ell+{\rm h.o.t}\ .
\eeqn

The second term in the numerator evaluates to

\beqn
&=& -2D\int_{\bq,\Omega}\frac{\gamma^2Kq_x^2q_\bot^2(\Omega^2-\Pi(\bq))}{[(\Omega^2 - \Pi(\bq))^2 + \Omega^2\Gamma(\bq)^2][(\Omega^2 - \Pi(\bq))^2 + \Omega^2\Gamma(\bq)^2]}\ ,
\\
&=& -2D\int_{\bq}\frac{\gamma^2Kq_x^2q_\bot^2}{4}\left[\frac{1}{\Pi(\bq)\Gamma(\bq)^3}-\frac{\Pi(\bq)+\Gamma(\bq)^2}{\Pi(\bq)^2\Gamma(\bq)^3}\right]\ ,
\\
&=& 2D\int_{\bq}\frac{1}{4}\frac{\gamma^2Kq_x^2q_\bot^2}{\Pi(\bq)^2\Gamma(\bq)}\ ,
\eeqn

This integral will be irrelevant as it does not produce $\mathcal{O}(\Lambda^{d-4})$ terms.

Therefore the overall integral evaluates to 

\beqn
\int_{\bq,\Omega} = \frac{1}{8}\frac{\gamma D}{\kappa_1\mu_x^{1/2}\mu_\bot^{3/2}}\frac{S_{d-1}}{(2\pi)^{d-1}}\Lambda^{d-4}d\ell\ .
\eeqn

For the integral in the second row, the numerator of its integrand becomes 

\beqn
&=&\gamma^2q_x^2(\ii\Omega+Kq_\bot^2)
\eeqn

Therefore, all integrals in variation VII ultimately adhere to the same structure such that they evaluate as 

\beqn
I_7 = \frac{1}{8}\frac{\gamma D}{\kappa_1\mu_x^{1/2}\mu_\bot^{3/2}}\frac{S_{d-1}}{(2\pi)^{d-1}}\Lambda^{d-4}d\ell\ .
\eeqn

\section{Calculations of 3-vertex diagrams}
\label{app3-vertex}.

For all 3 vertex diagrams, we can immediately take the hydrodynamic limit as all $\pp_x$ diagrams already have an external derivative outside the loop structure. This greatly simplifies the integrand structure as we can immediately set $\bk=0$ inside the integrand. Moreover, we have intentionally drawn the three vertex diagrams in sets of triplets as these will cancel each other out numerically. These can be expressed as,

\subsection{Set I}

\beqn
I_8 &=& 2D\int_{\tilde{\bq}}-q_x^2G_\phi(\tilde{\bq})G_\phi(\tilde{\bq})G_\phi(-\tilde{\bq})G_\phi(-\tilde{\bq})\ ,
\\
&=&-2D\int_{\bq,\Omega}\frac{\Omega^4+2K^2\Omega^2q_\bot^4+K^4q_\bot^8}{[(\Omega^2 - \Pi(\bq))^2 + \Omega^2\Gamma(\bq)^2][(\Omega^2 - \Pi(\bq))^2 + \Omega^2\Gamma(\bq)^2]}\ ,
\\
&\approx&-2D\int_{\bq,\Omega}\frac{q_x^2}{4}\frac{1}{\Gamma(\bq)^3}\ ,
\\
&=& -\frac{1}{32}\frac{D}{\mu_x^{3/2}\mu_\bot^{3/2}}\frac{S_{d-1}}{(2\pi)^{d-1}}\Lambda^{d-4} + {\rm h.o.t}\ ,
\\
I_9 &=& 2D\int_{\tilde{\bq}}q_x^2G_\phi(\tilde{\bq})G_\phi(\tilde{\bq})G_\phi(\tilde{\bq})G_\phi(-\tilde{\bq}) 
\\
&=&2D\int_{\bq,\Omega}\frac{-\Omega^4-2\ii K\Omega^3q_\bot^4 - 2\ii K^3\Omega q_\bot^8+K^4q_\bot^8}{[(\Omega^2 - \Pi(\bq))^2 + \Omega^2\Gamma(\bq)^2][\Omega^2 - \Pi(\bq) + \ii\Omega\Gamma(\bq)]^2}\ ,
\\
&=& \frac{1}{64}\frac{D}{\mu_x^{3/2}\mu_\bot^{3/2}}\frac{S_{d-1}}{(2\pi)^{d-1}}\Lambda^{d-4}+ {\rm h.o.t}\ ,
\\
\nonumber
\\
&\implies& I_8 + 2I_9 = 0\ .
\eeqn

\subsection{Set II}

\beqn
I_{10} &=& 
2D\int_{\tilde{\bq}}-q_x^2G_\rho(\tilde{\bq})G_\rho(-\tilde{\bq})G_\phi(\tilde{\bq})G_\phi(-\tilde{\bq})\ ,
\\
&=&-2D\int_{\bq,\Omega}\frac{\gamma^2q_x^2(\Omega^4+K^2q_\bot^4)}{[(\Omega^2 - \Pi(\bq))^2 + \Omega^2\Gamma(\bq)^2][(\Omega^2 - \Pi(\bq))^2 + \Omega^2\Gamma(\bq)^2]}\ ,
\\
&=& -2D\int_{\bq,\Omega}\frac{\gamma^2q_x^2}{4}\left[\frac{1}{\Gamma(\bq)^3}+\frac{K^2q_\bot^4[\Pi(\bq)+\Gamma(\bq)^2]}{\Pi(\bq)^3\Gamma(\bq)^3}\right]
\\
&=&  -\frac{1}{32}\frac{\gamma D}{\kappa_1\mu_x^{3/2}\mu_\bot^{3/2}}\frac{S_{d-1}}{(2\pi)^{d-1}}\Lambda^{d-4}+{\rm h.o.t}\ ,
\\
I_{11} &=& 2D\int_{\tilde{\bq}}q_x^2G_\rho(\tilde{\bq})G_\phi(\tilde{\bq})G_\phi(\tilde{\bq})G_\rho(-\tilde{\bq})\ ,
\\
&=& 2D\int_{\tilde{\bq}}q_x^2G_\rho(\tilde{\bq})G_\rho(\tilde{\bq})G_\phi(\tilde{\bq})G_\phi(-\tilde{\bq})\ ,
\\
&=&-2D\int_{\bq,\Omega}\frac{\gamma^2q_x^4(\Omega^2+2\ii K\Omega q_\bot^2 -K^2q_\bot^4)}{[(\Omega^2 - \Pi(\bq))^2 + \Omega^2\Gamma(\bq)^2][\Omega^2 - \Pi(\bq) + \ii\Omega\Gamma(\bq)]^2}\ ,
\\
&\approx&2D\int_{\bq}\frac{\gamma^2q_x^4}{8}\frac{1}{\Pi(\bq)\Gamma(\bq)^3}\ ,
\\
&=& \frac{1}{64}\frac{\gamma D}{\kappa_1\mu_x^{3/2}\mu_\bot^{3/2}}\frac{S_{d-1}}{(2\pi)^{d-1}}\Lambda^{d-4}+{\rm h.o.t}\ ,
\\
\nonumber
\\
&\implies& I_{10} + 2I_{11} = 0\ .
\eeqn

\subsection{Set III}

\beqn
I_{12} &=& 2D\int_{\tilde{\bq}}-q_x^2G_\rho(\tilde{\bq})G_\rho(\tilde{\bq})G_\rho(-\tilde{\bq})G_\rho(-\tilde{\bq})\ ,
\\
&=& -2D\int_{\bq,\Omega}\frac{\gamma^4q_x^6}{[(\Omega^2 - \Pi(\bq))^2 + \Omega^2\Gamma(\bq)^2][(\Omega^2 - \Pi(\bq))^2 + \Omega^2\Gamma(\bq)^2]}\ ,
\\
&=& -2D\int_{\bq}\frac{\gamma^4q_x^6}{4}\left[\frac{\Pi(\bq)+\Gamma(\bq)^2}{\Pi(\bq)^3\Gamma(\bq)^3}\right]\ ,
\\
&\approx& -2D\int_{\bq}\frac{\gamma^2q_x^2}{4}\frac{1}{(\gamma\kappa_1+K\mu_xq_\bot^2)^2}\frac{1}{\Gamma(\bq)^3}
\\
&=& - \frac{1}{32}\frac{\gamma^2D}{\kappa_1^2\mu_x^{3/2}\mu_\bot^{3/2}}\frac{S_{d-1}}{(2\pi)^{d-1}}\Lambda^{d-4}+{\rm h.o.t}\ ,
\\
I_{13} &=& 2D\int_{\tilde{\bq}}q_x^2G_\rho(\tilde{\bq})G_\rho(\tilde{\bq})G_\rho(\tilde{\bq})G_\rho(-\tilde{\bq})\ ,
\\
&=& -2D\int_{\bq,\Omega}\frac{\gamma^4q_x^6}{[(\Omega^2 - \Pi(\bq))^2 + \Omega^2\Gamma(\bq)^2][\Omega^2 - \Pi(\bq) + \ii\Omega\Gamma(\bq)]^2}\ ,
\\
&=& -2D\int_{\bq}\frac{\gamma^4q_x^6}{8}\frac{\left[\mu_xq_x^4-(\gamma\kappa_1-2\mu_x\mu_\bot q_\bot^2 + Kq_\bot^2)q_x^2+(\mu_\bot^2-K^2)q_\bot^4\right]}{\Pi(\bq)^3\Gamma(\bq)^3}\ ,
\\
&\approx& -2D\int_{\bq}\frac{\gamma^4}{8}\frac{1}{(\gamma\kappa_1+K\mu_xq_\bot^2)^3}\frac{\left[\mu_xq_x^4-(\gamma\kappa_1-2\mu_x\mu_\bot q_\bot^2 + Kq_\bot^2)q_x^2+(\mu_\bot^2-K^2)q_\bot^4\right]}{\Gamma(\bq)^3}\ ,
\\
&\approx& D\int_{\bq_\bot}\frac{1}{64}\frac{\gamma^4}{\mu_x^{3/2}\mu_\bot^{3/2}q_\bot^3}\frac{(\gamma\kappa_1-2\mu_x\mu_\bot q_\bot^2 + Kq_\bot^2)}{(\gamma\kappa_1+K\mu_xq_\bot^2)^3}
\\
&=& \frac{1}{64}\frac{\gamma^2D}{\kappa_1^2\mu_x^{3/2}\mu_\bot^{3/2}}\frac{S_{d-1}}{(2\pi)^{d-1}}\Lambda^{d-4}d\ell+{\rm h.o.t}\ ,
\\
\nonumber
\\
&\implies& I_{12} + 2I_{13} = 0\ .
\eeqn


\begin{thebibliography}{25}%
\makeatletter
\providecommand \@ifxundefined [1]{%
 \@ifx{#1\undefined}
}%
\providecommand \@ifnum [1]{%
 \ifnum #1\expandafter \@firstoftwo
 \else \expandafter \@secondoftwo
 \fi
}%
\providecommand \@ifx [1]{%
 \ifx #1\expandafter \@firstoftwo
 \else \expandafter \@secondoftwo
 \fi
}%
\providecommand \natexlab [1]{#1}%
\providecommand \enquote  [1]{``#1''}%
\providecommand \bibnamefont  [1]{#1}%
\providecommand \bibfnamefont [1]{#1}%
\providecommand \citenamefont [1]{#1}%
\providecommand \href@noop [0]{\@secondoftwo}%
\providecommand \href [0]{\begingroup \@sanitize@url \@href}%
\providecommand \@href[1]{\@@startlink{#1}\@@href}%
\providecommand \@@href[1]{\endgroup#1\@@endlink}%
\providecommand \@sanitize@url [0]{\catcode `\\12\catcode `\$12\catcode
  `\&12\catcode `\#12\catcode `\^12\catcode `\_12\catcode `\%12\relax}%
\providecommand \@@startlink[1]{}%
\providecommand \@@endlink[0]{}%
\providecommand \url  [0]{\begingroup\@sanitize@url \@url }%
\providecommand \@url [1]{\endgroup\@href {#1}{\urlprefix }}%
\providecommand \urlprefix  [0]{URL }%
\providecommand \Eprint [0]{\href }%
\providecommand \doibase [0]{https://doi.org/}%
\providecommand \selectlanguage [0]{\@gobble}%
\providecommand \bibinfo  [0]{\@secondoftwo}%
\providecommand \bibfield  [0]{\@secondoftwo}%
\providecommand \translation [1]{[#1]}%
\providecommand \BibitemOpen [0]{}%
\providecommand \bibitemStop [0]{}%
\providecommand \bibitemNoStop [0]{.\EOS\space}%
\providecommand \EOS [0]{\spacefactor3000\relax}%
\providecommand \BibitemShut  [1]{\csname bibitem#1\endcsname}%
\let\auto@bib@innerbib\@empty
\bibitem [{\citenamefont {Wilson}\ and\ \citenamefont
  {Fisher}(1972)}]{wilson_prl72}%
  \BibitemOpen
  \bibfield  {author} {\bibinfo {author} {\bibfnamefont {K.~G.}\ \bibnamefont
  {Wilson}}\ and\ \bibinfo {author} {\bibfnamefont {M.~E.}\ \bibnamefont
  {Fisher}},\ }\bibfield  {title} {\bibinfo {title} {Critical {Exponents} in
  3.99 {Dimensions}},\ }\href {https://doi.org/10.1103/PhysRevLett.28.240}
  {\bibfield  {journal} {\bibinfo  {journal} {Physical Review Letters}\
  }\textbf {\bibinfo {volume} {28}},\ \bibinfo {pages} {240} (\bibinfo {year}
  {1972})}\BibitemShut
  {NoStop}%
\bibitem [{\citenamefont {Marchetti}\ \emph {et~al.}(2013)\citenamefont
  {Marchetti}, \citenamefont {Joanny}, \citenamefont {Ramaswamy}, \citenamefont
  {Liverpool}, \citenamefont {Prost}, \citenamefont {Rao},\ and\ \citenamefont
  {Simha}}]{marchetti_rmp13}%
  \BibitemOpen
  \bibfield  {author} {\bibinfo {author} {\bibfnamefont {M.~C.}\ \bibnamefont
  {Marchetti}}, \bibinfo {author} {\bibfnamefont {J.~F.}\ \bibnamefont
  {Joanny}}, \bibinfo {author} {\bibfnamefont {S.}~\bibnamefont {Ramaswamy}},
  \bibinfo {author} {\bibfnamefont {T.~B.}\ \bibnamefont {Liverpool}}, \bibinfo
  {author} {\bibfnamefont {J.}~\bibnamefont {Prost}}, \bibinfo {author}
  {\bibfnamefont {M.}~\bibnamefont {Rao}},\ and\ \bibinfo {author}
  {\bibfnamefont {R.~A.}\ \bibnamefont {Simha}},\ }\bibfield  {title} {\bibinfo
  {title} {Hydrodynamics of soft active matter},\ }\href
  {https://doi.org/10.1103/RevModPhys.85.1143} {\bibfield  {journal} {\bibinfo
  {journal} {Reviews of Modern Physics}\ }\textbf {\bibinfo {volume} {85}},\
  \bibinfo {pages} {1143} (\bibinfo {year} {2013})}\BibitemShut {NoStop}%
\bibitem [{\citenamefont {Jülicher}\ \emph {et~al.}(2018)\citenamefont
  {Jülicher}, \citenamefont {Grill},\ and\ \citenamefont
  {Salbreux}}]{julicher_rpp18}%
  \BibitemOpen
  \bibfield  {author} {\bibinfo {author} {\bibfnamefont {F.}~\bibnamefont
  {Jülicher}}, \bibinfo {author} {\bibfnamefont {S.~W.}\ \bibnamefont
  {Grill}},\ and\ \bibinfo {author} {\bibfnamefont {G.}~\bibnamefont
  {Salbreux}},\ }\bibfield  {title} {\bibinfo {title} {Hydrodynamic theory of
  active matter},\ }\href {https://doi.org/10.1088/1361-6633/aab6bb} {\bibfield
   {journal} {\bibinfo  {journal} {Reports on Progress in Physics}\ }\textbf
  {\bibinfo {volume} {81}},\ \bibinfo {pages} {076601} (\bibinfo {year}
  {2018})}\BibitemShut {NoStop}%
\bibitem [{\citenamefont {Solon}\ and\ \citenamefont
  {Tailleur}(2013)}]{solon_prl13}%
  \BibitemOpen
  \bibfield  {author} {\bibinfo {author} {\bibfnamefont {A.~P.}\ \bibnamefont
  {Solon}}\ and\ \bibinfo {author} {\bibfnamefont {J.}~\bibnamefont
  {Tailleur}},\ }\bibfield  {title} {\bibinfo {title} {Revisiting the flocking
  transition using active spins.},\ }\href
  {https://doi.org/10.1103/PhysRevLett.111.078101} {\bibfield  {journal}
  {\bibinfo  {journal} {Physical review letters}\ }\textbf {\bibinfo {volume}
  {111}},\ \bibinfo {pages} {078101} (\bibinfo {year} {2013})}\BibitemShut {NoStop}%
\bibitem [{\citenamefont {Solon}\ and\ \citenamefont
  {Tailleur}(2015)}]{solon_pre15}%
  \BibitemOpen
  \bibfield  {author} {\bibinfo {author} {\bibfnamefont {A.~P.}\ \bibnamefont
  {Solon}}\ and\ \bibinfo {author} {\bibfnamefont {J.}~\bibnamefont
  {Tailleur}},\ }\bibfield  {title} {\bibinfo {title} {Flocking with discrete
  symmetry: {The} two-dimensional active {Ising} model.},\ }\href
  {https://doi.org/10.1103/PhysRevE.92.042119} {\bibfield  {journal} {\bibinfo
  {journal} {Physical review. E, Statistical, nonlinear, and soft matter
  physics}\ }\textbf {\bibinfo {volume} {92}},\ \bibinfo {pages} {042119}
  (\bibinfo {year} {2015})}\BibitemShut {NoStop}%
\bibitem [{\citenamefont {Scandolo}\ \emph {et~al.}(2023)\citenamefont
  {Scandolo}, \citenamefont {Pausch},\ and\ \citenamefont
  {Cates}}]{scandolo_epje23}%
  \BibitemOpen
  \bibfield  {author} {\bibinfo {author} {\bibfnamefont {M.}~\bibnamefont
  {Scandolo}}, \bibinfo {author} {\bibfnamefont {J.}~\bibnamefont {Pausch}},\
  and\ \bibinfo {author} {\bibfnamefont {M.~E.}\ \bibnamefont {Cates}},\
  }\bibfield  {title} {\bibinfo {title} {Active {Ising} {Models} of flocking: a
  field-theoretic approach},\ }\href
  {https://doi.org/10.1140/epje/s10189-023-00364-w} {\bibfield  {journal}
  {\bibinfo  {journal} {The European Physical Journal E}\ }\textbf {\bibinfo
  {volume} {46}},\ \bibinfo {pages} {103} (\bibinfo {year} {2023})}\BibitemShut
  {NoStop}%
\bibitem [{\citenamefont {Bandyopadhyay}\ \emph {et~al.}(2024)\citenamefont
  {Bandyopadhyay}, \citenamefont {Chatterjee}, \citenamefont {Dutta},
  \citenamefont {Karmakar}, \citenamefont {Rieger},\ and\ \citenamefont
  {Paul}}]{bandyopadhyay_pre24}%
  \BibitemOpen
  \bibfield  {author} {\bibinfo {author} {\bibfnamefont {S.}~\bibnamefont
  {Bandyopadhyay}}, \bibinfo {author} {\bibfnamefont {S.}~\bibnamefont
  {Chatterjee}}, \bibinfo {author} {\bibfnamefont {A.~K.}\ \bibnamefont
  {Dutta}}, \bibinfo {author} {\bibfnamefont {M.}~\bibnamefont {Karmakar}},
  \bibinfo {author} {\bibfnamefont {H.}~\bibnamefont {Rieger}},\ and\ \bibinfo
  {author} {\bibfnamefont {R.}~\bibnamefont {Paul}},\ }\bibfield  {title}
  {\bibinfo {title} {Ordering kinetics in the active {Ising} model},\ }\href
  {https://doi.org/10.1103/PhysRevE.109.064143} {\bibfield  {journal} {\bibinfo
   {journal} {Physical Review E}\ }\textbf {\bibinfo {volume} {109}},\ \bibinfo
  {pages} {064143} (\bibinfo {year} {2024})}\BibitemShut {NoStop}%
\bibitem [{\citenamefont {Vicsek}\ \emph {et~al.}(1995)\citenamefont {Vicsek},
  \citenamefont {Czirók}, \citenamefont {Ben-Jacob}, \citenamefont {Cohen},\
  and\ \citenamefont {Shochet}}]{vicsek_prl95}%
  \BibitemOpen
  \bibfield  {author} {\bibinfo {author} {\bibfnamefont {T.}~\bibnamefont
  {Vicsek}}, \bibinfo {author} {\bibfnamefont {A.}~\bibnamefont {Czirók}},
  \bibinfo {author} {\bibfnamefont {E.}~\bibnamefont {Ben-Jacob}}, \bibinfo
  {author} {\bibfnamefont {I.}~\bibnamefont {Cohen}},\ and\ \bibinfo {author}
  {\bibfnamefont {O.}~\bibnamefont {Shochet}},\ }\bibfield  {title} {\bibinfo
  {title} {Novel {Type} of {Phase} {Transition} in a {System} of
  {Self}-{Driven} {Particles}},\ }\href
  {https://doi.org/10.1103/PhysRevLett.75.1226} {\bibfield  {journal} {\bibinfo
   {journal} {Physical Review Letters}\ }\textbf {\bibinfo {volume} {75}},\
  \bibinfo {pages} {1226} (\bibinfo {year} {1995})}\BibitemShut {NoStop}%
\bibitem [{\citenamefont {Grégoire}\ and\ \citenamefont
  {Chaté}(2004)}]{gregoire_prl04}%
  \BibitemOpen
  \bibfield  {author} {\bibinfo {author} {\bibfnamefont {G.}~\bibnamefont
  {Grégoire}}\ and\ \bibinfo {author} {\bibfnamefont {H.}~\bibnamefont
  {Chaté}},\ }\bibfield  {title} {\bibinfo {title} {Onset of {Collective} and
  {Cohesive} {Motion}},\ }\href {https://doi.org/10.1103/PhysRevLett.92.025702}
  {\bibfield  {journal} {\bibinfo  {journal} {Physical Review Letters}\
  }\textbf {\bibinfo {volume} {92}},\ \bibinfo {pages} {025702} (\bibinfo
  {year} {2004})}\BibitemShut {NoStop}%
\bibitem [{\citenamefont {Chaté}\ \emph {et~al.}(2008)\citenamefont {Chaté},
  \citenamefont {Ginelli}, \citenamefont {Grégoire},\ and\ \citenamefont
  {Raynaud}}]{chate_pre08}%
  \BibitemOpen
  \bibfield  {author} {\bibinfo {author} {\bibfnamefont {H.}~\bibnamefont
  {Chaté}}, \bibinfo {author} {\bibfnamefont {F.}~\bibnamefont {Ginelli}},
  \bibinfo {author} {\bibfnamefont {G.}~\bibnamefont {Grégoire}},\ and\
  \bibinfo {author} {\bibfnamefont {F.}~\bibnamefont {Raynaud}},\ }\bibfield
  {title} {\bibinfo {title} {Collective motion of self-propelled particles
  interacting without cohesion},\ }\href
  {http://link.aps.org/doi/10.1103/PhysRevE.77.046113} {\bibfield  {journal}
  {\bibinfo  {journal} {Physical Review E}\ }\textbf {\bibinfo {volume} {77}},\
  \bibinfo {pages} {46113} (\bibinfo {year} {2008})}\BibitemShut {NoStop}%
\bibitem [{\citenamefont {Bertin}\ \emph {et~al.}(2006)\citenamefont {Bertin},
  \citenamefont {Droz},\ and\ \citenamefont {Gregoire}}]{bertin_pre06}%
  \BibitemOpen
  \bibfield  {author} {\bibinfo {author} {\bibfnamefont {E.}~\bibnamefont
  {Bertin}}, \bibinfo {author} {\bibfnamefont {M.}~\bibnamefont {Droz}},\ and\
  \bibinfo {author} {\bibfnamefont {G.}~\bibnamefont {Gregoire}},\ }\bibfield
  {title} {\bibinfo {title} {Boltzmann and hydrodynamic description for
  self-propelled particles},\ }\href
  {https://doi.org/10.1103/PhysRevE.74.022101} {\bibfield  {journal} {\bibinfo
  {journal} {Physical Review E}\ }\textbf {\bibinfo {volume} {74}},\ \bibinfo
  {pages} {022101} (\bibinfo {year} {2006})}\BibitemShut {NoStop}%
\bibitem [{\citenamefont {Nesbitt}\ \emph {et~al.}(2021)\citenamefont
  {Nesbitt}, \citenamefont {Pruessner},\ and\ \citenamefont
  {Lee}}]{nesbitt_njp21}%
  \BibitemOpen
  \bibfield  {author} {\bibinfo {author} {\bibfnamefont {D.}~\bibnamefont
  {Nesbitt}}, \bibinfo {author} {\bibfnamefont {G.}~\bibnamefont {Pruessner}},\
  and\ \bibinfo {author} {\bibfnamefont {C.~F.}\ \bibnamefont {Lee}},\
  }\bibfield  {title} {\bibinfo {title} {Uncovering novel phase transitions in
  dense dry polar active fluids using a lattice {Boltzmann} method},\ }\href
  {https://doi.org/10.1088/1367-2630/abd8c0} {\bibfield  {journal} {\bibinfo
  {journal} {New Journal of Physics}\ }\textbf {\bibinfo {volume} {23}},\
  \bibinfo {pages} {043047} (\bibinfo {year} {2021})}\BibitemShut {NoStop}%
\bibitem [{\citenamefont {Bertrand}\ and\ \citenamefont
  {Lee}(2022)}]{bertrand_prr22}%
  \BibitemOpen
  \bibfield  {author} {\bibinfo {author} {\bibfnamefont {T.}~\bibnamefont
  {Bertrand}}\ and\ \bibinfo {author} {\bibfnamefont {C.~F.}\ \bibnamefont
  {Lee}},\ }\bibfield  {title} {\bibinfo {title} {Diversity of phase
  transitions and phase separations in active fluids},\ }\href
  {https://doi.org/10.1103/PhysRevResearch.4.L022046} {\bibfield  {journal}
  {\bibinfo  {journal} {Physical Review Research}\ }\textbf {\bibinfo {volume}
  {4}},\ \bibinfo {pages} {L022046} (\bibinfo {year} {2022})}\BibitemShut
  {NoStop}%
\bibitem [{\citenamefont {Jentsch}\ and\ \citenamefont
  {Lee}(2023)}]{jentsch_prr23}%
  \BibitemOpen
  \bibfield  {author} {\bibinfo {author} {\bibfnamefont {P.}~\bibnamefont
  {Jentsch}}\ and\ \bibinfo {author} {\bibfnamefont {C.~F.}\ \bibnamefont
  {Lee}},\ }\bibfield  {title} {\bibinfo {title} {Critical phenomena in
  compressible polar active fluids: {Dynamical} and functional renormalization
  group studies},\ }\href {https://doi.org/10.1103/PhysRevResearch.5.023061}
  {\bibfield  {journal} {\bibinfo  {journal} {Physical Review Research}\
  }\textbf {\bibinfo {volume} {5}},\ \bibinfo {pages} {023061} (\bibinfo {year}
  {2023})}\BibitemShut {NoStop}%
\bibitem [{\citenamefont {Schnyder}\ \emph {et~al.}(2017)\citenamefont
  {Schnyder}, \citenamefont {Molina}, \citenamefont {Tanaka},\ and\
  \citenamefont {Yamamoto}}]{schnyder_scirep17}%
  \BibitemOpen
  \bibfield  {author} {\bibinfo {author} {\bibfnamefont {S.~K.}\ \bibnamefont
  {Schnyder}}, \bibinfo {author} {\bibfnamefont {J.~J.}\ \bibnamefont
  {Molina}}, \bibinfo {author} {\bibfnamefont {Y.}~\bibnamefont {Tanaka}},\
  and\ \bibinfo {author} {\bibfnamefont {R.}~\bibnamefont {Yamamoto}},\
  }\bibfield  {title} {\bibinfo {title} {Collective motion of cells crawling on
  a substrate: roles of cell shape and contact inhibition},\ }\href
  {https://doi.org/10.1038/s41598-017-05321-0} {\bibfield  {journal} {\bibinfo
  {journal} {Scientific Reports}\ }\textbf {\bibinfo {volume} {7}},\ \bibinfo
  {pages} {5163} (\bibinfo {year} {2017})}\BibitemShut {NoStop}%
\bibitem [{\citenamefont {Toner}\ and\ \citenamefont {Tu}(1995)}]{toner_prl95}%
  \BibitemOpen
  \bibfield  {author} {\bibinfo {author} {\bibfnamefont {J.}~\bibnamefont
  {Toner}}\ and\ \bibinfo {author} {\bibfnamefont {Y.}~\bibnamefont {Tu}},\
  }\bibfield  {title} {\bibinfo {title} {Long-{Range} {Order} in a
  {Two}-{Dimensional} {Dynamical} XY {Model}: {How} {Birds} {Fly}
  {Together}},\ }\href {https://doi.org/10.1103/PhysRevLett.75.4326} {\bibfield
   {journal} {\bibinfo  {journal} {Physical Review Letters}\ }\textbf {\bibinfo
  {volume} {75}},\ \bibinfo {pages} {4326} (\bibinfo {year} {1995})}\BibitemShut {NoStop}%
\bibitem [{\citenamefont {Toner}\ and\ \citenamefont {Tu}(1998)}]{toner_pre98}%
  \BibitemOpen
  \bibfield  {author} {\bibinfo {author} {\bibfnamefont {J.}~\bibnamefont
  {Toner}}\ and\ \bibinfo {author} {\bibfnamefont {Y.}~\bibnamefont {Tu}},\
  }\bibfield  {title} {\bibinfo {title} {Flocks, herds, and schools: {A}
  quantitative theory of flocking},\ }\href
  {https://doi.org/10.1103/PhysRevE.58.4828} {\bibfield  {journal} {\bibinfo
  {journal} {Physical Review E}\ }\textbf {\bibinfo {volume} {58}},\ \bibinfo
  {pages} {4828} (\bibinfo {year} {1998})}\BibitemShut {NoStop}%
\bibitem [{\citenamefont {Toner}(2012)}]{toner_pre12}%
  \BibitemOpen
  \bibfield  {author} {\bibinfo {author} {\bibfnamefont {J.}~\bibnamefont
  {Toner}},\ }\bibfield  {title} {\bibinfo {title} {Reanalysis of the
  hydrodynamic theory of fluid, polar-ordered flocks},\ }\href
  {https://doi.org/10.1103/PhysRevE.86.031918} {\bibfield  {journal} {\bibinfo
  {journal} {Physical Review E}\ }\textbf {\bibinfo {volume} {86}},\ \bibinfo
  {pages} {031918} (\bibinfo {year} {2012})}\BibitemShut {NoStop}%
\bibitem [{\citenamefont {Chen}\ \emph {et~al.}(2025)\citenamefont {Chen},
  \citenamefont {Jentsch}, \citenamefont {Lee}, \citenamefont {Maitra},
  \citenamefont {Ramaswamy},\ and\ \citenamefont {Toner}}]{chen_a25}%
  \BibitemOpen
  \bibfield  {author} {\bibinfo {author} {\bibfnamefont {L.}~\bibnamefont
  {Chen}}, \bibinfo {author} {\bibfnamefont {P.}~\bibnamefont {Jentsch}},
  \bibinfo {author} {\bibfnamefont {C.~F.}\ \bibnamefont {Lee}}, \bibinfo
  {author} {\bibfnamefont {A.}~\bibnamefont {Maitra}}, \bibinfo {author}
  {\bibfnamefont {S.}~\bibnamefont {Ramaswamy}},\ and\ \bibinfo {author}
  {\bibfnamefont {J.}~\bibnamefont {Toner}},\ }\href
  {https://doi.org/10.48550/arXiv.2503.17064} {\bibinfo {title} {The
  inconvenient truth about flocks}} (\bibinfo {year} {2025}),\ \bibinfo {note}
  {Eprint: arXiv:2503.17064}\BibitemShut {NoStop}%
\bibitem [{\citenamefont {Hohenberg}\ and\ \citenamefont
  {Halperin}(1977)}]{hohenberg_rmp77}%
  \BibitemOpen
  \bibfield  {author} {\bibinfo {author} {\bibfnamefont {P.~C.}\ \bibnamefont
  {Hohenberg}}\ and\ \bibinfo {author} {\bibfnamefont {B.~I.}\ \bibnamefont
  {Halperin}},\ }\bibfield  {title} {\bibinfo {title} {Theory of dynamic
  critical phenomena},\ }\href {https://doi.org/10.1103/RevModPhys.49.435}
  {\bibfield  {journal} {\bibinfo  {journal} {Reviews of Modern Physics}\
  }\textbf {\bibinfo {volume} {49}},\ \bibinfo {pages} {435} (\bibinfo {year}
  {1977})}\BibitemShut
  {NoStop}%
\bibitem [{\citenamefont {Agranov}\ \emph {et~al.}(2024)\citenamefont
  {Agranov}, \citenamefont {Jack}, \citenamefont {Cates},\ and\ \citenamefont
  {Fodor}}]{agranov_njp24}%
  \BibitemOpen
  \bibfield  {author} {\bibinfo {author} {\bibfnamefont {T.}~\bibnamefont
  {Agranov}}, \bibinfo {author} {\bibfnamefont {R.~L.}\ \bibnamefont {Jack}},
  \bibinfo {author} {\bibfnamefont {M.~E.}\ \bibnamefont {Cates}},\ and\
  \bibinfo {author} {\bibfnamefont {\'{E}.}~\bibnamefont {Fodor}},\ }\bibfield
  {title} {\bibinfo {title} {Thermodynamically consistent flocking: from
  discontinuous to continuous transitions},\ }\href
  {https://doi.org/10.1088/1367-2630/ad4dd6} {\bibfield  {journal} {\bibinfo
  {journal} {New Journal of Physics}\ }\textbf {\bibinfo {volume} {26}},\
  \bibinfo {pages} {063006} (\bibinfo {year} {2024})}\BibitemShut {NoStop}%
\bibitem [{\citenamefont {Bassler}\ and\ \citenamefont
  {Schmittmann}(1994)}]{bassler_prl94}%
  \BibitemOpen
  \bibfield  {author} {\bibinfo {author} {\bibfnamefont {K.~E.}\ \bibnamefont
  {Bassler}}\ and\ \bibinfo {author} {\bibfnamefont {B.}~\bibnamefont
  {Schmittmann}},\ }\bibfield  {title} {\bibinfo {title} {Critical {Dynamics}
  of {Nonconserved} {Ising}-{Like} {Systems}},\ }\href
  {https://doi.org/10.1103/PhysRevLett.73.3343} {\bibfield  {journal} {\bibinfo
   {journal} {Physical Review Letters}\ }\textbf {\bibinfo {volume} {73}},\
  \bibinfo {pages} {3343} (\bibinfo {year} {1994})}\BibitemShut {NoStop}%
  \bibitem [{\citenamefont {Kardar}(2007)}]{kardar_b07}%
  \BibitemOpen
  \bibfield  {author} {\bibinfo {author} {\bibfnamefont {M.}~\bibnamefont
  {Kardar}},\ }\href {https://doi.org/10.1017/CBO9780511815881} {\emph
  {\bibinfo {title} {Statistical {Physics} of {Fields}}}}  (\bibinfo  {publisher} {Cambridge University Press},\ \bibinfo
  {year} {2007})\BibitemShut {NoStop}%
\bibitem [{\citenamefont {Forster}\ \emph {et~al.}(1977)\citenamefont
  {Forster}, \citenamefont {Nelson},\ and\ \citenamefont
  {Stephen}}]{forster_pra77}%
  \BibitemOpen
  \bibfield  {author} {\bibinfo {author} {\bibfnamefont {D.}~\bibnamefont
  {Forster}}, \bibinfo {author} {\bibfnamefont {D.~R.}\ \bibnamefont
  {Nelson}},\ and\ \bibinfo {author} {\bibfnamefont {M.~J.}\ \bibnamefont
  {Stephen}},\ }\bibfield  {title} {\bibinfo {title} {Large-distance and
  long-time properties of a randomly stirred fluid},\ }\href
  {https://doi.org/10.1103/PhysRevA.16.732} {\bibfield  {journal} {\bibinfo
  {journal} {Physical Review A}\ }\textbf {\bibinfo {volume} {16}},\ \bibinfo
  {pages} {732} (\bibinfo {year} {1977})}\BibitemShut {NoStop}%
\bibitem [{\citenamefont {Toner}(2024)}]{toner_b24}%
  \BibitemOpen
  \bibfield  {author} {\bibinfo {author} {\bibfnamefont {J.}~\bibnamefont
  {Toner}},\ }\href {https://doi.org/10.1017/9781108993623} {\emph {\bibinfo
  {title} {The {Physics} of {Flocking}: {Birth}, {Death}, and {Flight} in
  {Active} {Matter}}}},\ \bibinfo {edition} {1st}\ ed.\ (\bibinfo  {publisher}
  {Cambridge University Press},\ \bibinfo {year} {2024})\BibitemShut {NoStop}%
\bibitem [{\citenamefont {Jentsch}\ and\ \citenamefont
  {Lee}(2024)}]{jentsch_prl24}%
  \BibitemOpen
  \bibfield  {author} {\bibinfo {author} {\bibfnamefont {P.}~\bibnamefont
  {Jentsch}}\ and\ \bibinfo {author} {\bibfnamefont {C.~F.}\ \bibnamefont
  {Lee}},\ }\bibfield  {title} {\bibinfo {title} {New {Universality} {Class}
  {Describes} {Vicsek}’s {Flocking} {Phase} in {Physical} {Dimensions}},\
  }\href {https://doi.org/10.1103/PhysRevLett.133.128301} {\bibfield  {journal}
  {\bibinfo  {journal} {Physical Review Letters}\ }\textbf {\bibinfo {volume}
  {133}},\ \bibinfo {pages} {128301} (\bibinfo {year} {2024})}\BibitemShut
  {NoStop}%
  \bibitem [{\citenamefont {Hwa}\ and\ \citenamefont {Kardar}(1989)}]{hwa_prl89}%
  \BibitemOpen
  \bibfield  {author} {\bibinfo {author} {\bibfnamefont {T.}~\bibnamefont
  {Hwa}}\ and\ \bibinfo {author} {\bibfnamefont {M.}~\bibnamefont {Kardar}},\
  }\bibfield  {title} {\bibinfo {title} {Dissipative transport in open systems:
  {An} investigation of self-organized criticality},\ }\href
  {http://link.aps.org/doi/10.1103/PhysRevLett.62.1813} {\bibfield  {journal}
  {\bibinfo  {journal} {Physical Review Letters}\ }\textbf {\bibinfo {volume}
  {62}},\ \bibinfo {pages} {1813} (\bibinfo {year} {1989})}\BibitemShut {NoStop}%
\bibitem [{\citenamefont {Hwa}\ and\ \citenamefont {Kardar}(1992)}]{hwa_pra92}%
  \BibitemOpen
  \bibfield  {author} {\bibinfo {author} {\bibfnamefont {T.}~\bibnamefont
  {Hwa}}\ and\ \bibinfo {author} {\bibfnamefont {M.}~\bibnamefont {Kardar}},\
  }\bibfield  {title} {\bibinfo {title} {Avalanches, hydrodynamics, and
  discharge events in models of sandpiles},\ }\href
  {https://doi.org/10.1103/PhysRevA.45.7002} {\bibfield  {journal} {\bibinfo
  {journal} {Physical Review A}\ }\textbf {\bibinfo {volume} {45}},\ \bibinfo
  {pages} {7002} (\bibinfo {year} {1992})}\BibitemShut {NoStop}%
  \bibitem{dupuis_physrep21}
  N. Dupuis, L. Canet, A. Eichhorn, W. Metzner, J.M. Pawlowski, M. Tissier, and N. Wschebor, The nonperturbative functional renormalization group and its applications, 
  \href {https://doi.org/10.1016/j.physrep.2021.01.001} {\bibfield  {journal}
  {\bibinfo  {journal} {Physics Reports}\ }\textbf {\bibinfo {volume}
  {910}},\ \bibinfo {pages} {1} (\bibinfo {year} {2021})}\BibitemShut
  {NoStop}%
  \bibitem{SM}
See Supplemental Material at ... for further
analytical details.
\end{thebibliography}

\begin{thebibliography}{7}%
\makeatletter
\providecommand \@ifxundefined [1]{%
 \@ifx{#1\undefined}
}%
\providecommand \@ifnum [1]{%
 \ifnum #1\expandafter \@firstoftwo
 \else \expandafter \@secondoftwo
 \fi
}%
\providecommand \@ifx [1]{%
 \ifx #1\expandafter \@firstoftwo
 \else \expandafter \@secondoftwo
 \fi
}%
\providecommand \natexlab [1]{#1}%
\providecommand \enquote  [1]{``#1''}%
\providecommand \bibnamefont  [1]{#1}%
\providecommand \bibfnamefont [1]{#1}%
\providecommand \citenamefont [1]{#1}%
\providecommand \href@noop [0]{\@secondoftwo}%
\providecommand \href [0]{\begingroup \@sanitize@url \@href}%
\providecommand \@href[1]{\@@startlink{#1}\@@href}%
\providecommand \@@href[1]{\endgroup#1\@@endlink}%
\providecommand \@sanitize@url [0]{\catcode `\\12\catcode `\$12\catcode
  `\&12\catcode `\#12\catcode `\^12\catcode `\_12\catcode `\%12\relax}%
\providecommand \@@startlink[1]{}%
\providecommand \@@endlink[0]{}%
\providecommand \url  [0]{\begingroup\@sanitize@url \@url }%
\providecommand \@url [1]{\endgroup\@href {#1}{\urlprefix }}%
\providecommand \urlprefix  [0]{URL }%
\providecommand \Eprint [0]{\href }%
\providecommand \doibase [0]{https://doi.org/}%
\providecommand \selectlanguage [0]{\@gobble}%
\providecommand \bibinfo  [0]{\@secondoftwo}%
\providecommand \bibfield  [0]{\@secondoftwo}%
\providecommand \translation [1]{[#1]}%
\providecommand \BibitemOpen [0]{}%
\providecommand \bibitemStop [0]{}%
\providecommand \bibitemNoStop [0]{.\EOS\space}%
\providecommand \EOS [0]{\spacefactor3000\relax}%
\providecommand \BibitemShut  [1]{\csname bibitem#1\endcsname}%
\let\auto@bib@innerbib\@empty
\bibitem [{\citenamefont {Toner}\ and\ \citenamefont {Tu}(1995)}]{Stoner_prl95}%
  \BibitemOpen
  \bibfield  {author} {\bibinfo {author} {\bibfnamefont {J.}~\bibnamefont
  {Toner}}\ and\ \bibinfo {author} {\bibfnamefont {Y.}~\bibnamefont {Tu}},\
  }\bibfield  {title} {\bibinfo {title} {Long-{Range} {Order} in a
  {Two}-{Dimensional} {Dynamical} XY {Model}: {How} {Birds} {Fly}
  {Together}},\ }\href {https://doi.org/10.1103/PhysRevLett.75.4326} {\bibfield
   {journal} {\bibinfo  {journal} {Physical Review Letters}\ }\textbf {\bibinfo
  {volume} {75}},\ \bibinfo {pages} {4326} (\bibinfo {year} {1995})}
  \BibitemShut {NoStop}%
\bibitem [{\citenamefont {Toner}\ and\ \citenamefont {Tu}(1998)}]{Stoner_pre98}%
  \BibitemOpen
  \bibfield  {author} {\bibinfo {author} {\bibfnamefont {J.}~\bibnamefont
  {Toner}}\ and\ \bibinfo {author} {\bibfnamefont {Y.}~\bibnamefont {Tu}},\
  }\bibfield  {title} {\bibinfo {title} {Flocks, herds, and schools: {A}
  quantitative theory of flocking},\ }\href
  {https://doi.org/10.1103/PhysRevE.58.4828} {\bibfield  {journal} {\bibinfo
  {journal} {Physical Review E}\ }\textbf {\bibinfo {volume} {58}},\ \bibinfo
  {pages} {4828} (\bibinfo {year} {1998})}\BibitemShut {NoStop}%
\bibitem [{\citenamefont {Toner}(2012)}]{Stoner_pre12}%
  \BibitemOpen
  \bibfield  {author} {\bibinfo {author} {\bibfnamefont {J.}~\bibnamefont
  {Toner}},\ }\bibfield  {title} {\bibinfo {title} {Reanalysis of the
  hydrodynamic theory of fluid, polar-ordered flocks},\ }\href
  {https://doi.org/10.1103/PhysRevE.86.031918} {\bibfield  {journal} {\bibinfo
  {journal} {Physical Review E}\ }\textbf {\bibinfo {volume} {86}},\ \bibinfo
  {pages} {031918} (\bibinfo {year} {2012})}\BibitemShut {NoStop}%
\bibitem [{\citenamefont {Chen}\ \emph {et~al.}(2025)\citenamefont {Chen},
  \citenamefont {Jentsch}, \citenamefont {Lee}, \citenamefont {Maitra},
  \citenamefont {Ramaswamy},\ and\ \citenamefont {Toner}}]{Schen_a25}%
  \BibitemOpen
  \bibfield  {author} {\bibinfo {author} {\bibfnamefont {L.}~\bibnamefont
  {Chen}}, \bibinfo {author} {\bibfnamefont {P.}~\bibnamefont {Jentsch}},
  \bibinfo {author} {\bibfnamefont {C.~F.}\ \bibnamefont {Lee}}, \bibinfo
  {author} {\bibfnamefont {A.}~\bibnamefont {Maitra}}, \bibinfo {author}
  {\bibfnamefont {S.}~\bibnamefont {Ramaswamy}},\ and\ \bibinfo {author}
  {\bibfnamefont {J.}~\bibnamefont {Toner}},\ }\href
  {https://doi.org/10.48550/arXiv.2503.17064} {\bibinfo {title} {The
  inconvenient truth about flocks}} (\bibinfo {year} {2025}),\ \bibinfo {note}
  {Eprint: arXiv:2503.17064}\BibitemShut {NoStop}%
\bibitem [{\citenamefont {Nesbitt}\ \emph {et~al.}(2021)\citenamefont
  {Nesbitt}, \citenamefont {Pruessner},\ and\ \citenamefont
  {Lee}}]{Snesbitt_njp21}%
  \BibitemOpen
  \bibfield  {author} {\bibinfo {author} {\bibfnamefont {D.}~\bibnamefont
  {Nesbitt}}, \bibinfo {author} {\bibfnamefont {G.}~\bibnamefont {Pruessner}},\
  and\ \bibinfo {author} {\bibfnamefont {C.~F.}\ \bibnamefont {Lee}},\
  }\bibfield  {title} {\bibinfo {title} {Uncovering novel phase transitions in
  dense dry polar active fluids using a lattice {Boltzmann} method},\ }\href
  {https://doi.org/10.1088/1367-2630/abd8c0} {\bibfield  {journal} {\bibinfo
  {journal} {New Journal of Physics}\ }\textbf {\bibinfo {volume} {23}},\
  \bibinfo {pages} {043047} (\bibinfo {year} {2021})}\BibitemShut {NoStop}%
\bibitem [{\citenamefont {Bertrand}\ and\ \citenamefont
  {Lee}(2022)}]{Sbertrand_prr22}%
  \BibitemOpen
  \bibfield  {author} {\bibinfo {author} {\bibfnamefont {T.}~\bibnamefont
  {Bertrand}}\ and\ \bibinfo {author} {\bibfnamefont {C.~F.}\ \bibnamefont
  {Lee}},\ }\bibfield  {title} {\bibinfo {title} {Diversity of phase
  transitions and phase separations in active fluids},\ }\href
  {https://doi.org/10.1103/PhysRevResearch.4.L022046} {\bibfield  {journal}
  {\bibinfo  {journal} {Physical Review Research}\ }\textbf {\bibinfo {volume}
  {4}},\ \bibinfo {pages} {L022046} (\bibinfo {year} {2022})}\BibitemShut
  {NoStop}%
\bibitem [{\citenamefont {Kardar}(2007)}]{Skardar_b07}%
  \BibitemOpen
  \bibfield  {author} {\bibinfo {author} {\bibfnamefont {M.}~\bibnamefont
  {Kardar}},\ }\href {https://doi.org/10.1017/CBO9780511815881} {\emph
  {\bibinfo {title} {Statistical {Physics} of {Fields}}}} (\bibinfo  {publisher} {Cambridge University Press},\ \bibinfo
  {year} {2007})\BibitemShut {NoStop}%
\end{thebibliography}
\end{document}